\documentclass[prx,aps, twocolumn, showpacs,groupedaddress,10pt,nofootinbib]{revtex4-2} 

\usepackage[T1]{fontenc}
\usepackage{bm,dcolumn}
\usepackage{amsmath,amsfonts,amssymb,amsthm,amscd} 
\usepackage{graphicx}
\usepackage{verbatim}
\usepackage[section]{placeins}
\usepackage{color}
\usepackage{soul}
\usepackage{xparse}
\usepackage{physics}
\usepackage{siunitx}
\usepackage[colorlinks,linkcolor=blue,citecolor=blue,anchorcolor=blue]{hyperref}

\graphicspath{ {./Figures/} }
\usepackage{pdfpages} 
\usepackage{pgffor}

\makeatletter
\AtBeginDocument{\let\LS@rot\@undefined}
\makeatother

\begin{document}

\title{Field-deployable Quantum Memory for Quantum Networking}

\author{Yang Wang\normalfont\textsuperscript{$\dagger$,}}
\email{Corresponding author. yang@quconn.com}

\author{Alexander N. Craddock}
\thanks{These two authors contributed equally}
\author{Rourke Sekelsky}
\author{Mael Flament}
\author{Mehdi Namazi}
\affiliation{Qunnect Inc. 141 Flushing Ave, Ste 1110, Brooklyn, NY 11205-1005}

\date{\today}

\begin{abstract}
High-performance quantum memories are an essential component for regulating temporal events in quantum networks. As a component in quantum-repeaters, they have the potential to support the distribution of entanglement beyond the physical limitations of fiber loss.  This will enable key applications such as quantum key distribution, network-enhanced quantum sensing, and distributed quantum computing. Here, we present a quantum memory engineered to meet real-world deployment and scaling challenges.  The memory technology utilizes a warm rubidium vapor as the storage medium, and operates at room temperature, without the need for vacuum- and/or cryogenic- support. 
We demonstrate performance specifications of high-fidelity retrieval (95\%) and low operation error $(10^{-2})$ at a storage time of 160 $\mu s$ for single-photon level quantum memory operations. We further show a substantially improved storage time (with classical-level light) of up to 1ms by suppressing atomic diffusions.  
The device is housed in an enclosure with a standard 2U rackmount form factor, and can robustly operate on a day scale in a noisy environment. This result marks an important step toward implementing quantum networks in the field. 
\end{abstract}

\maketitle

\section{Introduction}

Quantum information science holds the key to unprecedented information integration, processing, and distribution capabilities \cite{Simon2017,Wehner2018,Pirandola2016}. Similar to the evolution of the classical internet, where our modern-day applications were unimaginable in the earliest demonstrations of networking. The ``quantum internet'' \cite{Kimble2008,Awschalom2021} has the potential to enable revolutionary applications such as unconditionally-secure secret key exchange \cite{Gisin2002,Ekert2014,Scarani2009,Xu2020}, distributed quantum computation \cite{Cirac1999,Cacciapuoti2020}, enhanced quantum metrology \cite{Komar2014,Gottesman2021}, and tests of fundamental physics \cite{Jasminder2021,Guo2020}. 
Given the expansive telecommunications fiber infrastructure, and ease of information processing with photons, fiber networks are a prime candidate for hosting a global network. However, despite decades of optimization for digital communication, the transmission loss (> 0.2 dB/km) still presents considerable challenges to quantum networking since the quantum states cannot be copied or amplified (no-cloning theorem \cite{Dieks1982,Wootters1982}). This constraint fundamentally limits the distances over which remote parties can be directly networked at a reasonable rate. The concept of a quantum repeater was theorized \cite{Ekert1991,Briegel1998,Duan2001,Sangouard2011} to overcome this challenge, where one divides the long channel into many elementary links and uses pair-wise entanglement swapping to distribute entanglement between two remote parties.
%where entanglement is distributed over distances greater than the fiber loss by performing entanglement swapping. 

For most common quantum repeater schemes, a quantum memory (QM) is a core enabling device, which allows single photons to be temporarily stored on a long-lived matter state and retrieved on-demand \cite{Hedges2010, Julsgaard2004}, enabling the storage of entanglement generated over elementary links. It thus provides an edge in entanglement swapping (since the elementary entanglement process is probabilistic) \cite{Lukin2003,Hammerer2010}, significantly increases the entanglement distribution rate, serving as a foundation for global-scale quantum networks.

\begin{figure*}[t!]
\includegraphics[width=17.5cm]{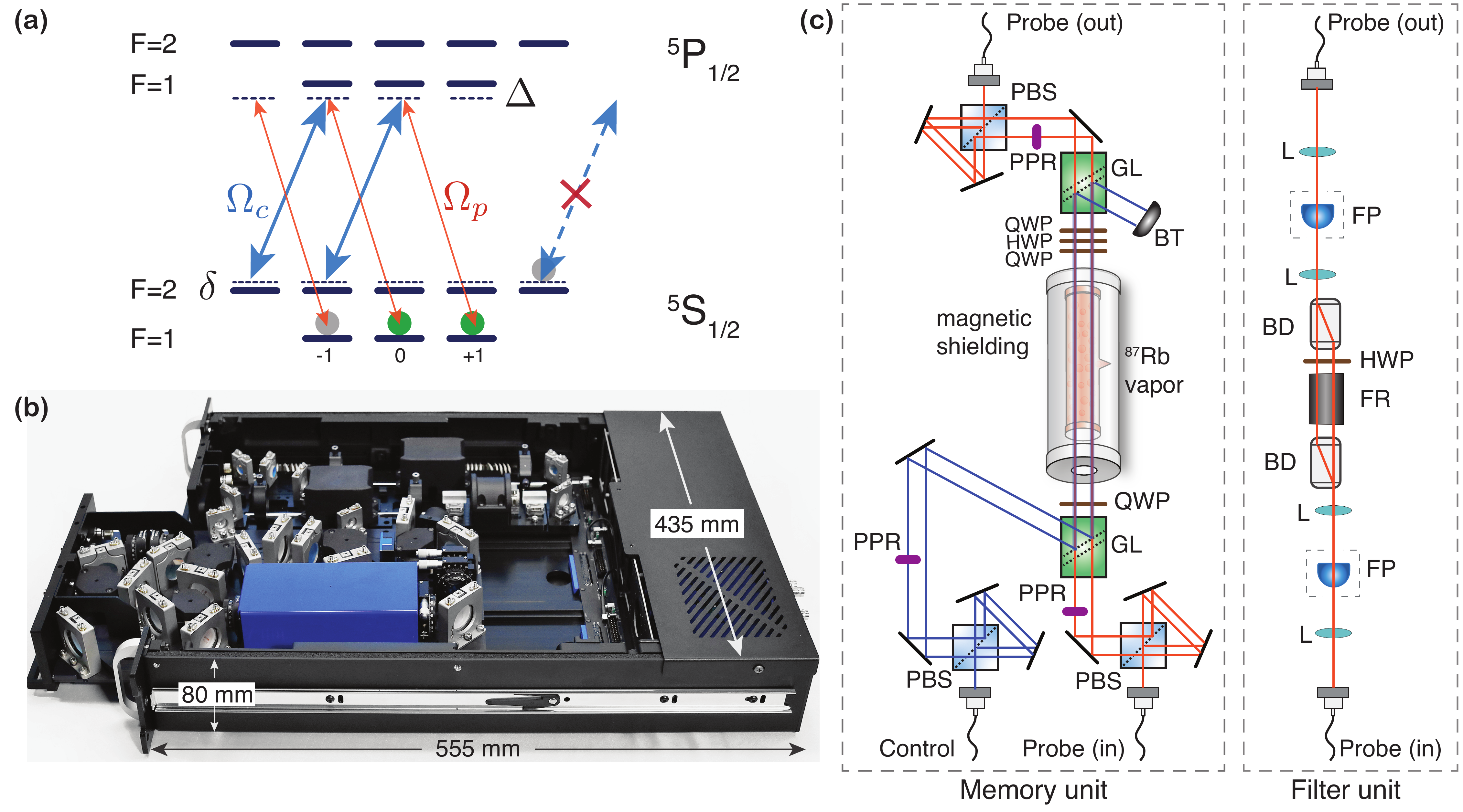}
\caption{Operating principle, implementation and design of the quantum memory.
\textbf{(a)} Energy levels and laser coupling schemes for realizing our QM based on electromagnetic induced transparency (EIT) with $^{87}$Rb. A strong $\sigma^+$-polarized control field (blue) and a weak $\sigma^-$-polarized probe field (orange) couple a $\Lambda$-system composed of F=2 and F=1 atoms. The single-photon detuning $\Delta$ and two-photon detuning $\delta$ can be independently controlled.
\textbf{(b)} A photo of the device with the memory unit (see (c)) slid out halfway. An enclosure (cap removed), with a standard 19-inch 2U rackmount form-factor, hosts the memory unit, the filter unit, and supporting electronic systems.
\textbf{(c)} Schematics of QM's optical layout, divided into two physical modules based on the relevant functions. In the memory unit, a photon carrying quantum information in its polarization degree of freedom is mapped onto a collective spin state of a warm vapor of $^{87}$Rb atoms and later retrieved using a strong control field. In the filter unit, the recalled photon is spectrally filtered with two high-finesse etalons to remove excess noise photons to achieve high fidelity. PBS: polarization beamsplitter; PPR: polarization plane rotator; GL: Glan-Laser prism; QWP: quarter waveplate; HWP: half waveplate; BT: beam trap; L: lens; FP: Fabry-P\'erot cavity; BD: beam displacer; FR: Faraday rotator.} 
\label{fig:1}
\end{figure*}

Here, we report the performance of our QM, a self-contained, rackmount solution optimized towards field deployment. 
We demonstrate a simple, robust, and scalable light-matter interface, based on a warm atomic vapor, which is versatile for a broad range of quantum optics applications. 
We push the QM performance while maintaining technical practicality. 
We package a tabletop quantum optics experiment with supporting electronic and mechanical systems into a compact module that fits a standard 2U 19-inch rack system (Fig. \ref{fig:1}).
Finally, we demonstrate the robustness of the device in noisy environments (stray electromagnetic fields, mechanical vibrations, and temperature drifts). 
%We explore the performance of the system under conditions consistent with technical practicality. We optimize the optical, electrical and mechanical design, and the device fits a standard 2U 19-inch rack system (Fig. \ref{fig:1}). 

Our device boasts high fidelity retrieval (95\%) at 5\% storage efficiency and a long storage time (up to 1 ms). This is consistent with the performance of well-controlled lab-based setups in a rackmount form factor. The acceptance bandwidth (typically a few MHz) is readily compatible with other atomic-based technologies, such as neutral atom quantum computers \cite{Graham2022,QuEra}, trapped ions \cite{Monroe2014} and NV centers \cite{NVcenter}. In addition, the wavelength (795 nm) permits high-fidelity one-step frequency conversion to telecom bands (e.g., O-/C-) using nonlinear crystals \cite{Hannegan2021} or atomic media \cite{Radnaev2010}. 
These features put our QM at an advantageous position in rapid field deployment and connecting different quantum nodes at the metropolitan scale ($\sim$ 100 km). We have identified promising technical routes to upscale the device performances further, extending the network distance to inter-city and ultimately global level.

%-------------------------------------------FIGURE 1-------------------------
%%-------------------------------------------FIGURE 1------------------

The manuscript is structured as follows. In Sec. \ref{FDQM} we discuss the need and requirements for a field-deployable quantum memory. Sec. \ref{TECH} details our research and engineering efforts to realize a compact and robust QM with high performance. The performance of our QM in its final form is benchmarked in Sec. \ref{RESULTS}. We conclude in Sec. \ref{OUTLOOK} and discuss future research directions.

%-----------------------------------------------------------------------------------------
\section{Field-deployable Quantum Memories\label{FDQM}}

QMs are devices capable of storing and retrieving photonic qubits on-demand \cite{Simon2010}. This primary function can be characterized by three parameters: fidelity $\mathcal{F}$, which evaluates the degradation of the input quantum state of the photons; storage efficiency $\eta$, defined as the probability of storing and retrieving photons; and storage time $T$, which assesses the time it takes for the storage efficiency to decay significantly. For a QM in a quantum network, these parameters jointly determine the entanglement distribution rate and scaling over distances \cite{Sangouard2011}, and can thus be considered as the ``quantum performance''. However, the real potential and applications of QMs are only unleashed when they can be successfully integrated into the existing telecommunication infrastructure, where a further set of metrics apply: fiber-hub compatibility, which prefers devices low in size, weight, and power consumption (SWaP); robustness against environmental noises (electromagnetic, thermal, mechanical); and scalability for mass deployment. We consider these requirements as the ``hardware performance''. A field-deployable quantum memory serving a large-scale quantum network needs to satisfy both criteria sets \cite{NASA2019}.

Different physical systems have been researched to realize a QM, such as ensembles of atoms \cite{Yang2016}, single atoms and ions \cite{Bimbard2014,Vittorini2014}, rare-earth-doped crystals \cite{Riedmatten2008,Ma2021_rareearth}, defects in diamonds \cite{Maximilian2021,Rozp2019} (and similar structures\cite{Bhaskar2020}).
The \emph{quantum performance} has been brought to high levels in various systems. For example, 
a one-hour storage time of classical optical pulses has been reported in a solid-state system \cite{Ma2021_rareearth} (although such memory has not been demonstrated to work in the quantum regime). 92 \% efficiency can be achieved in a cold atom ensemble \cite{Hsiao2018}; and >99 \% fidelity $\mathcal{F}$ has been demonstrated \cite{Wang2019_RAMAN,Saglamyurek2021}. However, from the perspective of \emph{hardware performance}, the resources and technology required to achieve such high quantum performance, such as cryogenic cooling, ultra-high vacuum system, and sophisticated laser cooling and trapping schemes (e.g, optical lattices \cite{Yang2016}),  
are prohibitive to be made field-deployable and scalable.

Here, we focus on warm atomic-vapor systems, which have shown promising results in both quantum and hardware performances. Warm vapor systems provide a simple and robust physical platform that operates at, or above, room temperature (cryogenic-free and vacuum-free), with access to the rich internal properties of atoms such as fine/hyperfine structures for useful light-matter coupling mechanisms. 
Therefore, warm-vapor systems are prime candidates for practical quantum optics applications that require robustness, low-SWaP, and high scalability. Successful products have already been developed with widespread deployment, such as chip-scale atomic clocks \cite{Knappe2004}, atomic magnetometers \cite{Limes2020}, atomic interference gravimeters \cite{Wu2019}, etc. 
There have been a plethora of research activities on warm-vapor based QM, and very promising memory performances have been demonstrated, such as 87\% efficiency \cite{Hosseini2011_NatComm}, 1 second storage time \cite{Katz2018} (with classical optical pulses), and close-to-unity fidelity \cite{ORCA,FLAME}. 

Diving into the details of warm-vapor-based QMs, there exist several types of storage mechanisms: electromagnetic induced transparency (EIT) with near-resonant fields \cite{Fleischhauer2000,Eisaman2005,Connor2015,Ma2017},  far-off-resonant Raman transitions \cite{Reim2011}, and gradient echo memory (GEM) \cite{Hosseini2011,Hosseini2011_NatComm}. Although they each work under very distinct conditions, the inherent physics and performance can be obtained in a similar way by modeling a three-level system \cite{Gorshkov2007,Hammerer2010}.
While the high quantum performance of the individual parameters of \{$\eta, \mathcal{F}, T$\} have been demonstrated in these systems \cite{Hosseini2011_NatComm,Katz2018,ORCA,FLAME}, simultaneous high performance of all three parameters has not been realized. This often-overlooked gap in the research space is due to the conflicting physical processes which arise during the optimization of the experimental system.  
For example, getting high fidelity requires exquisite filtering of the noise photons due to the strong control field (an external optical field modifies the optical response of the atomic medium). This can be achieved by angling the control field in a $\Lambda$-system \cite{Zhao2009} or having two fields counter-propagating in a ladder-system \cite{FLAME,ORCA}. However, the storage time is severely limited in both of those approaches, due to either the short spin-wave (SW) wavelength ($\Lambda$-system) or rapid dephasing due to the excited state (ladder-system). However, since this is often due to the details of a specific implementation rather than fundamental physics, accessing the complete \{$\eta, \mathcal{F}, T$\} parameter space is possible.

Here, we will describe a field-deployable QM using EIT for storage in a warm atomic vapor. We demonstrate optimization of the three critical performance parameters in one experimental setup, which is then packaged into a rack-mounted modular system fitting standard fiber hub environments. An innovative engineering design provides robust performance and stability.  The combined high quantum and hardware performances makes our QM the backbone of a memory-backed quantum repeater  for real-world applications.

%--------------------------------------------------------------------

\section{technical implementation\label{TECH}}

In this section, we describe our research and engineering efforts towards a field-deployable, low-SWaP quantum memory optimized for quantum networking.

\subsection{Quantum Performance}

\subsubsection{operating principle\label{principle}}
Our quantum memory, shown in Fig. \ref{fig:1}, is designed to store photonic polarization qubits that are near-resonant with the $F=1$ to $F'=1$ transition of the rubidium D1 line.
As is typical for an EIT based quantum memory\cite{Connor2015,Namazi2017},
a strong control field $\Omega_c$, addressing the $F=2$ to $F'=1$ transition of the rubidium D1 line, opens an EIT window under which a probe photon, $\Omega_p$, on two-photon resonance can propagate. 
By turning $\Omega_c$ off when the probe photon is within the atomic medium, we coherently map it onto a hyperfine $F=1,2$ SW on Rb atoms.
Restoring $\Omega_c$ back on at a later time maps the SW back into a propagating photon.

In order to store polarization qubits of arbitrary polarization, we separate the two orthogonal polarization modes ($\ket{H}$ and $\ket{V}$) into two spatial ``rails''. \cite{Connor2015,Namazi2017}
These rails propagate through the Rb vapor in parallel, each with their own overlapping control field.
After exiting the vapor cell, the rails are re-combined into a single spatial mode to recover the initial polarization qubit.

\subsubsection{storage time $T$}

%-------------------------------------------FIGURE 2--------------
\begin{figure}[!t]
\includegraphics[width=8.6 cm]{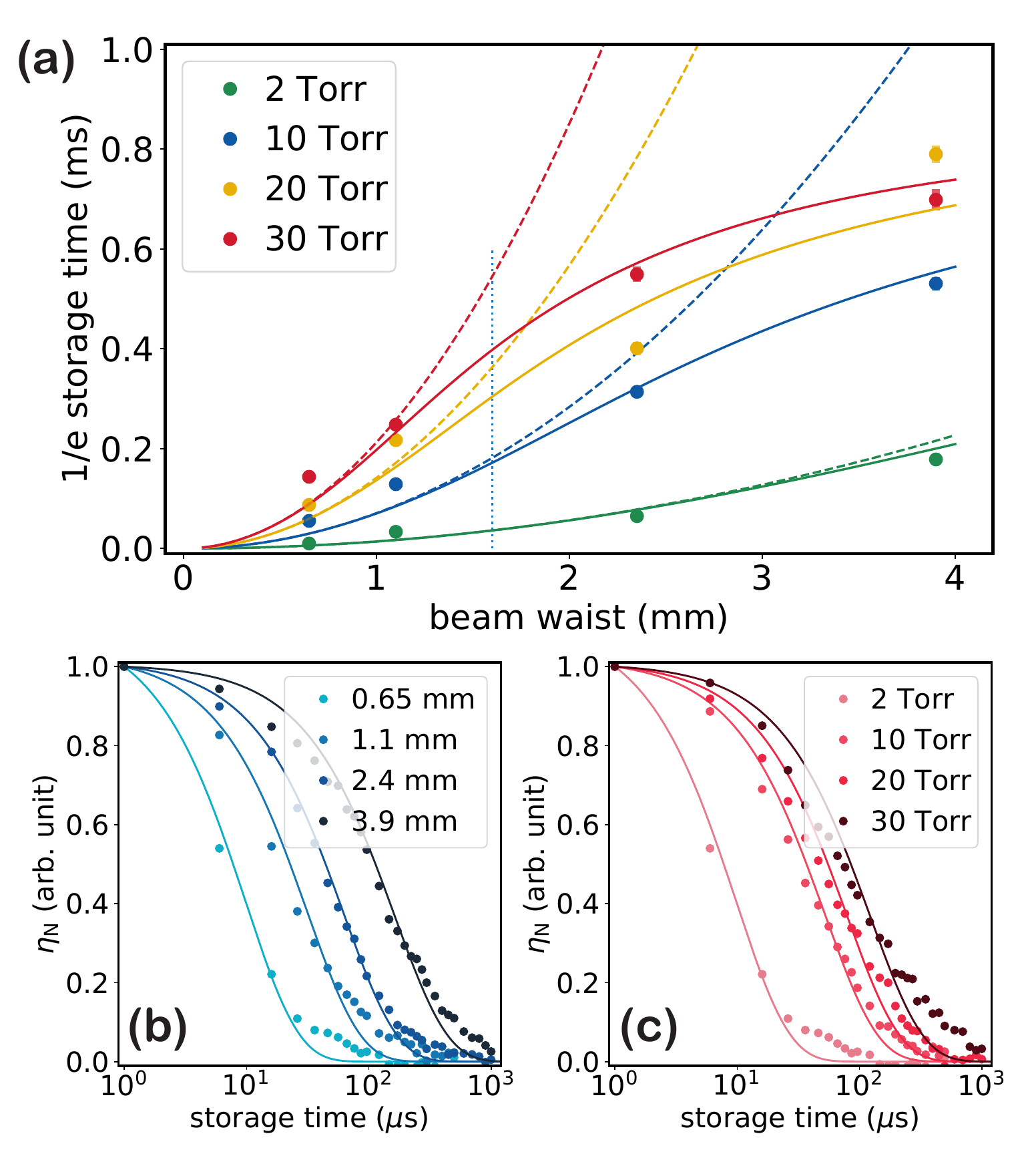}
\caption{Storage time study.
\textbf{(a)} We plot $T$ (defined as 1/e decay time of the storage efficiency) as a function of the probe field beam waist under different buffer gas (Ne) pressures. The solid lines are from a collective fit over all data points with two free parameters: Rb atom diffusive constant (D$_0$) and longitudinal magnetic field gradient. The dashed lines represent the inferred storage time without a magnetic field gradient. The actual beam waist in our final QM product is 1.6 mm (vertical dotted line). The experiment is performed with weak laser pulses as the probe field (< 1 $\mu $W).   
\textbf{(b)} Semi-log plot of the normalized memory efficiency $\eta$ as a function of time for an Rb vapor with 2 Torr Ne buffer gas with different probe beam waists. Solid lines are exponential fits. 
\textbf{(c)} Semi-log plot of the normalized memory efficiency $\eta$ as a function of time for an Rb vapor with 0.65 mm probe beam waist under different buffer gas (Ne) pressures. Solid lines are exponential fits. }
\label{fig:2}
\end{figure}
%%-------------------------------------------FIGURE 2----------------------

In a quantum repeater network, the storage time of a QM heavily limits the total network length \cite{Sangouard2011}, 
%the storage time of a QM defines the length of time a link of the network can store the photon while waiting for the sequential transaction \cite{Sangouard2011} 
and is therefore a critical metric of memory performance. In our QM, photons are stored in the form of an SW on Rb atoms. The primary decoherence mechanism of this SW is the atomic free motion and inhomogeneous magnetic fields \cite{Novikova2012}. Here we systematically study the decoherence mechanism and devise corresponding strategies to achieve a $1/e$ storage time of $ \sim$ 0.8 ms.

Magnetic fields cause dephasing of the quantum state of atoms by shifting their Zeeman levels. 
To suppress the ambient magnetic field,
we use a magnetic shield made of $\mu$-metal material. The Zeeman splitting of Rb atoms 
connects the characteristic dephasing time with the magnetic field. %For example, a 10-ms scale storage time would require a maximum field change along the SW to be < 7$\times$10$^{-6}$ mT, which corresponds to > 40 dB suppression. 
A detailed calculation can be found in Appendix. \ref{dephase}. The design of magnetic shielding has been extensively studied elsewhere \cite{Mager1970,Altarev2015}. Here, due to space limitations (height under 2U) and optical access requirements, we design a cylindrical magnetic shield of two concentric layers with 18-mm diameter center holes on the end caps for optical access. The ratio between the diameters is optimized to maximize the shielding effect in the transverse direction. \cite{Mager1970}
We construct a Helmholtz coil to measure the shielding performance. However, within our testing capability, we can only set a lower bound of >30 dB in each directions, which allows us to achieve > 1 ms storage time (when other decoherence mechanisms are not the limiting factor).

Since the quantum state of the photon is encoded on the atomic vapor with finite size, free atomic motion causes decoherence.  
By co-propagating $\Omega_c$ and $\Omega_p$ we achieve a maximum SW wavelength of $\lambda_{\rm{SW}}$ = 4.3 cm in the longitudinal direction. Transversely, the SW is defined by the probe field profile. 
Since the typical beam size (a few millimeters) is $\ll\lambda_{\rm{SW}}$, the diffusion of atoms in the radial direction is the primary dephasing mechanism, while the longitudinal, dephasing is secondary.   
By adding buffer gas (Ne) to the system, this diffusion is significantly slowed, with a diffusion constant $D$ determined by the buffer gas pressure according to $D=D_0 \frac{P_0}{P}$ ($P_0$ and $D_0$ defined at 760 Torr). 
We model the decay of the storage efficiency due to the diffusion process in Appendix \ref{dephase}, which predicts the dependence of $T$ on buffer gas pressure $P_{\rm{Ne}}$ and beam size $w$.

In Fig. \ref{fig:2} (a) we show the measured $T$, defined as $1/e$ decay time of the retrieved signal, as a function of $P_{\rm{Ne}}$ and $w$, while in (b) and (c) we show the raw dephasing data with either fixed pressure (2 Torr) or beam size (0.65 mm). In those cases ((b) and (c)) $T$ scales quadratically with $w$ and linearly with $P_{\rm{Ne}}$, because the radial diffusion is the dominant dephasing mechanism. In (a) we plot the theoretical prediction of $T$ when such dephasing is the only source (dashed color lines), which fits the data in the low $P_{\rm{Ne}}$ and small $w$ regime. However, in the high $P_{\rm{Ne}}$ and large $w$ limit, we see the divergence between the simple theory and the experimental results, which suggests that the $T$ is now bottlenecked by some mechanism other than the atomic diffusion. We note that this long-storage time performance is only achieved when the system is well degaussed, and we measure slightly different results after each degaussing process. 

We attribute this additional dephasing mechanism to a slight magnetic field gradient in the longitudinal direction. The finite optical access inevitably compromises the shielding capability along this direction, and the long SW makes the memory more sensitive to small gradients. This is in contrast to a cold-atom based approach, where the small size of the atomic cloud (e.g., 100 $\mu$m \cite{Hsiao2018}) makes it possible to actively cancel the residual field precisely (and gradients).  
We take this magnetic dephasing into account and plot the predictions in Fig. \ref{fig:2} (a) as the solid color lines, which matches with experimental results well. %From the fitting we get a diffusion constant D$_0$ = 1.8 cm$^2$/s and residual magnetic field change of 9$\times 10^{-5}$ mT along the SW. 
The inferred magnetic field change from the fitting suggests > 30 dB suppression, which matches with the lower bound of our measurement.

Despite the imperfect magnetic shielding, we have achieved $T$ = 0.8 ms, which could support an metropolitan scale quantum network. 
We anticipate that better magnetic shielding, particularly in the longitudinal direction, should improve the storage time to a few ms levels while using modest beam sizes and buffer gas pressures (high pressure has other negative consequences, explained later).   
%D_0 = 0.18 (mm)^2/ms
% ?? t_magnetic = 1.59 ms

\subsubsection{fidelity $\mathcal{F}$}

The memory fidelity $\mathcal{F}$ can be parsed into the operational fidelity $\mathcal{F}_{\rm{o}}$ and measurement fidelity $\mathcal{F}_{\rm{m}}$ via $\mathcal{F}=\mathcal{F}_{\rm{o}}\times\mathcal{F}_{\rm{m}}$. Here, $\mathcal{F}_{\rm{o}}$ 
refers to the degradation of the quantum state as the probe field travels through the device without storage, and the measurement fidelity $\mathcal{F}_{\rm{m}}$ reflects how a finite signal-to-noise ratio (SNR) affects the quantum state measurement. In our case, the former is an engineering effort and can be made close to unity (see data in Sec. \ref{RESULTS}). Thus, we mainly focus on $\mathcal{F}_{\rm{m}}$, which is directly related to SNR via $\mathcal{F}_{\rm{m}}=1-\frac{1}{2(1+\rm{SNR})}$ for a single incoming photon (see Appendix \ref{SNR}). 

We now focus on the SNR. The signal is directly related to storage efficiency $\eta$, while the noise is related to the strong control field $\Omega_c$. We observe noise photons due to $\Omega_c$ with two distinct origins. The first is due to $\Omega_c$ not involving atomic transitions, and can be measured when the Rb vapor is removed from the optical path. This noise is called ``technical noise''. The second is due to the interaction between $\Omega_c$ and Rb atoms, and is defined as the excess noise photons when Rb vapor is put back. We term this ``atomic noise''. 

In our system, we attribute atomic noise to two atomic processes: spontaneous Raman scattering (SRS) \cite{Raymer1985} and four-wave mixing (FWM) \cite{Lauk2013}. Here, the SRS is due to control field scattering of atoms in the $F=2$ state. FWM comes as two consecutive, phase-matched-enhanced SRS due to $F=1$ population, and is enhanced by the optical depth (OD) and control field strength. We note that the majority of atomic noise cannot be filtered as they have the same frequency as the $\Omega_p$ field.  To reduce (or eliminate) both types of noise, we leverage a selection rule by using orthogonal circular polarization for $\Omega_c$ and $\Omega_p$ field \cite{Walther2007,Zhang2014}, with a near-resonant $\Omega_c$ field to provide effective optical pumping. As a result, we have observed a lower noise rate compared with the linear polarization scheme \cite{Wolters2017,Namazi2017} under similar conditions.

Technical noise is composed of the strong, narrowband $\Omega_c$ photons and weak, broadband photons due to the processes of generating and delivering the control light.
Here, $\Omega_c$ is generated by a diode laser exhibiting broadband amplified spontaneous emission (ASE) noise, and delivered via optical fibers in which Raman scattering occurs \cite{Blow1989}. In principle, technical noise can be eliminated without compromising the memory performance. 

\subsubsection{storage efficiency $\eta$}

%-------------------------------------------FIGURE 3 -----
\begin{figure}[b]
\includegraphics[width=8.6cm]{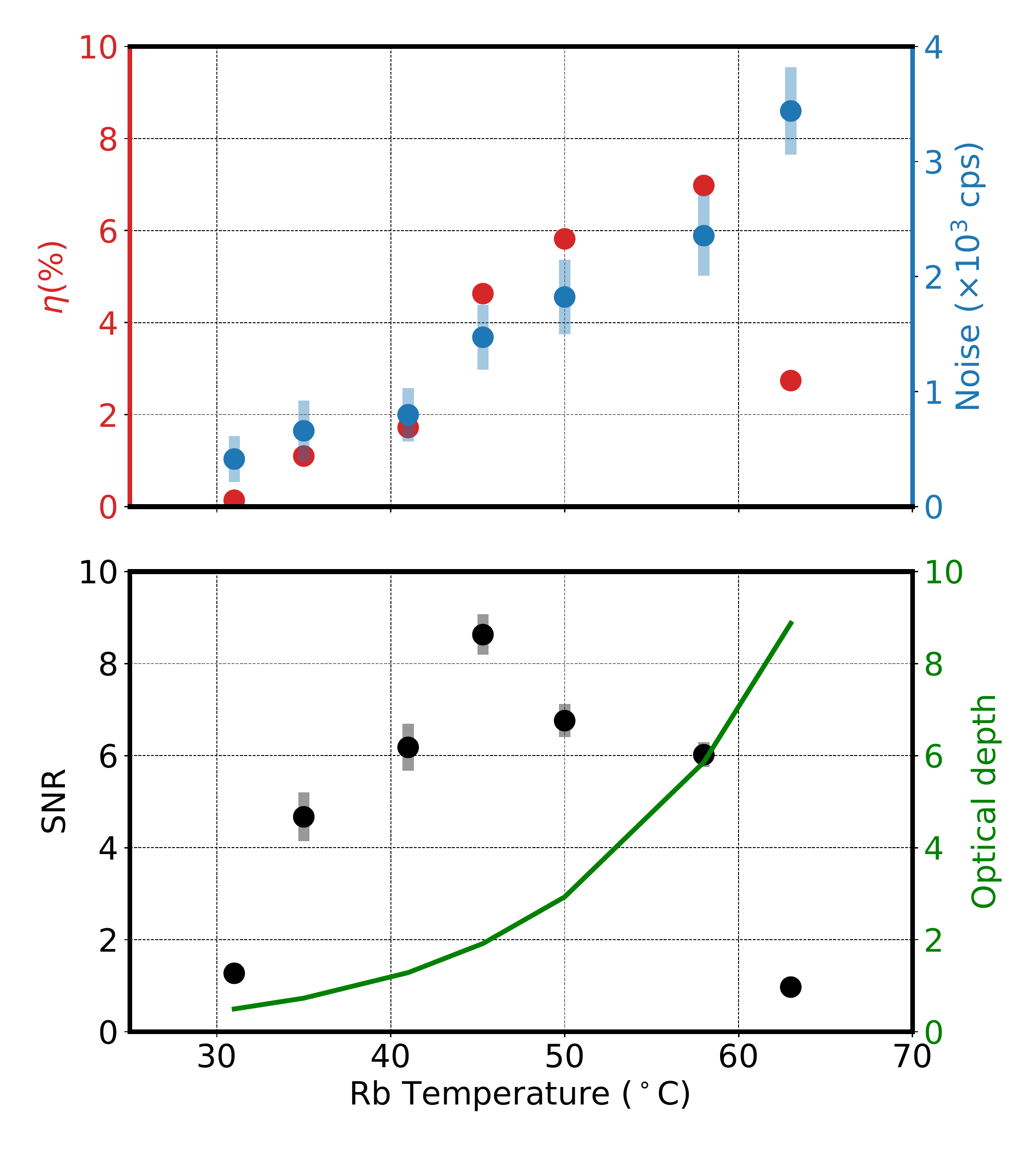}
\caption{SNR optimization.
In the upper panel we plot the storage efficiency (red circles) and noise (blue circles) under different vapor temperatures with fixed control power (20 mW).
In the lower panel, we show the corresponding SNR (black circles) normalized to single-photon input and the measured optical density of the atomic ensemble (green line) at each vapor temperature. The input probe pulse contains $\bar{n}= 2.74(1)$ photons. 
Error bars represent one standard deviations in all plots.}
\label{fig:3}
\end{figure}
%-------------------------------------------FIGURE-----

For a given EIT-system, there exists an optimal solution of the photon shape that maximizes the storage efficiency \cite{Gorshkov2007}. It has been shown theoretically and experimentally \cite{Novikova2007,Phillips2008}, that this $\eta_{\rm{max}}$ depends on the OD of the atomic medium, due to the collective enhancement. In a warm-vapor system, OD can be conveniently adjusted by changing the vapor temperature. Here, in Fig. \ref{fig:3} we show a typical result of $\eta$ as we increase OD. For each Rb temperature, we use an iterative approach (\cite{Novikova2007} and Appendix. \ref{method}) to find the input photon shape that gives maximum efficiency. At low OD, the efficiency increases monotonically. However, at high OD, the efficiency decreases, which we attribute to the increased dephasing (such as density-dependent Rb relaxation \cite{KleinThesis}) and increased absorption \cite{peyronel2013quantum}. The highest efficiency reaches 7.0(2)\%, which is limited by the destructive interference between two hyperfine excited levels ($F=1,2$) \cite{Vurgaftman2013}. 

\subsubsection{parameter optimization}

\begin{figure}[t!]
\includegraphics[width=8.6cm]{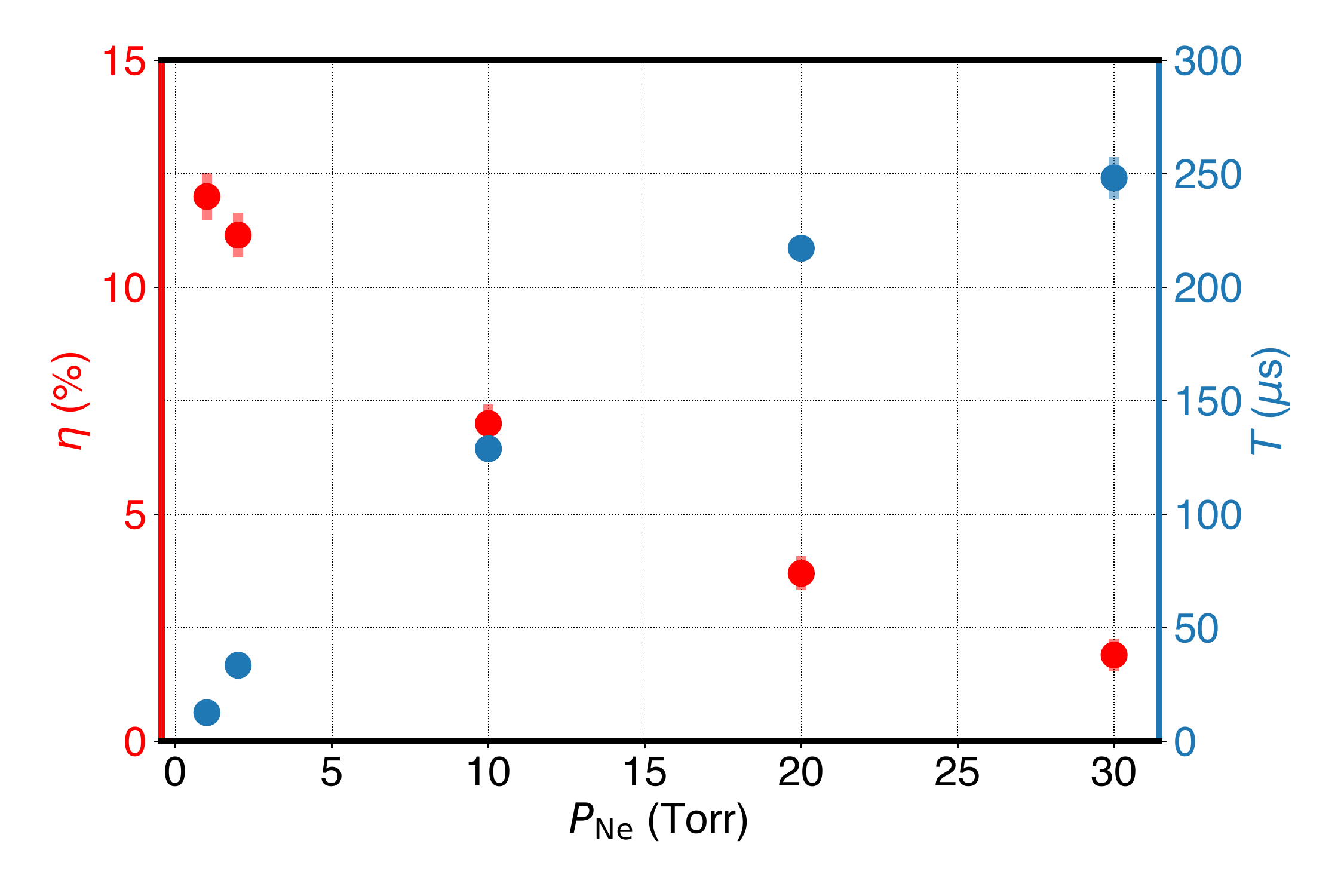}
\caption{Storage efficiency (red) and storage time (blue) for a $w$= 1.1 mm $\Omega_p$ beam as a function of buffer gas pressure. For each buffer gas pressure, we optimize experimental conditions (e.g., OD and pulse shaping) to find the maximum storage efficiency. We note that since $\eta$ is measured at $t = 5$ $ \mu$s (instead of $ t= 0$), it is underestimated for low $P_{\rm{Ne}}$ conditions. The error bars are one standard deviations.}
\label{fig:4}
\end{figure}

We now explore the systematic optimization in the \{$\eta,T,\mathcal{F}$\} space. 
The first parameter we consider is the buffer gas pressure, which affects both $T$ and $\eta$. As shown in Fig. \ref{fig:2}, the buffer gas significantly improves the storage time by suppressing Rb diffusion. However, the collision with Ne atoms also broadens the transitions to the hyperfine states \cite{Matthew1997}, which induces stronger destructive interference. This results in a lower $\eta$. We measure this effect using a $w$ = 1.1 mm probe beam. In Fig. \ref{fig:4}, the storage time and highest storage efficiency are plotted as a function of buffer gas pressure. $P_{\rm{Ne}}$ = 10 Torr achieves a sufficient balance between $T$ and $\eta$. Depending on a user’s specific application (e.g., short link length and detection efficiency) one can choose the buffer gas pressure to maximize the overall performance (e.g., higher efficiency but lower storage time).

Next, we consider $\mathcal{F}$ and $\eta$. The use of orthogonal circular polarization greatly suppresses the atomic noise by the selection rule \cite{Walther2007}. However, as pointed out in Ref \cite{Vurgaftman2013} it also limits the maximum achievable $\eta$ due to the destructive interference between two hyperfine excited levels ($F=1,2$).
We have observed a much improved SNR at the cost of a capped efficiency. One alternative is to use same circular polarization for both light fields, transforming the hyperfine levels interference from destructive to constructive, resulting in a high (> 40 \%) efficiency \cite{Phillips2008}. However, in the parallel polarization scheme used, FWM is no longer turned off by the selection rule, causing more atomic noise. Moreover, the lack of polarization filtering mechanism (up to 50 dB) makes it technically costly to remove the strong control field itself. Consequently, it is challenging to achieve high fidelity with this scheme. 

Finally, $\mathcal{F}$ is a function of $\eta$ and noise, both of which have similar dependence on OD, see Fig. \ref{fig:3}. We measure both parameters as a function of Rb vapor temperature, and plot the resulted SNR. 
Since control power remains fixed, the excess noise is entirely atomic noise (SRS), which scales with OD, as expected. 
Below 60 $^\circ$C, both $\eta$ and noise increase with temperature, but with slightly different rates. Consequently, the highest SNR (lower panel) is obtained at the cost of sub-optimal $\eta$. We note that in applications where slightly degraded fidelity is still acceptable, one can always gain by increasing memory efficiency. 
%-------------------------------------------FIGURE 4----------------
%%-------------------------------------------FIGURE -----------------

\subsection{Hardware Performance}

Informed by the optimization studies of the quantum performance \{$\eta, \mathcal{F}, T$\}, we design a QM solution suitable for field deployment, with an emphasis on achieving high \emph{hardware performance} without compromising its quantum performance. 

To this end, the QM design follows a \emph{modular} approach, permitting efficient multi-parameter optimization and independent upgrades while reducing system complexity and removing inter-dependence \cite{Pogorelov2021}. 
It is composed of a \emph{memory unit}, which hosts the light-matter interface, and a \emph{filter unit}, which provides broadband noise suppression. Both units are housed within a 19-inch 2U-rackmount \emph{enclosure} providing mechanical and electric support.  

\subsubsection{memory unit}

The memory unit (Fig.\ref{fig:1}(b,c)) contains the Rb vapor cell which acts as the medium for storing and retrieving $\Omega_p$ photons. 
The design objective is to reproduce the high quantum performance observed in tabletop experiments within the limited space of the enclosure with commercial off-the-shelf (COTS) optical components. 

A simplified schematic optical layout is shown in Fig.\ref{fig:1} (c), with details to follow. To ensure high quantum performance, we must: 1) recreate the beam profiles for light-matter interaction; 2) precisely align the probe field to minimize transmission losses; 3) precisely overlap $\Omega_c$ field with respect to $\Omega_p$ to ensure long SW wavelength \cite{Zhao2009}; and 4) ensure the polarization filtering can robustly achieved at > 50 dB suppression. Critically, the memory needs to be polarization agnostic, which is enabled by using a dual-rail setup (Appendix. \ref{method}). On a tabletop setup, one can easily satisfy these conditions. However, on a portable module with limited resources and spatial constraints, achieving these parameters is non-trivial.  

The first step of the design is to determine the beam size, which is closely related to $T$ as shown in Fig. \ref{fig:2}. A compromise is necessary because: 1) COTS optical components, such as fiber couplers, waveplates, mirrors, crystals, are limited by preset dimensions and performances (e.g., surface quality), 
and 2) there are diminishing returns when using larger beam diameters, due to the residual magnetic field gradients. Balancing the two, we use $w$ = 1.6 mm (2 mm) for the probe (control) field, at which point the larger beam sizes make sense only after we improve the magnetic field shielding. Nevertheless, with $P_{\rm{Ne}}$ = 10 Torr this arrangement ensures a $T$ > 150 $\mu s$ storage time. 

Next, we consider the implementation of the dual-rail system to efficiently map the two orthogonal polarization components of the $\Omega_p$ field onto two identical spatial rails with the same polarization. COTS beam displacers are incapable of providing the required performance, where two large-sized beams need to be created and widely separated to suppress crosstalk. 
Instead, we implement a Sagnac-like versatile beam displacement solution (Fig. \ref{fig:1}(c)) \cite{Salazar2015}, which creates two rails with large beam sizes and variable separation. The self-interference pathways ensure perfect parallelness of the rails and no optical path length difference \cite{Salazar2015}. These features allow us to efficiently create and recombine rails, and most importantly, perfectly overlap $\Omega_c$ and $\Omega_p$ (which is often an issue using COTS beam displacers).

Lastly, we finalize the selection of the remaining  optical and mechanical components. By using mirror mounts with vertical drives, we preserve adjustment capabilities in a minimal footprint. We identify an adjustment-free polarization plane rotator for applications where a fixed polarization rotation is needed (such as the dual-rail setup, see Fig. \ref{fig:1}). Compared with half-wave plates, these rotators greatly simplify the optical design. 
The polarization filtering requires a high level of spatial uniformity across the waveplates (recall we have two large-sized, widely separated beams). That is not a standard specification for COTS components. We screen batches of COTS components to select for our requirements. With these efforts, we successfully package the memory unit into a portable module engineered towards long-term stability. These results are discussed in Sec. \ref{RESULTS}.

 %-------------------------------------------FIGURE 5----------------
\begin{figure}[t!]
\includegraphics[width=8.6cm]{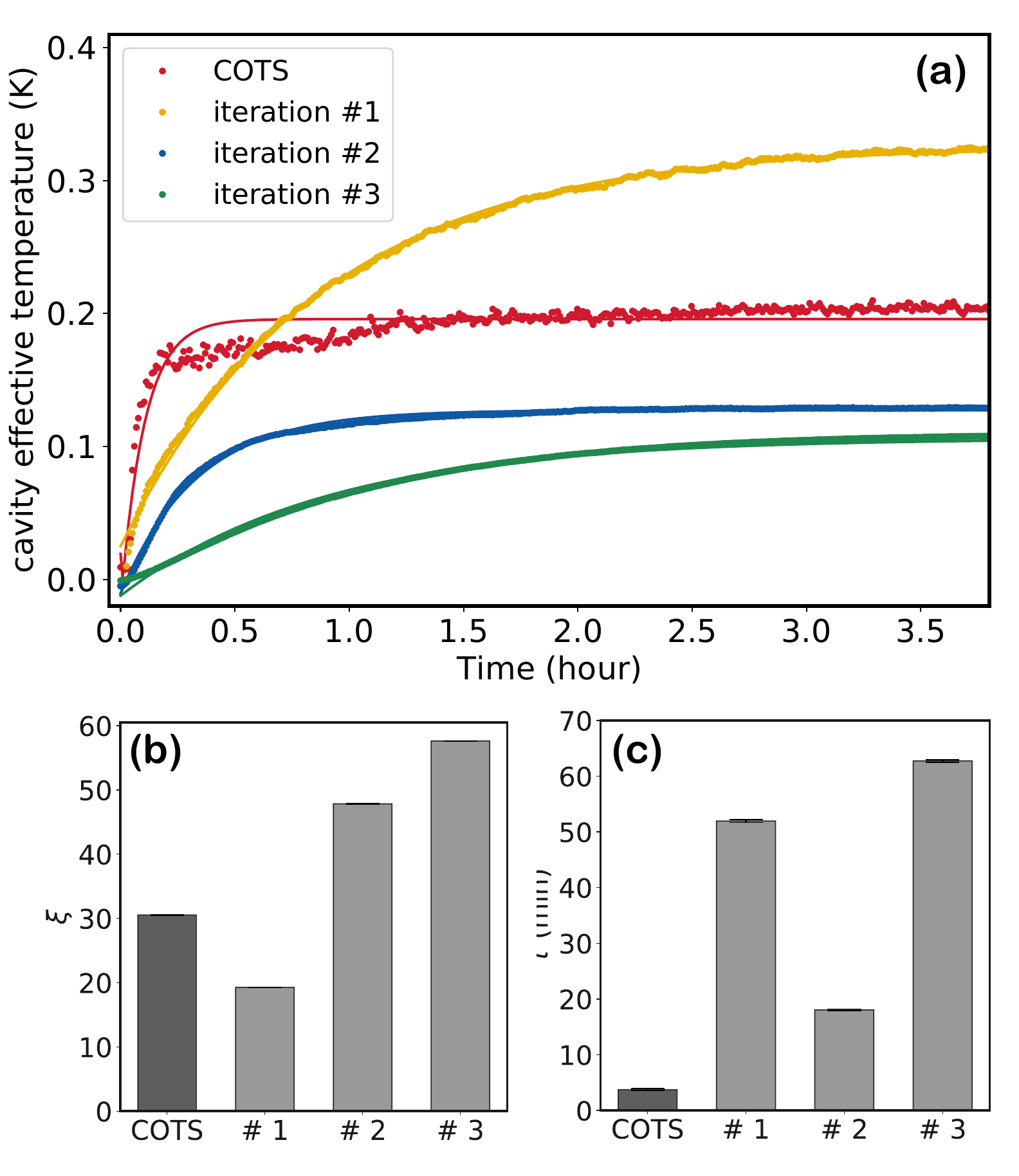}
\caption{Engineering etalon thermal stability. (a): the thermal response of an etalon subjecting to a sudden (< 10 minute) and large (6.3 K) environment temperature change. In each case (one commercial product and three in-house iterations), we fit the data with an exponential function. (b,c): the extracted isolation factor $\xi$ and time constant $\tau$ (see text for details) of the four cases. In each iteration we empirically change the housing material, mounting scheme, and mechanical assembly to improve the thermal performance, based on previous iterations feedback.}
\label{fig:5}
\end{figure}
%%-------------------------------------------FIGURE ----------------

\subsubsection{filter unit}
\label{filter}
High measurement fidelity, $\mathcal{F}_{\rm{m}}$, requires efficient filtering of excessive noise photons, specifically the technical noise due to the strong control field $\Omega_c$. There exist three types of (technical) noise photons: the control field $\Omega_c$, which is of the  order of $ 10^{10}$ photon per pulse; the fiber Raman scattering of $\Omega_c$; and the ASE noise from diode lasers. In the memory unit, we use polarization elements to combine and separate $\Omega_c$ and $\Omega_p$ with suppression of $\geq$ 50 dB. The remaining technical noise is further filtered spectrally in this filter unit. 

The filter unit is comprised of two high-finesse monolithic Fabry-P\'erot etalons, which are thermally tuned to be resonant with $\Omega_p$ field (with a coefficient of -2.4 MHz/mK). The temperature of etalons is actively stabilized via a feedback servo (with 1 mK resolution) by heating the etalon housing. A stable etalon temperature is essential to its performance in rejecting noise and transmitting qubits. Such temperature stability can be achieved with good thermal isolation, such as vacuum \cite{Dai2015}. 
However, due to our design goal of low-SWaP requirements, we instead engineer the thermal properties of the etalon housing to achieve high thermal insensitivity in a vacuum-free, small footprint system. 

In an open thermal system (the etalon plus its housing), the environment (bath) affects the actual temperature of the etalon. The temperature sensor and the servo create a virtual pivoting point setting a temperature gradient between the set-point and the environment. A finite-sized object (etalon) inevitably experiences this gradient shaped by the environment. Therefore, the \emph{effective temperature} (which determines the resonance peak locations) of an etalon changes as the environment changes, even though the thermal servo is effective. In order to have both short-term and long-term stability, a well-engineered system should have a weak and slow response to environmental temperature fluctuations. We characterize this property by probing the step response of a temperature-regulated etalon subjected to a sudden and large change in the environment \cite{Dai2015}. The effective temperature of an etalon changes exponentially with a time constant $\tau$, and the relative temperature change in the steady-state defines an isolation factor $\xi=\Delta T_{\rm{room}}/\Delta T_{\rm{cavity}}$. A thermal system with large $\xi$ and long $\tau$ will maintain a stable, effective temperature over fast and long-term environmental changes.

We tried several iterations of constructing a thermally stable etalon system, starting from a COTS heated lens tube. In each iteration, we change the housing material, sensor locations, and heating methods to improve both parameters, see Appendix \ref{method} for details. Our results are shown in Fig. \ref{fig:5}. To test the properties in each case, we first bring an etalon to a thermal equilibrium state. At t = 0, we introduce a large ($\sim$ 6.3 K) and sudden (< 10 minutes) change to the room temperature. The effective temperature of the etalon is inferred by measuring its transmission peak location. The results from three in-house iterations and an COTS module are shown in Fig.\ref{fig:5} (a), including exponential fits. The isolation factor $\xi$ and time constants $\tau$ are shown in (b) and (c). We successfully achieved a high isolation factor of 58 and a long relaxation time of 63 minutes, effectively creating a strong low-pass filter, making the etalon system immune to rapid temperature changes. Considering common HVAC regulated environments (e.g., a fiber hub) typically exhibit < 1 K change on a 10 $\sim$ 20-minutes time scale, our etalon system will reliably maintain high transmission (stay on resonance) for many hours without user intervention, which is critical for field-deployable quantum applications. A temperature fluctuation induced cavity transmission variation as low as 1.9\% (relative) on a day timescale is achieved with a room temperature fluctuation amplitude of $\sim$ 0.5 K; see Appendix \ref{method} for details. This well-characterized thermal response also warrants the possibility of further improvement of long-term performance by monitoring the environmental temperature.  

 \subsubsection{enclosure}
 \label{box}

The stable operation of any quantum optics apparatus relies on a well-controlled environment (e.g., temperature, pressure, mechanical) and supporting equipment. Transitioning from a typical lab-environment into the field, we engineer a general-propose enclosure to provide a stable local environment and hardware support. This rack-compatible enclosure allows many modular quantum devices to be integrated into a node for quantum networking.  

First, based on the physical size of the memory and filter units, we determine a form factor for the enclosure to be 19-inch 2U rackmount (Fig. \ref{fig:1}). This comfortably accommodates most free-space optical elements and mounts (up to 1 inch), while maintaining enough vertical space for mechanical assembly. We then partition the space into quantum optics devices (front) and supporting electronics (rear). The former is equipped with a passively-dampened internal rail system bolted to the solid skeleton of the enclosure. The dampening mechanism  isolates vibrations from nearby oscillating equipment like fans present in rack cabinets at telecommunication hubs, which would otherwise cause beam misalignment/pointing instability over time. 
Both memory and filter units are built on monolithic tempered custom baseplates for mechanical stability and optical performance. These baseplates can slide on the internal mounting rails, such that a user can access each unit (e.g., to diagnose) without opening the enclosure. The units are further shielded from each other and the environment with internal hoods to regulate the temperature and airflow across the free-space optical parts.

The supporting electrical components reside in the rear of the enclosure. A motherboard with interchangeable slots allows a custom circuit board to control each quantum device (e.g., precision temperature  servos) via a Sullins Edgecard connector under each baseplate. This arrangement is optimized for easy maintenance and flexible upgrade, which are critical for field deployment with limited accessibility. This general-purpose electrical system allows the enclosure to host different quantum devices. The motherboard also provides Ethernet/USB interfaces with remote access capabilities, which, together with our software, allows users to monitor, control, and debug devices remotely through an API. The total power consumption is 15 W with the memory and filter units installed and in operation. 

%-----------------------------------------------------------------------------------------

\section{Memory Benchmarking\label{RESULTS}}
In this section we detail the results of the various characterizations performed.
Details on how each measurement was performed can be found in Appendix A.

\subsubsection{bandwidth engineering}
%-------------------------------------------FIGURE 6------------------------
\begin{figure}[t!]
\includegraphics[width=9.cm]{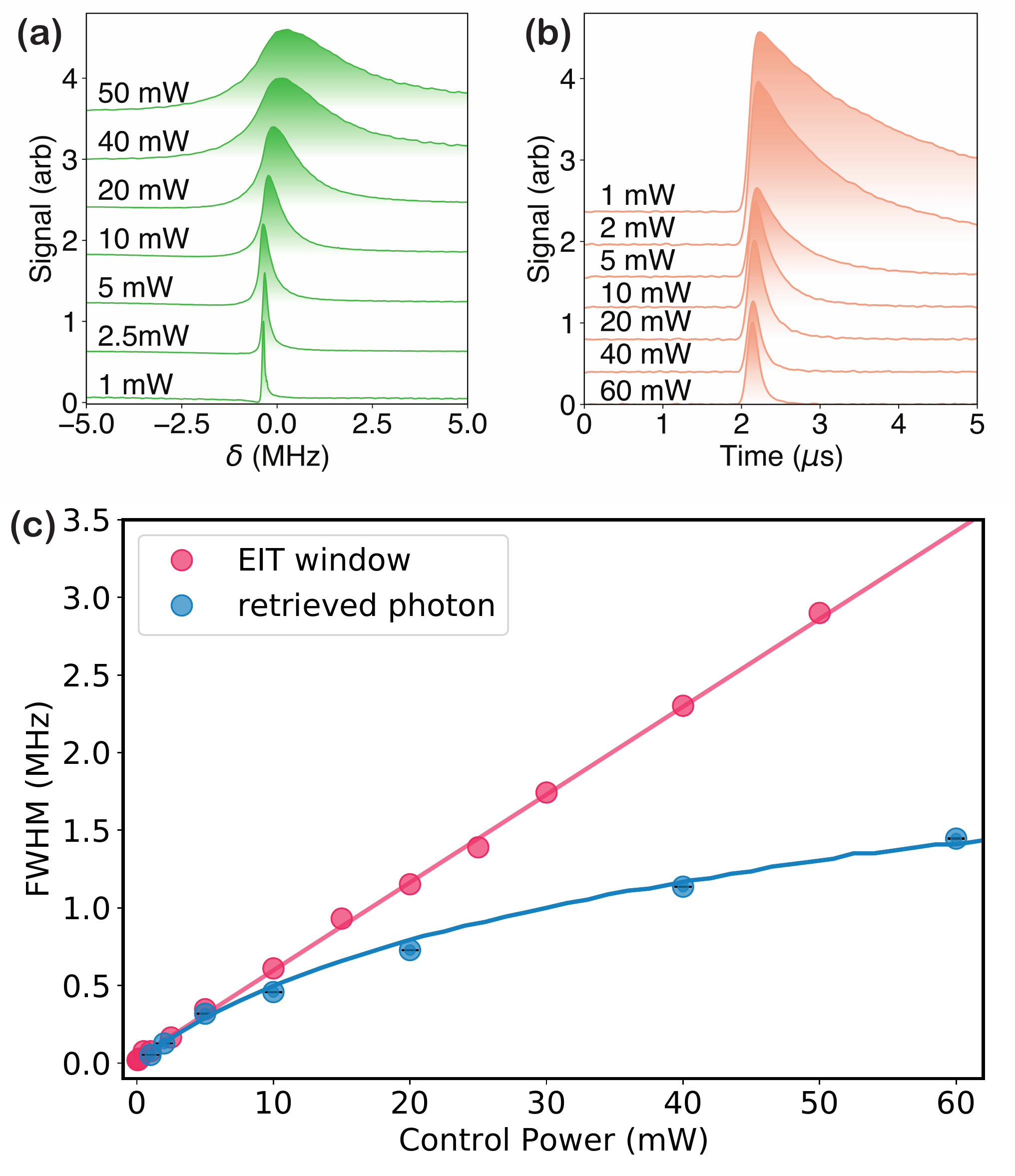}
\caption{Engineering photon bandwidth.
\textbf{(a)} EIT window with different control field powers. We plot the (normalized) transmissions of a weak probe field (5 $\mu$W) with different $\Omega_c$ powers (see labels) as a function of two-photon detuning $\delta$. The traces are offset vertically for easy visualisation. 
\textbf{(b)} The photon temporal profiles with optimal storage efficiency for different $\Omega_c$ powers (see labels). Traces are offset vertically. 
\textbf{(c)} We plot the full width at half maximum (FWHM) of the EIT window (red) as shown in (a), and the FWHM of spectral peak (blue) of the retrieved photon (after Fourier transforming the temporal traces in (b)) as a function of the control field power. The EIT window is well captured by theory (red solid line), and the retrieved photon is limited by the finite switching time of the control field, which is captured by a simple theory as well (blue solid line). The experimental data uncertainty is smaller than the symbols.}
\label{fig:6}
\end{figure}
%-------------------------------------------FIGURE ---------------------------

One benefit of EIT-based QMs is their ability to store photons of varying bandwidths to support the users' application. This is realized by adjusting the EIT window with the control field power, which affects the accepted spectral components of the atomic medium.  
In Fig. \ref{fig:6}, we show this effect by measuring the EIT signal and the bandwidths of optimal photons (photon profile with maximal $\eta$, see Appendix \ref{method}) for a weak coherent probe field at various control field powers.
The EIT peak width and the optimal photon bandwidth increase with increasing control power. 
The linear increase of 0.128(1) MHz/mW in the EIT window width is well explained theoretically as in the ``power-linear'' regime \cite{Wei2020}. 
Similarly, we expect the optimal photon bandwidth to increase linearly with control field power.
However, for technical reasons, $\Omega_c$ could not be turned off sufficiently fast, resulting in the sub-linear dependence shown in the plot, which is explained by an ad-hoc theory taking this experimental limitation into account (see Appendix.\ref{method}). We anticipate that faster optical switching should bridge the gap. 
Importantly, we note that changing the photon bandwidth does not affect other memory performance such as $\mathcal{F},\eta$, and $T$.

\subsubsection{stable \& high operation fidelity}
%-------------------------------------------FIGURE 7-----------------
\begin{figure}[t!]
\includegraphics[width=8.6cm]{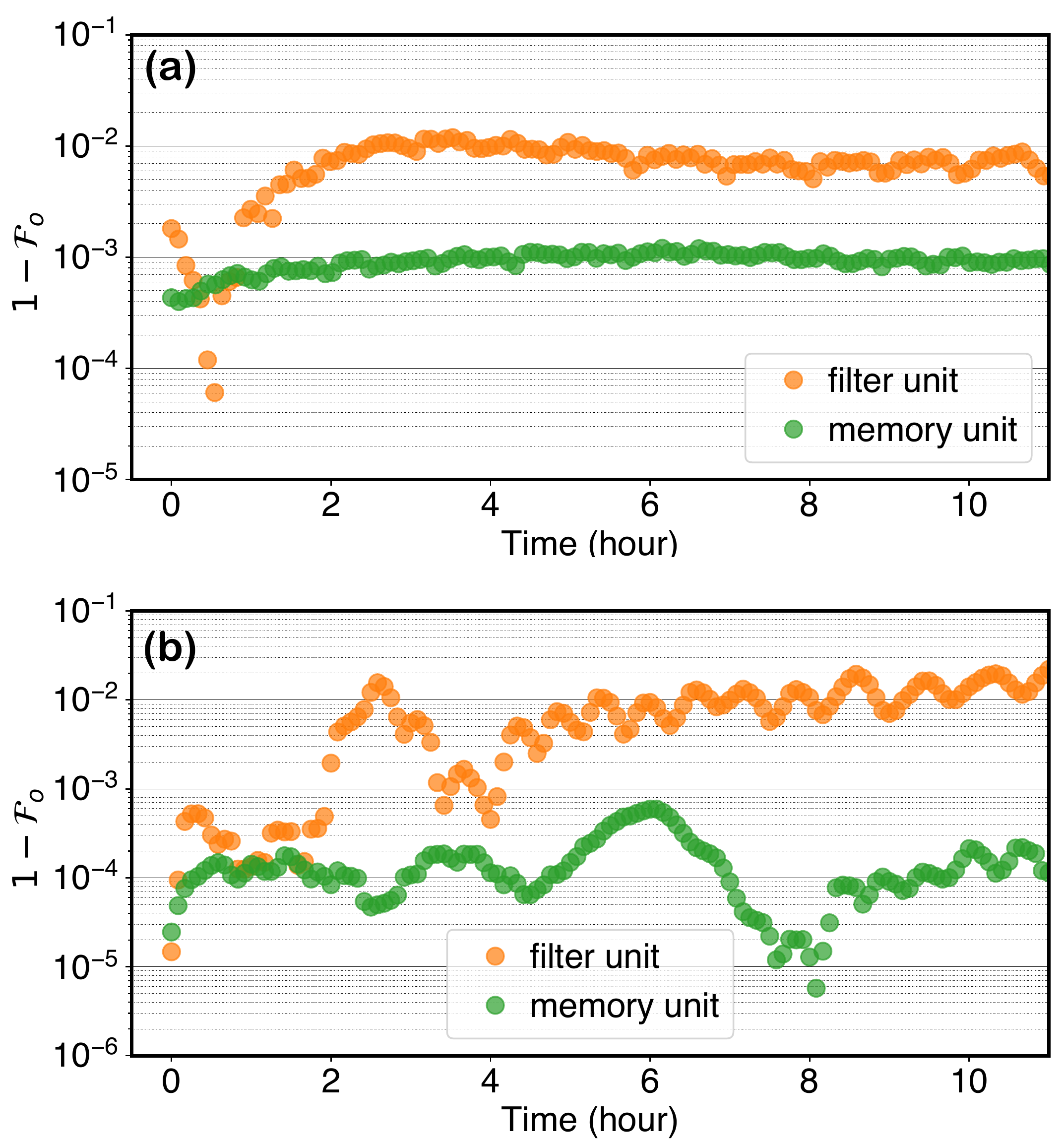}
\caption{High operation fidelity. We plot $1-\mathcal{F}_{\rm{o}}$ of both the memory unit and the filter unit on a day scale under typical experimental conditions ( <4 K environment temperature changes, 20$\sim$30 \% humidity, in a rack tower). The errors are caused by the transmission performances between rails, where the amplitude induced part is shown in the upper panel, and the phase induced part is shown in the lower panel. See main text for details.}
\label{fig:7}
\end{figure}
%-------------------------------------------FIGURE -------------------

Geared towards field deployment, our QM needs to have robust performance against environmental changes over many hours. Here, we present the results of the operation fidelity $\mathcal{F}_{\rm{o}}$ for both memory and filter units. 
Since our memory stores arbitrary polarization qubits using two spatial rails within the atomic vapor, $\mathcal{F}_{\rm{o}}$ is related to the mechanisms affecting the transmission of two rails, which can be broken into the amplitude and phase.

The fidelity loss due to differential transmission depends on the input state. Therefore, we consider the ``worst case'' scenario, in which a specific input state results in the lowest fidelity. 
The amplitude can be measured directly, where we define $T$ as the transmission ratio between two rails ($0\le T\le1$). An analytical solution of fidelity in the worst-case has the form: $\mathcal{F_{\rm{o}}}=\frac{4\sqrt{T}}{(1+\sqrt{T})^2}$, see Appendix. \ref{dual} for details. To access the phase, we adjust the input state to be an equal superposition of two modes and measure the output polarization state using a polarimeter. The parameters of the polarization ellipse (ellipticity and azimuth angle) are used to calculate this differential phase.

Figure \ref{fig:7} illustrates the measured operation error, defined as $1-\mathcal{F}_{\rm{o}}$, due to amplitude (a) and phase (b) as a function of time. These operation errors remain remarkably low: < 0.002 for the memory and < 0.02 for the filter over many hours. 
In the memory unit, the dominant source of operation error is attributed to the creation and recombination of two widely-separated and large-waist rails. While in the filter unit, it is mostly due to the birefringence of the etalon optics. We anticipate optical elements with better specifications and low-stress etalon mounting techniques \cite{Ahlrichs2013} should further lower this error in both cases to  $\leq 10^{-3}$ level.

\subsubsection{$\mathcal{F},\eta$, and $T$}

%-------------------------------------------FIGURE8----
\begin{figure}[b!]
\includegraphics[width=8.6cm]{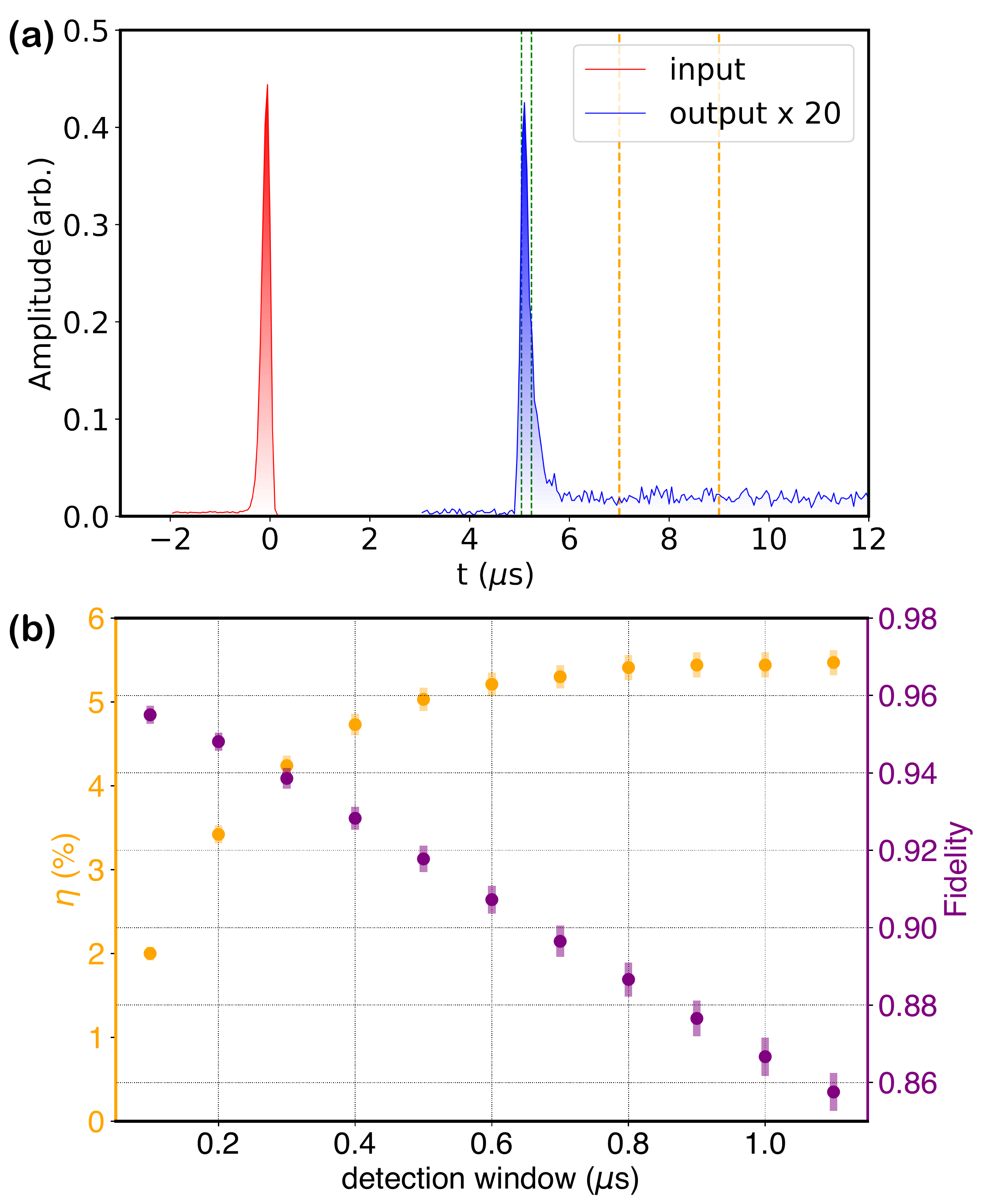}
\caption{High fidelity memory operation at single-photon level. 
\textbf{(a)} A coherent pulse with an arbitrary polarization containing $\bar{n}= 2.74(1)$ photons (red peak) is stored and retrieved (blue peak, scaled by a factor of 20). 
A 200-ns detection window (dashed green lines) is chosen for analyzing the signal, while a 2-$\mu s$ window far after the retrieval (dashed orange lines) is chosen to analyze the noise.
\textbf{(b)} The storage efficiency $\eta$ and fidelity $\mathcal{F}$ at single-photon level as a function of the detection window size.}
\label{fig:8}
\end{figure}
%-------------------------------------------FIGURE-----

To measure the single-photon level performance of the memory, we perform storage and retrieval of a coherent pulse with a random polarization and mean photon number of order one.
Fig. \ref{fig:8} (a) plots a typical result of memory operation, where the input pulse has an FWHM duration of 218(1) ns (bandwidth: $2\pi\times 0.770(2) $MHz), and the Rb vapor temperature is 45 $^{\circ}$C with an experimentally measured OD of 2.0(1). 
To evaluate the SNR, we select a detection window (green dashed lines) of order the input pulse size to analyze the signal. 
We then select a larger window (orange dashed lines), taken after the photon retrieval (where $\Omega_c$ remains on), to determine the noise rate with better photon statistics, which is consistent with the measured noise under the detection window when the input pulse is blocked, where the unconditional noise floor is $1.9(1)\times 10^{-3}$ photons per storage trial.
The SNR is then calculated and scaled to obtain the SNR for an input of exactly one photon. 
The choice of the detection window size is a trade-off between the number of successful events and read-out fidelity. We show in Fig. \ref{fig:8} (b) the dependence of efficiency and fidelity on the width of the detection window. Depending on the details of the applications (e.g., link length, bandwidth mismatch, detector efficiency/dark count, etc), the user can choose a suitable window size to optimize specific functions. Using a 200-ns window that is matched to the input pulse size, we find $\rm{SNR}_{\bar{n}=1}$ = 8.63(44) under polarization-agnostic operation.
The corresponding measurement fidelity is $\mathcal{F}_{\rm{m}} = 94.8(2)\%$. For applications with known input photon polarization state, higher SNR and fidelity can be straightforwardly obtained by using only one rail, leading to $\rm{SNR}_{\bar{n}=1}=17.3$. For a generous window size that encloses the entire pulse, our QM offers > 5\% efficiency at $\mathcal{F}$ > 0.9. The above high fidelity proves our memory works in the quantum regime, even when compared to a modified threshold for a weak coherent state with non-unity device efficiency \cite{Mustafa2012,Specht2011}: 85.3\% (with mean photon number of one and including all the optical loss of the device).

\begin{figure}[t!]
\includegraphics[width=8.6cm]{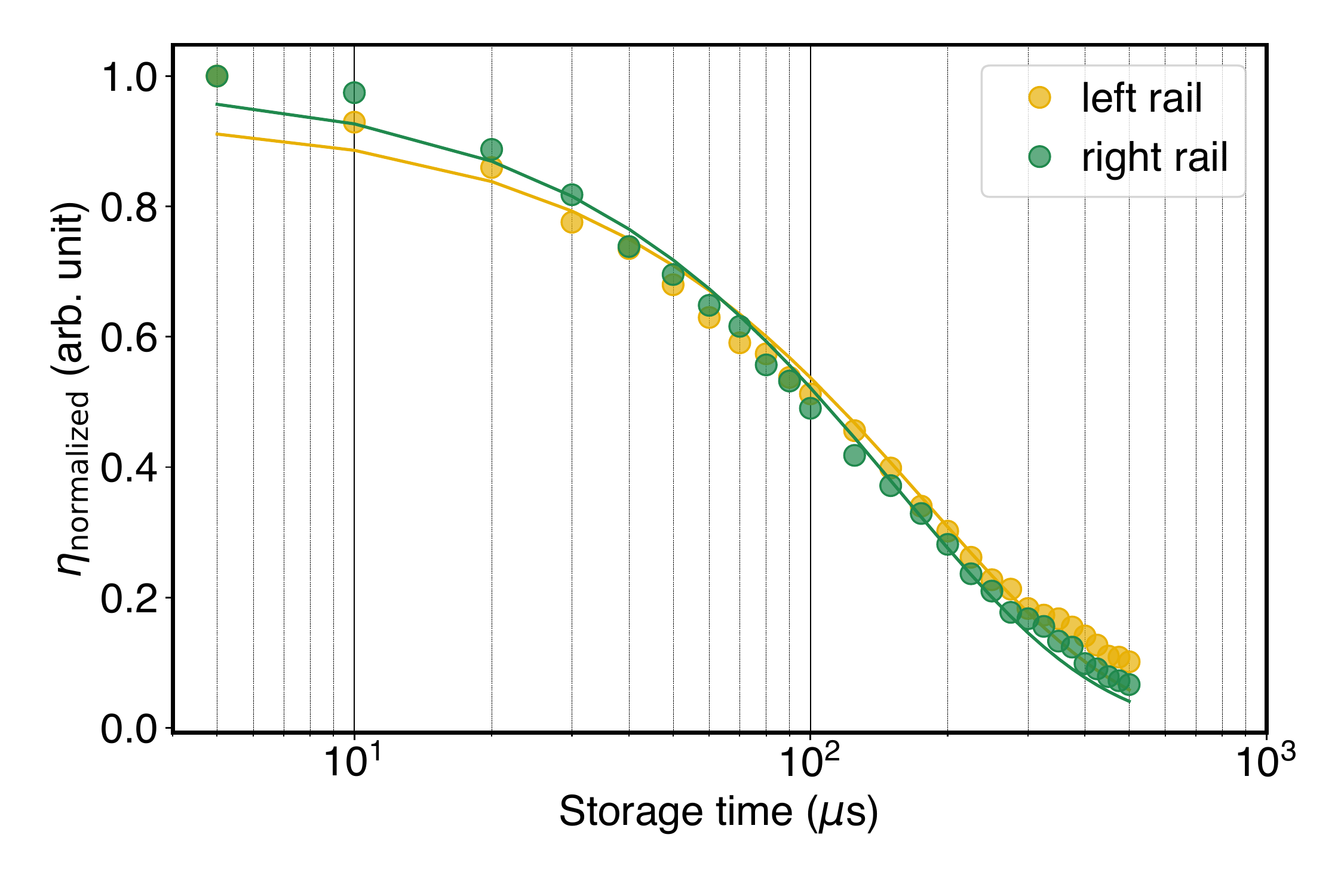}
\caption{Storage time. We plot the normalized storage efficiency $\eta$ as a function of storage time for the two rails. The colored lines represent exponential fits, with 1/e decay constants of 180(6) $\mu s$ (left rail) and 157(4) $\mu s$ (right rail), respectively.}
\label{fig:9}
\end{figure}

We also measure the storage time $T$, see Fig. \ref{fig:9}. To demonstrate polarization-agnostic operation, we plot the (normalized) efficiencies as a function of storage time for both rails, which are very close to each other.
% which show very little variance across the entire time. 
Both rails exhibit $T > 150$ $\mu$s.  
%-----------------------------------------------------------------------------------------
\section{conclusion and outlook\label{OUTLOOK}}

%-------------------------------------------FIGURE9-----
%-------------------------------------------FIGURE9-----

To conclude, we have systematically studied and optimized a quantum memory based on a warm atomic (Rb) vapor fully supported by scalable room-temperature technologies. We designed and packaged the experimental quantum optics system into a 2U-rackmount-sized product. The design is robust against environmentally-induced noises while preserving high quantum performance ($\mathcal{F} \sim 95\%$, $T > 160 \mu$s, and $\eta > 5\%$) for single photons carrying quantum information of arbitrary polarization. With further diffusion suppression, the storage time can be increased to 1 ms (demonstrated using classical-level light). 

As the design continues to evolve, we will explore other methods to further improve \{$\eta,\mathcal{F},T$\} within warm-vapor-based QMs. First, the storage time $T$ can be lengthened through several approaches: the longitudinal magnetic shielding can be improved if the magnetic field direction is different from the optical path. One way to realize this is to place mirrors inside of the shielding. The use of an anti-relaxation coated vapor cell could also help maintain the long storage time with modest-sized beams. However, its compatibility with the dual-rail configuration remains an open question. After surpassing the millisecond-level storage time, one will fight higher-order dephasing mechanisms, where the use of decoherence-free subspace \cite{Katz2018} with Zeeman states provides a promising direction. Second, we consider the storage efficiency $\eta$. It has been proposed \cite{Yan2018} and demonstrated \cite{Saunders2016,Ma2021,Dideriksen2021,Bao2012} that an optical cavity outside of an atomic vapor could enhance the light-matter interaction while, at the same time, suppressing the noise generation, which yields high $\eta$ without sacrificing $\mathcal{F}$. Third, there are methods to further suppress the noise to achieve even higher fidelity, such as the judicious use of ``magic detuning'' \cite{Dideriksen2021}. 

In parallel to the research agenda, we will also advance the hardware performance towards building more robust, compact, and cost-effective product, ready for mass production. We will further explore different approaches to miniaturization, for example, the use of integrated photonics. Efforts are also ongoing to design a companion product containing the control field laser and electronics with a rackmount form factor. 

Apart from this product-oriented agenda, we plan to study and optimize the interfacing of this QM with a Rb vapor-based entangled photon-pair source \cite{Willis2010}, currently under development. In this type of experiment, the finite SNR (e.g., noise floor) of a QM and the heralding efficiency of the photon-pair source jointly determine the interfacing performance (e.g., nonclassical correlations). The high SNR (17, single rail) of our QM would require a heralding efficiency as low as 5.7\% to exceed the classical bound (e.g., Cauchy-Schwarz parameter $R\ge1$), which is very doable compared with other similar experiments \cite{Lukas2021,Buser2022,Seri2017}. The combination of those two components not only form the basic building blocks of a quantum repeater, but also patches the memory to the existing telecom infrastructure for direct use in quantum networks.

\section*{Acknowledgments}
We thank Gabriel Bello Portmann for help with control software development, Anita Richardson for supporting electronics, and Noel Goddard for manuscript preparation. 
This work is supported by the U.S. Department of Energy under award no. DE-SC0019702.

\appendix

%--------------------------------------------------------------
\section{Method \label{method}}

\textbf{General.} Our quantum memory is based on electromagnetic induced transparency (EIT) in a warm $^{87}$Rb atom vapor.
A strong, $\sigma^+$ polarized control field, $\Omega_c$, addresses the $^5S_{1/2}$ $F=2$ to $^5P_{1/2}$ $F=1$ transition, while a weak, $\sigma^-$ polarized probe field (the photonic qubit to be stored), $\Omega_p$, addresses the $^5S_{1/2}$ $F=1$ to $^5P_{1/2}$ $F=1$ transition. 
We reference the $\Omega_c$ field to the $^{87}$Rb transition using saturated absorption spectroscopy.
We beatnote lock the $\Omega_p$ field to be detuned from the $\Omega_c$ field by approximately the ground-state hyperfine splitting.
For this work we fix the single photon detuning, $\Delta= -2\pi\times 120\,\mathrm{MHz}$, but vary the two-photon detuning, $\delta$ according to the specific operation. 

The $\Omega_c$ field originates from a tapered amplifier (TOPTICA TA pro).
In order to suppress ASE noise from the TA we use a pre-filter comprising 
of a narrow-band ($\sim2\pi$ $\times$ 40 GHz) volume-Bragg-grating (Coherent ASE-794.98) with up to 70 dB out-of-band suppression, and a flat etalon with a $\sim2\pi$ $\times$ 500 MHz bandwidth (LightMachinery OP-7423-6743-2). 

The Rb vapor (>99\% enriched $^{87}$Rb) is contained in a cylindrical glass (quartz) cell that is 80 mm long and 25.4 mm in diameter. A set of pre-made vapor cells with different buffer gas (Ne) pressures (2 - 30 Torr) were employed in our study. These buffer gas atoms homogeneously broaden the Rb resonance lines, effectively increasing the inverse lifetime of the $^5P_{1/2}, F=1$ state with a scaling of 9.8 MHz/Torr\cite{Matthew1997}. 

We construct a double-layer cylindrical magnetic shields using $\mu$-metal (Ad-Vance Magnetics, AD-MU-80). Outside the outer layer we use parallel radiant heat films to regulate the cell temperature while introducing negligible magnetic field. Outside the heater we have a saturation layer made of different $\mu$-metal material (Ad-Vance Magnetics, AD-MU-00) meant to protect the double-layer shields. 

\textbf{The memory unit.} For polarization-agnostic operation, we convert two orthogonal polarization components $|H\rangle$ and $|V\rangle$ into two spatial modes of the same polarization using a Sagnac-like beam displacing setup.
We then use a polarization plane rotator to rotate the polarization of one rail by 90$^{\circ}$, making two rails identical. We combine the rails of the orthogonally polarized control and probe fields using a Glan-Laser (GL) prism.
A quarter waveplate turns the subsequent polarization to $\sigma^{+,-}$ before these fields interact with a Rb vapor.
After the cell a set of waveplates and a GL are used to filter the control from the probe field, with a suppression of >50 dB.
We then use another polarization rotator on one of the rails and another Sagnac setup to recombine the probe's two spatial rails back into its original polarization form (single rail).

\textbf{The filter unit.} The filter unit is comprised of two monolithic, 0.5-inch diameter, plano-convex high-finesse Fabry-P\'erot etalons (LightMachinery, custom made) with incommensurate free spectral ranges (FSR). This combination of high finesse (200 $\sim$ 300), and FSR (13 GHz and 21 GHz), provide >80 dB suppression at the control field photon and > 40 dB suppression to the broadband noise. The corresponding linewidths are $2\pi$ $\times$ 40 and $2\pi$ $\times$ 100 MHz, respectively. With proper mode matching practice we can routinely achieve >80\% transmission for single cavity, and >50\% through the filter unit. 

To prevent the back reflection from the second etalon causing interference we construct a polarization-agnostic optical isolator composed of two calcite beam displacers with 2.7mm separation, a Faraday rotator and half waveplate. This provides >40 dB rejection for light propagating backwards. %todo emphsize small? 

\textbf{Memory operation.} A typical memory operation starts with 50 $\mu$s optical pumping with $\Omega_c$ field, resulting in a stead-state 
distribution with atoms nearly evenly divided among these four states (Fig.\ref{fig:1} (c)): $|2,+2\rangle, |1,+1\rangle, |1,0\rangle, |1,-1\rangle$. Only the last two states participate in the memory operation.

For storage, a pulse of $\Omega_p$ field enters the atomic medium under EIT created by the $\Omega_c$ field. The pulse is ``compressed'' as it propagates due to the reduced group velocity inside the Rb vapor. The $\Omega_c$ field is rapidly switched off, mapping this pulse onto a SW. To retrieve, $\Omega_c$ is turned back on, mapping the SW back to the $\Omega_p$ field in a time-reversal manner.    

To characterize (diagnose) the QM performance (such as $\eta$ or $T$), we send a probe pulse that contains a large number of photons >$10^6$, and measure the output using an amplified photodiode (Thorlab APD430A). To obtain the memory $\mathcal{F}$ at single photon level, we optically attenuate the probe field so that, on average, each pulse contains of order one photon (a coherent state). We then use a single photon counting module (EXCELITAS, SPCM-AQRH-15-FC) with a quantum efficiency of 70\% to measure the output. The single photon events are time-tagged with a time-to-digital converter (qutools GmbH, quTAU) with 81 ps resolution. The data shown in Fig:\ref{fig:8} is accumulated over 5 minutes at a repetition rate of 200 Hz. 

\textbf{Iterative pulse shaping} 
It has been shown theoretically and experimentally \cite{Novikova2007,Phillips2008} that an optimal photon storage scheme for a three-level system exists and can be accessed via an iterative approach. We implement this method in our experiment: we first send in a probe pulse with pre-determined profile; the retrieved pulse (which may be very different from the pulse we send it) is measured, which is closer to the optimal shape; and a time-reversed copy of this pulse is sent to the memory for the next iteration. Three such iterations typically result in convergence. This iterative process automatically maximizes $\eta$ since it matches the photon's spectrum to that of the EIT medium, which depends on the experimental details such as OD, $\Omega_c$, $\Delta$, and so on.

\textbf{Deviation between EIT and photon bandwidth}
The finite turn-off and turn-on time of control light is responsible for the bandwidth mismatch of retrieved photon and EIT window Fig. \ref{fig:6}. The control light is intensity modulated by a free space acoustic-optical-modulator (AOM) with a rise time of 100-ns, which is comparable to the pulse duration. This finite rise time creates an effective $\Omega_c$ that is lower than its peak value, reducing the storage efficiency of high-frequency spectral components. We model this behavior by numerically integrating the optical Bloch equation and find satisfying agreement (blue solid line). We anticipate that with faster switching (e.g., using an EOM) of $\Omega_c$ the retrieved photon linewidth at high control field power will increase to eventually overlap with the EIT window, which has been shown elsewhere \cite{Wei2020}.  

\textbf{Etalon housing iterations} The first iteration is inspired by one COTS solution  (Thorlab, SM1L10HR) we use in our study. It consists of a simple Al-6061 lens tube embedded in an Al-6061 mount, bolted directly on the baseplate. The lens tube is temperature stabilized with a ceramic ring-shaped resistive heater at the end, without any thermal decoupling. Effectively, the heater can affect the mounting and the environment, or vise-versa. The second iteration adds isolation between the etalon and the baseplate/assembly by changing the material of the mount. The same lens tube is used but held in a Delrin block mount bolted to the baseplate. Because of the mount material change, the ceramic heater is replaced with a Kapton resistive heating strip. While the isolation factor improves significantly, the time constant decreases due to the smaller heat capacity of the thermal bath (Delrin). The third iteration improves upon both factors by replacing the lens tube with a larger solid cylinder of Tempered Al-7075 with small apertures. This part is wrapped in Aerogel and Teflon tape to isolate it from the environment and then held in place concentrically in a Stainless-Steel 316 armature bolted to the baseplate. Only the part containing the etalon is temperature regulated. This provides a more stable bath, better environmental isolation, and solid mounting to the baseplate with minimal thermal coupling.

\begin{figure}[h]
\includegraphics[width=8.6cm]{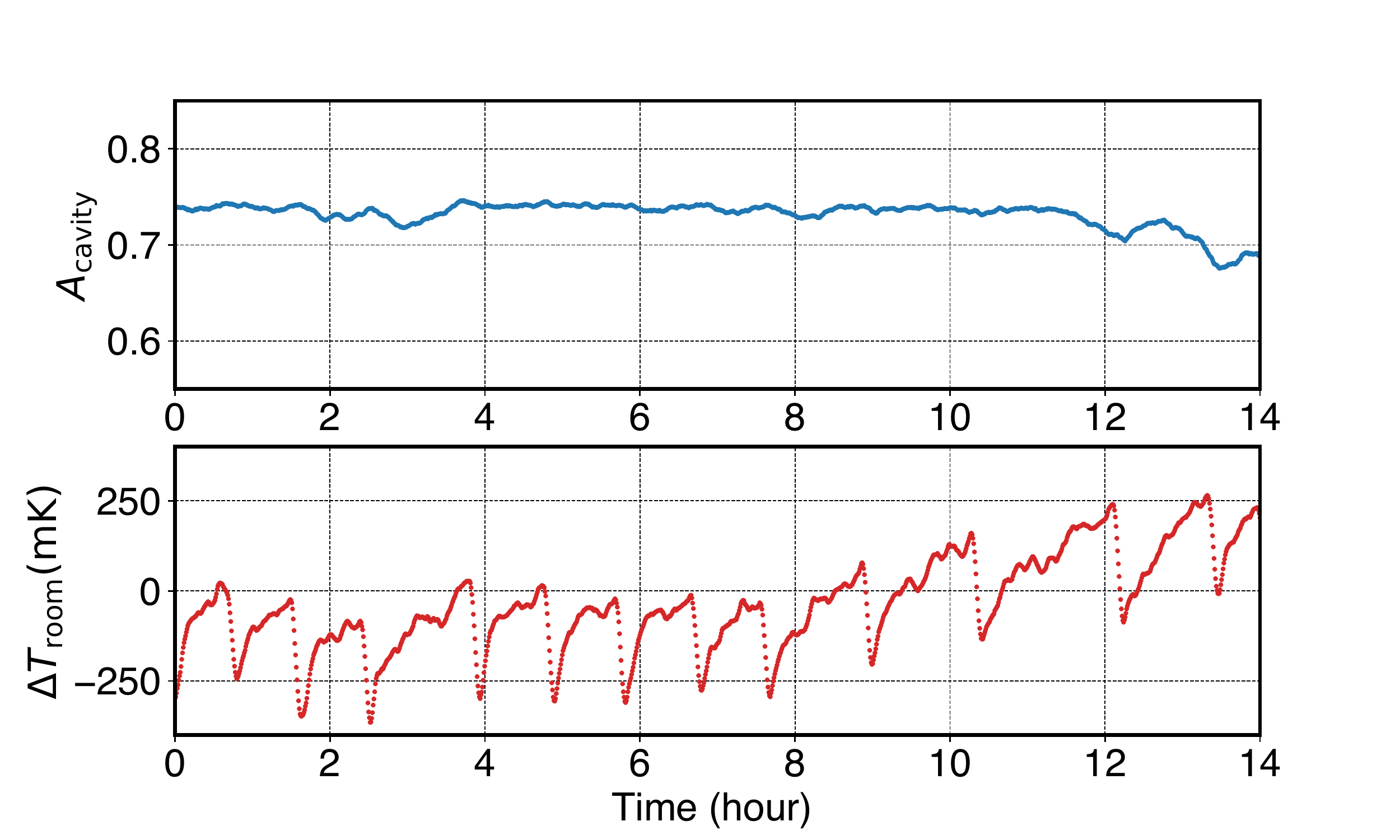}
\caption{Etalon thermal stability. Top: cavity transmission. Bottom: room temperature with fast HVAC fluctuations plus slow drifts. Our cavity is robust against rapid temperature changes. And a feed-forward style correction based on the well-characterized cavity response would improve the long-term cavity performance. }
\label{fig:10}
\end{figure}

\textbf{Etalon stability}
In Fig. \ref{fig:10} we show a characteristic etalon operation under a fluctuating thermal environment. We stabilize the frequency of the probe field to a Rb reference, and measure the transmission of it through a typical high-finesse etalon for many hours (upper panel); we also monitor the room temperature, maintained by an active air conditioning system exhibiting fast ($\sim$ 10 minutes) temperature swings ($\pm$ 250 mK. For a continuous 14-hour operation the mean cavity transmission is 73.1\% with a standard deviation of 1.4\%. The well engineered thermal properties of the etalon housing makes the etalon insensitive to rapid temperature changes.

%-------------------------------------------FIGURE10-----
%-------------------------------------------FIGURE-----

%\textbf{Characterizing quantum memory with coherent pulses}
%write up something
%-------------------------------------------FIGURE10-----
%\begin{figure}[h]
%\includegraphics[width=8.6cm]{Fig11}
%\caption{Fidelity VS efficiency. }
%\label{fig:11}
%\end{figure}
%-------------------------------------------FIGURE-----

%--------------------------------------------------------------
\section{Spin Wave Dephasing\label{dephase}}

Multiple processes can cause dephasing, or effective dephasing of the stored spin wave.
In general, the storage efficiency is proportional to the overlap integral of the initial spin wave, $S(r)$, and that at the retrieval time, $S'(r)$
\begin{equation}
    \eta \propto \abs{\int dr\, S^*(r) S'(r) }^2.
\end{equation}
Here, we consider several dephasing processes and their effect on the storage efficiency.

\subsection{Radial Expansion of the Mode Volume}

In warm atom systems a major contributor to dephasing is the motion of atoms after storage.
In our experimental geometry the radial size of the SW (typically a few mm) is significantly smaller than the wavelength of the SW ($\lambda_{\rm{SW}}$ = 4.3 cm).
Therefore, radial (as opposed to longitudinal) motion of the atoms, is the dominant process to consider. Its effect is calculated here.

When the SW is stored we assume that it has a 2D Gaussian profile identical to that of the probe beam profile:
\begin{equation}
    \abs{S(x,y)}^2 = \frac{2}{\pi w^2}e^{-2\frac{x^2+y^2}{w^2}},
\end{equation}
where $w$ is the $1/e^2$ radius of the probe field.

After a time, $t$, the transverse spatial profile of the SW is given by
\begin{equation}
\begin{aligned}
    \left|S'(x,y)\right|^2 = &  \int\int dx' dx'' dy' dy''  \left|S(x',y')\right|^2C(x'')C(y'') \\ 
    & \delta(x-(x'+x''))\delta(y-(y'+y'')),
\end{aligned}
\end{equation}
where
\begin{equation}
    C(x) = \frac{1}{2\sqrt{\pi D\, t}} e^{-\frac{x^2}{4 D t}},
\end{equation}
is the spatial profile of diffusion from a point source, with a diffusion constant $D$.
Performing the integration, we find
\begin{equation}
\begin{aligned}
    \left|S'(x,y)\right|^2 &= \frac{2}{\pi \left(8 D t+w^2\right)}e^{-2\frac{x^2+y^2}{8 D t+w^2}}\\
    &= \frac{2}{\pi w'^2}e^{-2\frac{x^2+y^2}{w'^2}},
\end{aligned}
\end{equation}
where
\begin{equation}
    w'^2 = w^2+8D t.
\end{equation}
If we assume the expansions imparts no radial phase upon the atoms, then the storage efficiency is given by
\begin{equation}
\begin{aligned}
    \eta &\propto \frac{4w^2w'^2}{\left(w^2+w'^2\right)^2}\\
    &\propto 1-\frac{1}{\left(1+\frac{w^2}{4D t}\right)^2}.
\end{aligned}
\end{equation}

\subsection{Magnetic Field Gradients}

Due to the differing magnetic field sensitivities of the states involved in the spin wave, magnetic field gradients can cause different parts of the SW to precess at different rates, and therefore cause dephasing.
Here, we calculate the effect of magnetic field gradients on the SW coherence.

In our system the longitudinal extent of the spin wave is significantly larger than that in the radial direction.
% CHECK THIS STATEMENT
Additionally, the magnetic shielding means the gradient in the radial direction is significantly smaller than in the longitudinal direction.
Therefore, here we shall only consider a gradient along the axis of the spin wave.
After some storage time we can write the spin wave as
\begin{equation}
    S'(z) = S(z) e^{-i \Delta(z) t},
\end{equation}
where $\Delta(z)$ is the position dependent energy shift due to the magnetic field.
If we assume that the magnetic field has a linear gradient
\begin{equation}
    B(z) = B_0 + B'z,
\end{equation}
then we can write SW
\begin{equation}
    S'(z) = S(z) e^{-i \left(B_0 + B'z\right)\mathcal{E}_B t},
\end{equation}
where $\mathcal{E}_B$ is the shift per unit magnetic field.
If we further assume $S(z)=$constant then the storage efficiency is given by
\begin{equation}
\begin{aligned}
    \eta &\propto \abs{\int_0^L \frac{dz}{L}\, e^{-i \left(B_0 + B'z\right)\mathcal{E}_B t}}^2\\
    &\propto \mathrm{sinc}^2\left(\frac{B'\mathcal{E}_B L t}{2}\right)
\end{aligned}
\end{equation}
where $L$ is the spatial extent of the spin wave.

\section{SNR affect on fidelity\label{SNR}}

The finite signal-to-noise-ratio (SNR) of the memory is a limiting factor in our quantum memory.
Here we calculate how the finite SNR of the memory effects its fidelity.
We assume an noise term of the form
\begin{equation}
    \hat{\rho}_n = \frac{1}{2} \left(\ket{H}\bra{H} + \ket{V}\bra{V}\right).
\end{equation}
The signal (input photon), $|\psi\rangle= \alpha |H\rangle+\beta |V\rangle$, is a pure state
\begin{equation}
    \hat{\rho}_{s} = \ket{\psi}\bra{\psi}
\end{equation}
%and for an SNR we can write the total density matrix after the memory
the density matrix after the memory would have the form
\begin{equation}
    \hat{\rho} = A \hat{\rho}_s +  B \hat{\rho}_n.
\end{equation}
where $A+B=1$, and $A/B$=SNR. This means $A=\frac{SNR}{1+SNR}$, and $B=\frac{1}{1+SNR}$. The fidelity of the density matrix post memory due to the SNR is then given by
\begin{equation}
    \begin{split}
        \mathcal{F} &= \left(\mathrm{Tr}\sqrt{\sqrt{\hat{\rho}_{s}} \hat{\rho} \sqrt{\hat{\rho}_{s}}}\right)^2\\
         &=  1- \frac{1}{2(1 + SNR)}
    \end{split}
\end{equation}

%--------------------------------------------------------------
\section{Differing rail transmission affect on fidelity\label{dual}}

Another factor which can affect the fidelity of the memory is the differing transmission of the two rails.
To quantify this effect let us consider an arbitrary state
\begin{equation}
    \ket{\psi} = \alpha\ket{H}+\beta\ket{V},
\end{equation}
where $ \alpha$ and $\beta$ are complex numbers normalised such that $| \alpha|^2+|\beta|^2=1$.
We'll assume that one polarization experiences perfect transmission, while the other is imperfect with a real coefficient $0\leq t \leq 1$, resulting in a state
\begin{equation}
    \ket{\psi'}= \frac{1}{\sqrt{|\alpha|^2+|\beta|^2t^2}}\left(\alpha\ket{H}+\beta t\ket{V}\right),
\end{equation}
where we have re-normalized to account for the loss in the $\ket{V}$ mode.
The fidelity, given the loss, is then
\begin{equation}
    \begin{split}
        F&=\left|\bra{\psi}\ket{\psi'}\right|^2\\
        &= \frac{\left( 1+\mathcal{B}\left(t-1\right)\right)^2}{1+\mathcal{B}\left(t^2-1\right)},
    \end{split}
\end{equation}
where we've made use of the fact that $|\alpha|^2+|\beta|^2=1$ to eliminate $\alpha$ and set $\mathcal{B}=|\beta|^2$.
Differentiating this to find the extrema
\begin{equation}
    \frac{\partial F}{\partial \mathcal{B}} = 0 \rightarrow \mathcal{B} = \frac{1}{1+t},\frac{1}{1-t}.
\end{equation}
The second solution is unphysical, while the first solution corresponds to the state for which the fidelity is a minimum, thus
\begin{equation}
    F_{\rm{transmission}} \geq \frac{4 \sqrt{T}}{\left(1+\sqrt{T}\right)^2},
\end{equation}
where $T = t^2$ here is the transmission for the imperfectly transmitted rail.

%\begin{thebibliography}%
%\end{thebibliography}%

\bibliographystyle{apsrev4-2.bst}
\bibliography{library}

\providecommand{\noopsort}[1]{}\providecommand{\singleletter}[1]{#1}%
\begin{thebibliography}{91}%
\makeatletter
\providecommand \@ifxundefined [1]{%
 \@ifx{#1\undefined}
}%
\providecommand \@ifnum [1]{%
 \ifnum #1\expandafter \@firstoftwo
 \else \expandafter \@secondoftwo
 \fi
}%
\providecommand \@ifx [1]{%
 \ifx #1\expandafter \@firstoftwo
 \else \expandafter \@secondoftwo
 \fi
}%
\providecommand \natexlab [1]{#1}%
\providecommand \enquote  [1]{``#1''}%
\providecommand \bibnamefont  [1]{#1}%
\providecommand \bibfnamefont [1]{#1}%
\providecommand \citenamefont [1]{#1}%
\providecommand \href@noop [0]{\@secondoftwo}%
\providecommand \href [0]{\begingroup \@sanitize@url \@href}%
\providecommand \@href[1]{\@@startlink{#1}\@@href}%
\providecommand \@@href[1]{\endgroup#1\@@endlink}%
\providecommand \@sanitize@url [0]{\catcode `\\12\catcode `\$12\catcode
  `\&12\catcode `\#12\catcode `\^12\catcode `\_12\catcode `\%12\relax}%
\providecommand \@@startlink[1]{}%
\providecommand \@@endlink[0]{}%
\providecommand \url  [0]{\begingroup\@sanitize@url \@url }%
\providecommand \@url [1]{\endgroup\@href {#1}{\urlprefix }}%
\providecommand \urlprefix  [0]{URL }%
\providecommand \Eprint [0]{\href }%
\providecommand \doibase [0]{https://doi.org/}%
\providecommand \selectlanguage [0]{\@gobble}%
\providecommand \bibinfo  [0]{\@secondoftwo}%
\providecommand \bibfield  [0]{\@secondoftwo}%
\providecommand \translation [1]{[#1]}%
\providecommand \BibitemOpen [0]{}%
\providecommand \bibitemStop [0]{}%
\providecommand \bibitemNoStop [0]{.\EOS\space}%
\providecommand \EOS [0]{\spacefactor3000\relax}%
\providecommand \BibitemShut  [1]{\csname bibitem#1\endcsname}%
\let\auto@bib@innerbib\@empty
%</preamble>
\bibitem [{\citenamefont {Simon}(2017)}]{Simon2017}%
  \BibitemOpen
  \bibfield  {author} {\bibinfo {author} {\bibfnamefont {C.}~\bibnamefont
  {Simon}},\ }\href {https://doi.org/10.1038/s41566-017-0032-0} {\bibfield
  {journal} {\bibinfo  {journal} {Nature Photonics}\ }\textbf {\bibinfo
  {volume} {11}},\ \bibinfo {pages} {678} (\bibinfo {year} {2017})}\BibitemShut
  {NoStop}%
\bibitem [{\citenamefont {Wehner}\ \emph {et~al.}(2018)\citenamefont {Wehner},
  \citenamefont {Elkouss},\ and\ \citenamefont {Hanson}}]{Wehner2018}%
  \BibitemOpen
  \bibfield  {author} {\bibinfo {author} {\bibfnamefont {S.}~\bibnamefont
  {Wehner}}, \bibinfo {author} {\bibfnamefont {D.}~\bibnamefont {Elkouss}},\
  and\ \bibinfo {author} {\bibfnamefont {R.}~\bibnamefont {Hanson}},\ }\href
  {https://doi.org/10.1126/science.aam9288} {\bibfield  {journal} {\bibinfo
  {journal} {Science}\ }\textbf {\bibinfo {volume} {362}},\ \bibinfo {pages}
  {9288} (\bibinfo {year} {2018})}\BibitemShut {NoStop}%
\bibitem [{\citenamefont {Pirandola}\ and\ \citenamefont
  {Braunstein}(2016)}]{Pirandola2016}%
  \BibitemOpen
  \bibfield  {author} {\bibinfo {author} {\bibfnamefont {S.}~\bibnamefont
  {Pirandola}}\ and\ \bibinfo {author} {\bibfnamefont {S.~L.}\ \bibnamefont
  {Braunstein}},\ }\href {https://doi.org/10.1038/532169a} {\bibfield
  {journal} {\bibinfo  {journal} {Nature}\ }\textbf {\bibinfo {volume} {532}},\
  \bibinfo {pages} {169} (\bibinfo {year} {2016})}\BibitemShut {NoStop}%
\bibitem [{\citenamefont {Kimble}(2008)}]{Kimble2008}%
  \BibitemOpen
  \bibfield  {author} {\bibinfo {author} {\bibfnamefont {H.~J.}\ \bibnamefont
  {Kimble}},\ }\href {https://doi.org/10.1038/nature07127} {\bibfield
  {journal} {\bibinfo  {journal} {Nature}\ }\textbf {\bibinfo {volume} {453}},\
  \bibinfo {pages} {1023} (\bibinfo {year} {2008})}\BibitemShut {NoStop}%
\bibitem [{\citenamefont {Awschalom}\ \emph {et~al.}(2021)\citenamefont
  {Awschalom}, \citenamefont {Berggren}, \citenamefont {Bernien}, \citenamefont
  {Bhave}, \citenamefont {Carr}, \citenamefont {Davids}, \citenamefont
  {Economou}, \citenamefont {Englund}, \citenamefont {Faraon}, \citenamefont
  {Fejer}, \citenamefont {Guha}, \citenamefont {Gustafsson}, \citenamefont
  {Hu}, \citenamefont {Jiang}, \citenamefont {Kim}, \citenamefont {Korzh},
  \citenamefont {Kumar}, \citenamefont {Kwiat}, \citenamefont
  {Lon\ifmmode~\check{c}\else \v{c}\fi{}ar}, \citenamefont {Lukin},
  \citenamefont {Miller}, \citenamefont {Monroe}, \citenamefont {Nam},
  \citenamefont {Narang}, \citenamefont {Orcutt}, \citenamefont {Raymer},
  \citenamefont {Safavi-Naeini}, \citenamefont {Spiropulu}, \citenamefont
  {Srinivasan}, \citenamefont {Sun}, \citenamefont {Vu\ifmmode \check{c}\else
  \v{c}\fi{}kovi\ifmmode~\acute{c}\else \'{c}\fi{}}, \citenamefont {Waks},
  \citenamefont {Walsworth}, \citenamefont {Weiner},\ and\ \citenamefont
  {Zhang}}]{Awschalom2021}%
  \BibitemOpen
  \bibfield  {author} {\bibinfo {author} {\bibfnamefont {D.}~\bibnamefont
  {Awschalom}}, \bibinfo {author} {\bibfnamefont {K.~K.}\ \bibnamefont
  {Berggren}}, \bibinfo {author} {\bibfnamefont {H.}~\bibnamefont {Bernien}},
  \bibinfo {author} {\bibfnamefont {S.}~\bibnamefont {Bhave}}, \bibinfo
  {author} {\bibfnamefont {L.~D.}\ \bibnamefont {Carr}}, \bibinfo {author}
  {\bibfnamefont {P.}~\bibnamefont {Davids}}, \bibinfo {author} {\bibfnamefont
  {S.~E.}\ \bibnamefont {Economou}}, \bibinfo {author} {\bibfnamefont
  {D.}~\bibnamefont {Englund}}, \bibinfo {author} {\bibfnamefont
  {A.}~\bibnamefont {Faraon}}, \bibinfo {author} {\bibfnamefont
  {M.}~\bibnamefont {Fejer}}, \bibinfo {author} {\bibfnamefont
  {S.}~\bibnamefont {Guha}}, \bibinfo {author} {\bibfnamefont {M.~V.}\
  \bibnamefont {Gustafsson}}, \bibinfo {author} {\bibfnamefont
  {E.}~\bibnamefont {Hu}}, \bibinfo {author} {\bibfnamefont {L.}~\bibnamefont
  {Jiang}}, \bibinfo {author} {\bibfnamefont {J.}~\bibnamefont {Kim}}, \bibinfo
  {author} {\bibfnamefont {B.}~\bibnamefont {Korzh}}, \bibinfo {author}
  {\bibfnamefont {P.}~\bibnamefont {Kumar}}, \bibinfo {author} {\bibfnamefont
  {P.~G.}\ \bibnamefont {Kwiat}}, \bibinfo {author} {\bibfnamefont
  {M.}~\bibnamefont {Lon\ifmmode~\check{c}\else \v{c}\fi{}ar}}, \bibinfo
  {author} {\bibfnamefont {M.~D.}\ \bibnamefont {Lukin}}, \bibinfo {author}
  {\bibfnamefont {D.~A.}\ \bibnamefont {Miller}}, \bibinfo {author}
  {\bibfnamefont {C.}~\bibnamefont {Monroe}}, \bibinfo {author} {\bibfnamefont
  {S.~W.}\ \bibnamefont {Nam}}, \bibinfo {author} {\bibfnamefont
  {P.}~\bibnamefont {Narang}}, \bibinfo {author} {\bibfnamefont {J.~S.}\
  \bibnamefont {Orcutt}}, \bibinfo {author} {\bibfnamefont {M.~G.}\
  \bibnamefont {Raymer}}, \bibinfo {author} {\bibfnamefont {A.~H.}\
  \bibnamefont {Safavi-Naeini}}, \bibinfo {author} {\bibfnamefont
  {M.}~\bibnamefont {Spiropulu}}, \bibinfo {author} {\bibfnamefont
  {K.}~\bibnamefont {Srinivasan}}, \bibinfo {author} {\bibfnamefont
  {S.}~\bibnamefont {Sun}}, \bibinfo {author} {\bibfnamefont {J.}~\bibnamefont
  {Vu\ifmmode \check{c}\else \v{c}\fi{}kovi\ifmmode~\acute{c}\else
  \'{c}\fi{}}}, \bibinfo {author} {\bibfnamefont {E.}~\bibnamefont {Waks}},
  \bibinfo {author} {\bibfnamefont {R.}~\bibnamefont {Walsworth}}, \bibinfo
  {author} {\bibfnamefont {A.~M.}\ \bibnamefont {Weiner}},\ and\ \bibinfo
  {author} {\bibfnamefont {Z.}~\bibnamefont {Zhang}},\ }\href
  {https://doi.org/10.1103/PRXQuantum.2.017002} {\bibfield  {journal} {\bibinfo
   {journal} {PRX Quantum}\ }\textbf {\bibinfo {volume} {2}},\ \bibinfo {pages}
  {017002} (\bibinfo {year} {2021})}\BibitemShut {NoStop}%
\bibitem [{\citenamefont {Gisin}\ \emph {et~al.}(2002)\citenamefont {Gisin},
  \citenamefont {Ribordy}, \citenamefont {Tittel},\ and\ \citenamefont
  {Zbinden}}]{Gisin2002}%
  \BibitemOpen
  \bibfield  {author} {\bibinfo {author} {\bibfnamefont {N.}~\bibnamefont
  {Gisin}}, \bibinfo {author} {\bibfnamefont {G.}~\bibnamefont {Ribordy}},
  \bibinfo {author} {\bibfnamefont {W.}~\bibnamefont {Tittel}},\ and\ \bibinfo
  {author} {\bibfnamefont {H.}~\bibnamefont {Zbinden}},\ }\href
  {https://doi.org/10.1103/RevModPhys.74.145} {\bibfield  {journal} {\bibinfo
  {journal} {Rev. Mod. Phys.}\ }\textbf {\bibinfo {volume} {74}},\ \bibinfo
  {pages} {145} (\bibinfo {year} {2002})}\BibitemShut {NoStop}%
\bibitem [{\citenamefont {Ekert}\ and\ \citenamefont
  {Renner}(2014)}]{Ekert2014}%
  \BibitemOpen
  \bibfield  {author} {\bibinfo {author} {\bibfnamefont {A.}~\bibnamefont
  {Ekert}}\ and\ \bibinfo {author} {\bibfnamefont {R.}~\bibnamefont {Renner}},\
  }\href {https://doi.org/10.1038/nature13132} {\bibfield  {journal} {\bibinfo
  {journal} {Nature}\ }\textbf {\bibinfo {volume} {507}},\ \bibinfo {pages}
  {443} (\bibinfo {year} {2014})}\BibitemShut {NoStop}%
\bibitem [{\citenamefont {Scarani}\ \emph {et~al.}(2009)\citenamefont
  {Scarani}, \citenamefont {Bechmann-Pasquinucci}, \citenamefont {Cerf},
  \citenamefont {Du\ifmmode~\check{s}\else \v{s}\fi{}ek}, \citenamefont
  {L\"utkenhaus},\ and\ \citenamefont {Peev}}]{Scarani2009}%
  \BibitemOpen
  \bibfield  {author} {\bibinfo {author} {\bibfnamefont {V.}~\bibnamefont
  {Scarani}}, \bibinfo {author} {\bibfnamefont {H.}~\bibnamefont
  {Bechmann-Pasquinucci}}, \bibinfo {author} {\bibfnamefont {N.~J.}\
  \bibnamefont {Cerf}}, \bibinfo {author} {\bibfnamefont {M.}~\bibnamefont
  {Du\ifmmode~\check{s}\else \v{s}\fi{}ek}}, \bibinfo {author} {\bibfnamefont
  {N.}~\bibnamefont {L\"utkenhaus}},\ and\ \bibinfo {author} {\bibfnamefont
  {M.}~\bibnamefont {Peev}},\ }\href
  {https://doi.org/10.1103/RevModPhys.81.1301} {\bibfield  {journal} {\bibinfo
  {journal} {Rev. Mod. Phys.}\ }\textbf {\bibinfo {volume} {81}},\ \bibinfo
  {pages} {1301} (\bibinfo {year} {2009})}\BibitemShut {NoStop}%
\bibitem [{\citenamefont {Xu}\ \emph {et~al.}(2020)\citenamefont {Xu},
  \citenamefont {Ma}, \citenamefont {Zhang}, \citenamefont {Lo},\ and\
  \citenamefont {Pan}}]{Xu2020}%
  \BibitemOpen
  \bibfield  {author} {\bibinfo {author} {\bibfnamefont {F.}~\bibnamefont
  {Xu}}, \bibinfo {author} {\bibfnamefont {X.}~\bibnamefont {Ma}}, \bibinfo
  {author} {\bibfnamefont {Q.}~\bibnamefont {Zhang}}, \bibinfo {author}
  {\bibfnamefont {H.-K.}\ \bibnamefont {Lo}},\ and\ \bibinfo {author}
  {\bibfnamefont {J.-W.}\ \bibnamefont {Pan}},\ }\href
  {https://doi.org/10.1103/RevModPhys.92.025002} {\bibfield  {journal}
  {\bibinfo  {journal} {Rev. Mod. Phys.}\ }\textbf {\bibinfo {volume} {92}},\
  \bibinfo {pages} {025002} (\bibinfo {year} {2020})}\BibitemShut {NoStop}%
\bibitem [{\citenamefont {Cirac}\ \emph {et~al.}(1999)\citenamefont {Cirac},
  \citenamefont {Ekert}, \citenamefont {Huelga},\ and\ \citenamefont
  {Macchiavello}}]{Cirac1999}%
  \BibitemOpen
  \bibfield  {author} {\bibinfo {author} {\bibfnamefont {J.~I.}\ \bibnamefont
  {Cirac}}, \bibinfo {author} {\bibfnamefont {A.~K.}\ \bibnamefont {Ekert}},
  \bibinfo {author} {\bibfnamefont {S.~F.}\ \bibnamefont {Huelga}},\ and\
  \bibinfo {author} {\bibfnamefont {C.}~\bibnamefont {Macchiavello}},\ }\href
  {https://doi.org/10.1103/PhysRevA.59.4249} {\bibfield  {journal} {\bibinfo
  {journal} {Phys. Rev. A}\ }\textbf {\bibinfo {volume} {59}},\ \bibinfo
  {pages} {4249} (\bibinfo {year} {1999})}\BibitemShut {NoStop}%
\bibitem [{\citenamefont {Cacciapuoti}\ \emph {et~al.}(2020)\citenamefont
  {Cacciapuoti}, \citenamefont {Caleffi}, \citenamefont {Tafuri}, \citenamefont
  {Cataliotti}, \citenamefont {Gherardini},\ and\ \citenamefont
  {Bianchi}}]{Cacciapuoti2020}%
  \BibitemOpen
  \bibfield  {author} {\bibinfo {author} {\bibfnamefont {A.~S.}\ \bibnamefont
  {Cacciapuoti}}, \bibinfo {author} {\bibfnamefont {M.}~\bibnamefont
  {Caleffi}}, \bibinfo {author} {\bibfnamefont {F.}~\bibnamefont {Tafuri}},
  \bibinfo {author} {\bibfnamefont {F.~S.}\ \bibnamefont {Cataliotti}},
  \bibinfo {author} {\bibfnamefont {S.}~\bibnamefont {Gherardini}},\ and\
  \bibinfo {author} {\bibfnamefont {G.}~\bibnamefont {Bianchi}},\ }\href
  {https://doi.org/10.1109/MNET.001.1900092} {\bibfield  {journal} {\bibinfo
  {journal} {IEEE Network}\ }\textbf {\bibinfo {volume} {34}},\ \bibinfo
  {pages} {137} (\bibinfo {year} {2020})}\BibitemShut {NoStop}%
\bibitem [{\citenamefont {K{\'{o}}m{\'{a}}r}\ \emph {et~al.}(2014)\citenamefont
  {K{\'{o}}m{\'{a}}r}, \citenamefont {Kessler}, \citenamefont {Bishof},
  \citenamefont {Jiang}, \citenamefont {S{\o}rensen}, \citenamefont {Ye},\ and\
  \citenamefont {Lukin}}]{Komar2014}%
  \BibitemOpen
  \bibfield  {author} {\bibinfo {author} {\bibfnamefont {P.}~\bibnamefont
  {K{\'{o}}m{\'{a}}r}}, \bibinfo {author} {\bibfnamefont {E.~M.}\ \bibnamefont
  {Kessler}}, \bibinfo {author} {\bibfnamefont {M.}~\bibnamefont {Bishof}},
  \bibinfo {author} {\bibfnamefont {L.}~\bibnamefont {Jiang}}, \bibinfo
  {author} {\bibfnamefont {A.~S.}\ \bibnamefont {S{\o}rensen}}, \bibinfo
  {author} {\bibfnamefont {J.}~\bibnamefont {Ye}},\ and\ \bibinfo {author}
  {\bibfnamefont {M.~D.}\ \bibnamefont {Lukin}},\ }\href
  {https://doi.org/10.1038/nphys3000} {\bibfield  {journal} {\bibinfo
  {journal} {Nature Physics}\ }\textbf {\bibinfo {volume} {10}},\ \bibinfo
  {pages} {582} (\bibinfo {year} {2014})}\BibitemShut {NoStop}%
\bibitem [{\citenamefont {Gottesman}\ \emph {et~al.}(2012)\citenamefont
  {Gottesman}, \citenamefont {Jennewein},\ and\ \citenamefont
  {Croke}}]{Gottesman2021}%
  \BibitemOpen
  \bibfield  {author} {\bibinfo {author} {\bibfnamefont {D.}~\bibnamefont
  {Gottesman}}, \bibinfo {author} {\bibfnamefont {T.}~\bibnamefont
  {Jennewein}},\ and\ \bibinfo {author} {\bibfnamefont {S.}~\bibnamefont
  {Croke}},\ }\href {https://doi.org/10.1103/PhysRevLett.109.070503} {\bibfield
   {journal} {\bibinfo  {journal} {Phys. Rev. Lett.}\ }\textbf {\bibinfo
  {volume} {109}},\ \bibinfo {pages} {070503} (\bibinfo {year}
  {2012})}\BibitemShut {NoStop}%
\bibitem [{\citenamefont {Sidhu}\ \emph {et~al.}(2021)\citenamefont {Sidhu},
  \citenamefont {Joshi}, \citenamefont {Gündoğan}, \citenamefont {Brougham},
  \citenamefont {Lowndes}, \citenamefont {Mazzarella}, \citenamefont {Krutzik},
  \citenamefont {Mohapatra}, \citenamefont {Dequal}, \citenamefont {Vallone},
  \citenamefont {Villoresi}, \citenamefont {Ling}, \citenamefont {Jennewein},
  \citenamefont {Mohageg}, \citenamefont {Rarity}, \citenamefont {Fuentes},
  \citenamefont {Pirandola},\ and\ \citenamefont {Oi}}]{Jasminder2021}%
  \BibitemOpen
  \bibfield  {author} {\bibinfo {author} {\bibfnamefont {J.~S.}\ \bibnamefont
  {Sidhu}}, \bibinfo {author} {\bibfnamefont {S.~K.}\ \bibnamefont {Joshi}},
  \bibinfo {author} {\bibfnamefont {M.}~\bibnamefont {Gündoğan}}, \bibinfo
  {author} {\bibfnamefont {T.}~\bibnamefont {Brougham}}, \bibinfo {author}
  {\bibfnamefont {D.}~\bibnamefont {Lowndes}}, \bibinfo {author} {\bibfnamefont
  {L.}~\bibnamefont {Mazzarella}}, \bibinfo {author} {\bibfnamefont
  {M.}~\bibnamefont {Krutzik}}, \bibinfo {author} {\bibfnamefont
  {S.}~\bibnamefont {Mohapatra}}, \bibinfo {author} {\bibfnamefont
  {D.}~\bibnamefont {Dequal}}, \bibinfo {author} {\bibfnamefont
  {G.}~\bibnamefont {Vallone}}, \bibinfo {author} {\bibfnamefont
  {P.}~\bibnamefont {Villoresi}}, \bibinfo {author} {\bibfnamefont
  {A.}~\bibnamefont {Ling}}, \bibinfo {author} {\bibfnamefont {T.}~\bibnamefont
  {Jennewein}}, \bibinfo {author} {\bibfnamefont {M.}~\bibnamefont {Mohageg}},
  \bibinfo {author} {\bibfnamefont {J.~G.}\ \bibnamefont {Rarity}}, \bibinfo
  {author} {\bibfnamefont {I.}~\bibnamefont {Fuentes}}, \bibinfo {author}
  {\bibfnamefont {S.}~\bibnamefont {Pirandola}},\ and\ \bibinfo {author}
  {\bibfnamefont {D.~K.~L.}\ \bibnamefont {Oi}},\ }\href
  {https://doi.org/https://doi.org/10.1049/qtc2.12015} {\bibfield  {journal}
  {\bibinfo  {journal} {IET Quantum Communication}\ }\textbf {\bibinfo {volume}
  {2}},\ \bibinfo {pages} {182} (\bibinfo {year} {2021})}\BibitemShut {NoStop}%
\bibitem [{\citenamefont {Guo}\ \emph {et~al.}(2020)\citenamefont {Guo},
  \citenamefont {Breum}, \citenamefont {Borregaard}, \citenamefont {Izumi},
  \citenamefont {Larsen}, \citenamefont {Gehring}, \citenamefont {Christandl},
  \citenamefont {Neergaard-Nielsen},\ and\ \citenamefont {Andersen}}]{Guo2020}%
  \BibitemOpen
  \bibfield  {author} {\bibinfo {author} {\bibfnamefont {X.}~\bibnamefont
  {Guo}}, \bibinfo {author} {\bibfnamefont {C.~R.}\ \bibnamefont {Breum}},
  \bibinfo {author} {\bibfnamefont {J.}~\bibnamefont {Borregaard}}, \bibinfo
  {author} {\bibfnamefont {S.}~\bibnamefont {Izumi}}, \bibinfo {author}
  {\bibfnamefont {M.~V.}\ \bibnamefont {Larsen}}, \bibinfo {author}
  {\bibfnamefont {T.}~\bibnamefont {Gehring}}, \bibinfo {author} {\bibfnamefont
  {M.}~\bibnamefont {Christandl}}, \bibinfo {author} {\bibfnamefont {J.~S.}\
  \bibnamefont {Neergaard-Nielsen}},\ and\ \bibinfo {author} {\bibfnamefont
  {U.~L.}\ \bibnamefont {Andersen}},\ }\href
  {https://doi.org/10.1038/s41567-019-0743-x} {\bibfield  {journal} {\bibinfo
  {journal} {Nature Physics}\ }\textbf {\bibinfo {volume} {16}},\ \bibinfo
  {pages} {281} (\bibinfo {year} {2020})}\BibitemShut {NoStop}%
\bibitem [{\citenamefont {Dieks}(1982)}]{Dieks1982}%
  \BibitemOpen
  \bibfield  {author} {\bibinfo {author} {\bibfnamefont {D.}~\bibnamefont
  {Dieks}},\ }\href {https://doi.org/10.1016/0375-9601(82)90084-6} {\bibfield
  {journal} {\bibinfo  {journal} {Phys. Lett. A}\ }\textbf {\bibinfo {volume}
  {92}},\ \bibinfo {pages} {271} (\bibinfo {year} {1982})}\BibitemShut
  {NoStop}%
\bibitem [{\citenamefont {Wootters}\ and\ \citenamefont
  {Zurek}(1982)}]{Wootters1982}%
  \BibitemOpen
  \bibfield  {author} {\bibinfo {author} {\bibfnamefont {W.~K.}\ \bibnamefont
  {Wootters}}\ and\ \bibinfo {author} {\bibfnamefont {W.~H.}\ \bibnamefont
  {Zurek}},\ }\href {https://doi.org/10.1038/299802a0} {\bibfield  {journal}
  {\bibinfo  {journal} {Nature}\ }\textbf {\bibinfo {volume} {299}},\ \bibinfo
  {pages} {802} (\bibinfo {year} {1982})}\BibitemShut {NoStop}%
\bibitem [{\citenamefont {Ekert}(1991)}]{Ekert1991}%
  \BibitemOpen
  \bibfield  {author} {\bibinfo {author} {\bibfnamefont {A.~K.}\ \bibnamefont
  {Ekert}},\ }\href {https://doi.org/10.1103/PhysRevLett.67.661} {\bibfield
  {journal} {\bibinfo  {journal} {Phys. Rev. Lett.}\ }\textbf {\bibinfo
  {volume} {67}},\ \bibinfo {pages} {661} (\bibinfo {year} {1991})}\BibitemShut
  {NoStop}%
\bibitem [{\citenamefont {Briegel}\ \emph {et~al.}(1998)\citenamefont
  {Briegel}, \citenamefont {D\"ur}, \citenamefont {Cirac},\ and\ \citenamefont
  {Zoller}}]{Briegel1998}%
  \BibitemOpen
  \bibfield  {author} {\bibinfo {author} {\bibfnamefont {H.-J.}\ \bibnamefont
  {Briegel}}, \bibinfo {author} {\bibfnamefont {W.}~\bibnamefont {D\"ur}},
  \bibinfo {author} {\bibfnamefont {J.~I.}\ \bibnamefont {Cirac}},\ and\
  \bibinfo {author} {\bibfnamefont {P.}~\bibnamefont {Zoller}},\ }\href
  {https://doi.org/10.1103/PhysRevLett.81.5932} {\bibfield  {journal} {\bibinfo
   {journal} {Phys. Rev. Lett.}\ }\textbf {\bibinfo {volume} {81}},\ \bibinfo
  {pages} {5932} (\bibinfo {year} {1998})}\BibitemShut {NoStop}%
\bibitem [{\citenamefont {Duan}\ \emph {et~al.}(2001)\citenamefont {Duan},
  \citenamefont {Lukin}, \citenamefont {Cirac},\ and\ \citenamefont
  {Zoller}}]{Duan2001}%
  \BibitemOpen
  \bibfield  {author} {\bibinfo {author} {\bibfnamefont {L.-M.}\ \bibnamefont
  {Duan}}, \bibinfo {author} {\bibfnamefont {M.~D.}\ \bibnamefont {Lukin}},
  \bibinfo {author} {\bibfnamefont {J.~I.}\ \bibnamefont {Cirac}},\ and\
  \bibinfo {author} {\bibfnamefont {P.}~\bibnamefont {Zoller}},\ }\href
  {https://doi.org/10.1038/35106500} {\bibfield  {journal} {\bibinfo  {journal}
  {Nature}\ }\textbf {\bibinfo {volume} {414}},\ \bibinfo {pages} {413}
  (\bibinfo {year} {2001})}\BibitemShut {NoStop}%
\bibitem [{\citenamefont {Sangouard}\ \emph {et~al.}(2011)\citenamefont
  {Sangouard}, \citenamefont {Simon}, \citenamefont {de~Riedmatten},\ and\
  \citenamefont {Gisin}}]{Sangouard2011}%
  \BibitemOpen
  \bibfield  {author} {\bibinfo {author} {\bibfnamefont {N.}~\bibnamefont
  {Sangouard}}, \bibinfo {author} {\bibfnamefont {C.}~\bibnamefont {Simon}},
  \bibinfo {author} {\bibfnamefont {H.}~\bibnamefont {de~Riedmatten}},\ and\
  \bibinfo {author} {\bibfnamefont {N.}~\bibnamefont {Gisin}},\ }\href
  {https://doi.org/10.1103/RevModPhys.83.33} {\bibfield  {journal} {\bibinfo
  {journal} {Rev. Mod. Phys.}\ }\textbf {\bibinfo {volume} {83}},\ \bibinfo
  {pages} {33} (\bibinfo {year} {2011})}\BibitemShut {NoStop}%
\bibitem [{\citenamefont {Hedges}\ \emph {et~al.}(2010)\citenamefont {Hedges},
  \citenamefont {Longdell}, \citenamefont {Li},\ and\ \citenamefont
  {Sellars}}]{Hedges2010}%
  \BibitemOpen
  \bibfield  {author} {\bibinfo {author} {\bibfnamefont {M.~P.}\ \bibnamefont
  {Hedges}}, \bibinfo {author} {\bibfnamefont {J.~J.}\ \bibnamefont
  {Longdell}}, \bibinfo {author} {\bibfnamefont {Y.}~\bibnamefont {Li}},\ and\
  \bibinfo {author} {\bibfnamefont {M.~J.}\ \bibnamefont {Sellars}},\ }\href
  {https://doi.org/10.1038/nature09081} {\bibfield  {journal} {\bibinfo
  {journal} {Nature}\ }\textbf {\bibinfo {volume} {465}},\ \bibinfo {pages}
  {1052} (\bibinfo {year} {2010})}\BibitemShut {NoStop}%
\bibitem [{\citenamefont {Julsgaard}\ \emph {et~al.}(2004)\citenamefont
  {Julsgaard}, \citenamefont {Sherson}, \citenamefont {Cirac}, \citenamefont
  {Fiur{\'{a}}{\v{s}}ek},\ and\ \citenamefont {Polzik}}]{Julsgaard2004}%
  \BibitemOpen
  \bibfield  {author} {\bibinfo {author} {\bibfnamefont {B.}~\bibnamefont
  {Julsgaard}}, \bibinfo {author} {\bibfnamefont {J.}~\bibnamefont {Sherson}},
  \bibinfo {author} {\bibfnamefont {J.~I.}\ \bibnamefont {Cirac}}, \bibinfo
  {author} {\bibfnamefont {J.}~\bibnamefont {Fiur{\'{a}}{\v{s}}ek}},\ and\
  \bibinfo {author} {\bibfnamefont {E.~S.}\ \bibnamefont {Polzik}},\ }\href
  {https://doi.org/10.1038/nature03064} {\bibfield  {journal} {\bibinfo
  {journal} {Nature}\ }\textbf {\bibinfo {volume} {432}},\ \bibinfo {pages}
  {482} (\bibinfo {year} {2004})}\BibitemShut {NoStop}%
\bibitem [{\citenamefont {Lukin}(2003)}]{Lukin2003}%
  \BibitemOpen
  \bibfield  {author} {\bibinfo {author} {\bibfnamefont {M.~D.}\ \bibnamefont
  {Lukin}},\ }\href {https://doi.org/10.1103/RevModPhys.75.457} {\bibfield
  {journal} {\bibinfo  {journal} {Rev. Mod. Phys.}\ }\textbf {\bibinfo {volume}
  {75}},\ \bibinfo {pages} {457} (\bibinfo {year} {2003})}\BibitemShut
  {NoStop}%
\bibitem [{\citenamefont {Hammerer}\ \emph {et~al.}(2010)\citenamefont
  {Hammerer}, \citenamefont {S\o{}rensen},\ and\ \citenamefont
  {Polzik}}]{Hammerer2010}%
  \BibitemOpen
  \bibfield  {author} {\bibinfo {author} {\bibfnamefont {K.}~\bibnamefont
  {Hammerer}}, \bibinfo {author} {\bibfnamefont {A.~S.}\ \bibnamefont
  {S\o{}rensen}},\ and\ \bibinfo {author} {\bibfnamefont {E.~S.}\ \bibnamefont
  {Polzik}},\ }\href {https://doi.org/10.1103/RevModPhys.82.1041} {\bibfield
  {journal} {\bibinfo  {journal} {Rev. Mod. Phys.}\ }\textbf {\bibinfo {volume}
  {82}},\ \bibinfo {pages} {1041} (\bibinfo {year} {2010})}\BibitemShut
  {NoStop}%
\bibitem [{\citenamefont {Graham}\ \emph {et~al.}(2022)\citenamefont {Graham},
  \citenamefont {Song}, \citenamefont {Scott}, \citenamefont {Poole},
  \citenamefont {Phuttitarn}, \citenamefont {Jooya}, \citenamefont {Eichler},
  \citenamefont {Jiang}, \citenamefont {Marra}, \citenamefont {Grinkemeyer},
  \citenamefont {Kwon}, \citenamefont {Ebert}, \citenamefont {Cherek},
  \citenamefont {Lichtman}, \citenamefont {Gillette}, \citenamefont {Gilbert},
  \citenamefont {Bowman}, \citenamefont {Ballance}, \citenamefont {Campbell},
  \citenamefont {Dahl}, \citenamefont {Crawford}, \citenamefont {Blunt},
  \citenamefont {Rogers}, \citenamefont {Noel},\ and\ \citenamefont
  {Saffman}}]{Graham2022}%
  \BibitemOpen
  \bibfield  {author} {\bibinfo {author} {\bibfnamefont {T.~M.}\ \bibnamefont
  {Graham}}, \bibinfo {author} {\bibfnamefont {Y.}~\bibnamefont {Song}},
  \bibinfo {author} {\bibfnamefont {J.}~\bibnamefont {Scott}}, \bibinfo
  {author} {\bibfnamefont {C.}~\bibnamefont {Poole}}, \bibinfo {author}
  {\bibfnamefont {L.}~\bibnamefont {Phuttitarn}}, \bibinfo {author}
  {\bibfnamefont {K.}~\bibnamefont {Jooya}}, \bibinfo {author} {\bibfnamefont
  {P.}~\bibnamefont {Eichler}}, \bibinfo {author} {\bibfnamefont
  {X.}~\bibnamefont {Jiang}}, \bibinfo {author} {\bibfnamefont
  {A.}~\bibnamefont {Marra}}, \bibinfo {author} {\bibfnamefont
  {B.}~\bibnamefont {Grinkemeyer}}, \bibinfo {author} {\bibfnamefont
  {M.}~\bibnamefont {Kwon}}, \bibinfo {author} {\bibfnamefont {M.}~\bibnamefont
  {Ebert}}, \bibinfo {author} {\bibfnamefont {J.}~\bibnamefont {Cherek}},
  \bibinfo {author} {\bibfnamefont {M.~T.}\ \bibnamefont {Lichtman}}, \bibinfo
  {author} {\bibfnamefont {M.}~\bibnamefont {Gillette}}, \bibinfo {author}
  {\bibfnamefont {J.}~\bibnamefont {Gilbert}}, \bibinfo {author} {\bibfnamefont
  {D.}~\bibnamefont {Bowman}}, \bibinfo {author} {\bibfnamefont
  {T.}~\bibnamefont {Ballance}}, \bibinfo {author} {\bibfnamefont
  {C.}~\bibnamefont {Campbell}}, \bibinfo {author} {\bibfnamefont {E.~D.}\
  \bibnamefont {Dahl}}, \bibinfo {author} {\bibfnamefont {O.}~\bibnamefont
  {Crawford}}, \bibinfo {author} {\bibfnamefont {N.~S.}\ \bibnamefont {Blunt}},
  \bibinfo {author} {\bibfnamefont {B.}~\bibnamefont {Rogers}}, \bibinfo
  {author} {\bibfnamefont {T.}~\bibnamefont {Noel}},\ and\ \bibinfo {author}
  {\bibfnamefont {M.}~\bibnamefont {Saffman}},\ }\href
  {https://doi.org/10.1038/s41586-022-04603-6} {\bibfield  {journal} {\bibinfo
  {journal} {Nature}\ }\textbf {\bibinfo {volume} {604}},\ \bibinfo {pages}
  {457} (\bibinfo {year} {2022})}\BibitemShut {NoStop}%
\bibitem [{\citenamefont {Bluvstein}\ \emph {et~al.}(2022)\citenamefont
  {Bluvstein}, \citenamefont {Levine}, \citenamefont {Semeghini}, \citenamefont
  {Wang}, \citenamefont {Ebadi}, \citenamefont {Kalinowski}, \citenamefont
  {Keesling}, \citenamefont {Maskara}, \citenamefont {Pichler}, \citenamefont
  {Greiner}, \citenamefont {Vuleti{\'c}},\ and\ \citenamefont {Lukin}}]{QuEra}%
  \BibitemOpen
  \bibfield  {author} {\bibinfo {author} {\bibfnamefont {D.}~\bibnamefont
  {Bluvstein}}, \bibinfo {author} {\bibfnamefont {H.}~\bibnamefont {Levine}},
  \bibinfo {author} {\bibfnamefont {G.}~\bibnamefont {Semeghini}}, \bibinfo
  {author} {\bibfnamefont {T.~T.}\ \bibnamefont {Wang}}, \bibinfo {author}
  {\bibfnamefont {S.}~\bibnamefont {Ebadi}}, \bibinfo {author} {\bibfnamefont
  {M.}~\bibnamefont {Kalinowski}}, \bibinfo {author} {\bibfnamefont
  {A.}~\bibnamefont {Keesling}}, \bibinfo {author} {\bibfnamefont
  {N.}~\bibnamefont {Maskara}}, \bibinfo {author} {\bibfnamefont
  {H.}~\bibnamefont {Pichler}}, \bibinfo {author} {\bibfnamefont
  {M.}~\bibnamefont {Greiner}}, \bibinfo {author} {\bibfnamefont
  {V.}~\bibnamefont {Vuleti{\'c}}},\ and\ \bibinfo {author} {\bibfnamefont
  {M.~D.}\ \bibnamefont {Lukin}},\ }\href
  {https://doi.org/10.1038/s41586-022-04592-6} {\bibfield  {journal} {\bibinfo
  {journal} {Nature}\ }\textbf {\bibinfo {volume} {604}},\ \bibinfo {pages}
  {451} (\bibinfo {year} {2022})}\BibitemShut {NoStop}%
\bibitem [{\citenamefont {Monroe}\ \emph {et~al.}(2014)\citenamefont {Monroe},
  \citenamefont {Raussendorf}, \citenamefont {Ruthven}, \citenamefont {Brown},
  \citenamefont {Maunz}, \citenamefont {Duan},\ and\ \citenamefont
  {Kim}}]{Monroe2014}%
  \BibitemOpen
  \bibfield  {author} {\bibinfo {author} {\bibfnamefont {C.}~\bibnamefont
  {Monroe}}, \bibinfo {author} {\bibfnamefont {R.}~\bibnamefont {Raussendorf}},
  \bibinfo {author} {\bibfnamefont {A.}~\bibnamefont {Ruthven}}, \bibinfo
  {author} {\bibfnamefont {K.~R.}\ \bibnamefont {Brown}}, \bibinfo {author}
  {\bibfnamefont {P.}~\bibnamefont {Maunz}}, \bibinfo {author} {\bibfnamefont
  {L.-M.}\ \bibnamefont {Duan}},\ and\ \bibinfo {author} {\bibfnamefont
  {J.}~\bibnamefont {Kim}},\ }\href
  {https://doi.org/10.1103/PhysRevA.89.022317} {\bibfield  {journal} {\bibinfo
  {journal} {Phys. Rev. A}\ }\textbf {\bibinfo {volume} {89}},\ \bibinfo
  {pages} {022317} (\bibinfo {year} {2014})}\BibitemShut {NoStop}%
\bibitem [{\citenamefont {Bradley}\ \emph {et~al.}(2019)\citenamefont
  {Bradley}, \citenamefont {Randall}, \citenamefont {Abobeih}, \citenamefont
  {Berrevoets}, \citenamefont {Degen}, \citenamefont {Bakker}, \citenamefont
  {Markham}, \citenamefont {Twitchen},\ and\ \citenamefont
  {Taminiau}}]{NVcenter}%
  \BibitemOpen
  \bibfield  {author} {\bibinfo {author} {\bibfnamefont {C.~E.}\ \bibnamefont
  {Bradley}}, \bibinfo {author} {\bibfnamefont {J.}~\bibnamefont {Randall}},
  \bibinfo {author} {\bibfnamefont {M.~H.}\ \bibnamefont {Abobeih}}, \bibinfo
  {author} {\bibfnamefont {R.~C.}\ \bibnamefont {Berrevoets}}, \bibinfo
  {author} {\bibfnamefont {M.~J.}\ \bibnamefont {Degen}}, \bibinfo {author}
  {\bibfnamefont {M.~A.}\ \bibnamefont {Bakker}}, \bibinfo {author}
  {\bibfnamefont {M.}~\bibnamefont {Markham}}, \bibinfo {author} {\bibfnamefont
  {D.~J.}\ \bibnamefont {Twitchen}},\ and\ \bibinfo {author} {\bibfnamefont
  {T.~H.}\ \bibnamefont {Taminiau}},\ }\href
  {https://doi.org/10.1103/PhysRevX.9.031045} {\bibfield  {journal} {\bibinfo
  {journal} {Phys. Rev. X}\ }\textbf {\bibinfo {volume} {9}},\ \bibinfo {pages}
  {031045} (\bibinfo {year} {2019})}\BibitemShut {NoStop}%
\bibitem [{\citenamefont {Hannegan}\ \emph {et~al.}(2021)\citenamefont
  {Hannegan}, \citenamefont {Saha}, \citenamefont {Siverns}, \citenamefont
  {Cassell}, \citenamefont {Waks},\ and\ \citenamefont
  {Quraishi}}]{Hannegan2021}%
  \BibitemOpen
  \bibfield  {author} {\bibinfo {author} {\bibfnamefont {J.}~\bibnamefont
  {Hannegan}}, \bibinfo {author} {\bibfnamefont {U.}~\bibnamefont {Saha}},
  \bibinfo {author} {\bibfnamefont {J.~D.}\ \bibnamefont {Siverns}}, \bibinfo
  {author} {\bibfnamefont {J.}~\bibnamefont {Cassell}}, \bibinfo {author}
  {\bibfnamefont {E.}~\bibnamefont {Waks}},\ and\ \bibinfo {author}
  {\bibfnamefont {Q.}~\bibnamefont {Quraishi}},\ }\href
  {https://doi.org/10.1063/5.0059966} {\bibfield  {journal} {\bibinfo
  {journal} {Applied Physics Letters}\ }\textbf {\bibinfo {volume} {119}},\
  \bibinfo {pages} {084001} (\bibinfo {year} {2021})}\BibitemShut {NoStop}%
\bibitem [{\citenamefont {Radnaev}\ \emph {et~al.}(2010)\citenamefont
  {Radnaev}, \citenamefont {Dudin}, \citenamefont {Zhao}, \citenamefont {Jen},
  \citenamefont {Jenkins}, \citenamefont {Kuzmich},\ and\ \citenamefont
  {Kennedy}}]{Radnaev2010}%
  \BibitemOpen
  \bibfield  {author} {\bibinfo {author} {\bibfnamefont {A.~G.}\ \bibnamefont
  {Radnaev}}, \bibinfo {author} {\bibfnamefont {Y.~O.}\ \bibnamefont {Dudin}},
  \bibinfo {author} {\bibfnamefont {R.}~\bibnamefont {Zhao}}, \bibinfo {author}
  {\bibfnamefont {H.~H.}\ \bibnamefont {Jen}}, \bibinfo {author} {\bibfnamefont
  {S.~D.}\ \bibnamefont {Jenkins}}, \bibinfo {author} {\bibfnamefont
  {A.}~\bibnamefont {Kuzmich}},\ and\ \bibinfo {author} {\bibfnamefont
  {T.~A.~B.}\ \bibnamefont {Kennedy}},\ }\href
  {https://doi.org/10.1038/nphys1773} {\bibfield  {journal} {\bibinfo
  {journal} {Nature Physics}\ }\textbf {\bibinfo {volume} {6}},\ \bibinfo
  {pages} {894} (\bibinfo {year} {2010})}\BibitemShut {NoStop}%
\bibitem [{\citenamefont {Simon}\ \emph {et~al.}(2010)\citenamefont {Simon},
  \citenamefont {Afzelius}, \citenamefont {Appel}, \citenamefont {Boyer de~la
  Giroday}, \citenamefont {Dewhurst}, \citenamefont {Gisin}, \citenamefont
  {Hu}, \citenamefont {Jelezko}, \citenamefont {Kr{\"o}ll}, \citenamefont
  {M{\"u}ller}, \citenamefont {Nunn}, \citenamefont {Polzik}, \citenamefont
  {Rarity}, \citenamefont {De~Riedmatten}, \citenamefont {Rosenfeld},
  \citenamefont {Shields}, \citenamefont {Sk{\"o}ld}, \citenamefont
  {Stevenson}, \citenamefont {Thew}, \citenamefont {Walmsley}, \citenamefont
  {Weber}, \citenamefont {Weinfurter}, \citenamefont {Wrachtrup},\ and\
  \citenamefont {Young}}]{Simon2010}%
  \BibitemOpen
  \bibfield  {author} {\bibinfo {author} {\bibfnamefont {C.}~\bibnamefont
  {Simon}}, \bibinfo {author} {\bibfnamefont {M.}~\bibnamefont {Afzelius}},
  \bibinfo {author} {\bibfnamefont {J.}~\bibnamefont {Appel}}, \bibinfo
  {author} {\bibfnamefont {A.}~\bibnamefont {Boyer de~la Giroday}}, \bibinfo
  {author} {\bibfnamefont {S.~J.}\ \bibnamefont {Dewhurst}}, \bibinfo {author}
  {\bibfnamefont {N.}~\bibnamefont {Gisin}}, \bibinfo {author} {\bibfnamefont
  {C.~Y.}\ \bibnamefont {Hu}}, \bibinfo {author} {\bibfnamefont
  {F.}~\bibnamefont {Jelezko}}, \bibinfo {author} {\bibfnamefont
  {S.}~\bibnamefont {Kr{\"o}ll}}, \bibinfo {author} {\bibfnamefont {J.~H.}\
  \bibnamefont {M{\"u}ller}}, \bibinfo {author} {\bibfnamefont
  {J.}~\bibnamefont {Nunn}}, \bibinfo {author} {\bibfnamefont {E.~S.}\
  \bibnamefont {Polzik}}, \bibinfo {author} {\bibfnamefont {J.~G.}\
  \bibnamefont {Rarity}}, \bibinfo {author} {\bibfnamefont {H.}~\bibnamefont
  {De~Riedmatten}}, \bibinfo {author} {\bibfnamefont {W.}~\bibnamefont
  {Rosenfeld}}, \bibinfo {author} {\bibfnamefont {A.~J.}\ \bibnamefont
  {Shields}}, \bibinfo {author} {\bibfnamefont {N.}~\bibnamefont {Sk{\"o}ld}},
  \bibinfo {author} {\bibfnamefont {R.~M.}\ \bibnamefont {Stevenson}}, \bibinfo
  {author} {\bibfnamefont {R.}~\bibnamefont {Thew}}, \bibinfo {author}
  {\bibfnamefont {I.~A.}\ \bibnamefont {Walmsley}}, \bibinfo {author}
  {\bibfnamefont {M.~C.}\ \bibnamefont {Weber}}, \bibinfo {author}
  {\bibfnamefont {H.}~\bibnamefont {Weinfurter}}, \bibinfo {author}
  {\bibfnamefont {J.}~\bibnamefont {Wrachtrup}},\ and\ \bibinfo {author}
  {\bibfnamefont {R.~J.}\ \bibnamefont {Young}},\ }\href
  {https://doi.org/10.1140/epjd/e2010-00103-y} {\bibfield  {journal} {\bibinfo
  {journal} {The European Physical Journal D}\ }\textbf {\bibinfo {volume}
  {58}},\ \bibinfo {pages} {1} (\bibinfo {year} {2010})}\BibitemShut {NoStop}%
\bibitem [{\citenamefont {Hamilton}(2019)}]{NASA2019}%
  \BibitemOpen
  \bibfield  {author} {\bibinfo {author} {\bibfnamefont {S.~A.}\ \bibnamefont
  {Hamilton}},\ }\href
  {https://www.nasa.gov/sites/default/files/atoms/files/iac-19-b2.7.12_overview_of_nasa_nsql_program_paper.pdf}
  {\bibfield  {journal} {\bibinfo  {journal} {70th International Astronautical
  Congress}\ }\textbf {\bibinfo {volume} {IAC-19-B2.7.12}} (\bibinfo {year}
  {2019})}\BibitemShut {NoStop}%
\bibitem [{\citenamefont {Yang}\ \emph {et~al.}(2016)\citenamefont {Yang},
  \citenamefont {Wang}, \citenamefont {Bao},\ and\ \citenamefont
  {Pan}}]{Yang2016}%
  \BibitemOpen
  \bibfield  {author} {\bibinfo {author} {\bibfnamefont {S.-J.}\ \bibnamefont
  {Yang}}, \bibinfo {author} {\bibfnamefont {X.-J.}\ \bibnamefont {Wang}},
  \bibinfo {author} {\bibfnamefont {X.-H.}\ \bibnamefont {Bao}},\ and\ \bibinfo
  {author} {\bibfnamefont {J.-W.}\ \bibnamefont {Pan}},\ }\href
  {https://doi.org/10.1038/nphoton.2016.51} {\bibfield  {journal} {\bibinfo
  {journal} {Nature Photonics}\ }\textbf {\bibinfo {volume} {10}},\ \bibinfo
  {pages} {381} (\bibinfo {year} {2016})}\BibitemShut {NoStop}%
\bibitem [{\citenamefont {Bimbard}\ \emph {et~al.}(2014)\citenamefont
  {Bimbard}, \citenamefont {Boddeda}, \citenamefont {Vitrant}, \citenamefont
  {Grankin}, \citenamefont {Parigi}, \citenamefont {Stanojevic}, \citenamefont
  {Ourjoumtsev},\ and\ \citenamefont {Grangier}}]{Bimbard2014}%
  \BibitemOpen
  \bibfield  {author} {\bibinfo {author} {\bibfnamefont {E.}~\bibnamefont
  {Bimbard}}, \bibinfo {author} {\bibfnamefont {R.}~\bibnamefont {Boddeda}},
  \bibinfo {author} {\bibfnamefont {N.}~\bibnamefont {Vitrant}}, \bibinfo
  {author} {\bibfnamefont {A.}~\bibnamefont {Grankin}}, \bibinfo {author}
  {\bibfnamefont {V.}~\bibnamefont {Parigi}}, \bibinfo {author} {\bibfnamefont
  {J.}~\bibnamefont {Stanojevic}}, \bibinfo {author} {\bibfnamefont
  {A.}~\bibnamefont {Ourjoumtsev}},\ and\ \bibinfo {author} {\bibfnamefont
  {P.}~\bibnamefont {Grangier}},\ }\href
  {https://doi.org/10.1103/PhysRevLett.112.033601} {\bibfield  {journal}
  {\bibinfo  {journal} {Phys. Rev. Lett.}\ }\textbf {\bibinfo {volume} {112}},\
  \bibinfo {pages} {033601} (\bibinfo {year} {2014})}\BibitemShut {NoStop}%
\bibitem [{\citenamefont {Vittorini}\ \emph {et~al.}(2014)\citenamefont
  {Vittorini}, \citenamefont {Hucul}, \citenamefont {Inlek}, \citenamefont
  {Crocker},\ and\ \citenamefont {Monroe}}]{Vittorini2014}%
  \BibitemOpen
  \bibfield  {author} {\bibinfo {author} {\bibfnamefont {G.}~\bibnamefont
  {Vittorini}}, \bibinfo {author} {\bibfnamefont {D.}~\bibnamefont {Hucul}},
  \bibinfo {author} {\bibfnamefont {I.~V.}\ \bibnamefont {Inlek}}, \bibinfo
  {author} {\bibfnamefont {C.}~\bibnamefont {Crocker}},\ and\ \bibinfo {author}
  {\bibfnamefont {C.}~\bibnamefont {Monroe}},\ }\href
  {https://doi.org/10.1103/PhysRevA.90.040302} {\bibfield  {journal} {\bibinfo
  {journal} {Phys. Rev. A}\ }\textbf {\bibinfo {volume} {90}},\ \bibinfo
  {pages} {040302} (\bibinfo {year} {2014})}\BibitemShut {NoStop}%
\bibitem [{\citenamefont {de~Riedmatten}\ \emph {et~al.}(2008)\citenamefont
  {de~Riedmatten}, \citenamefont {Afzelius}, \citenamefont {Staudt},
  \citenamefont {Simon},\ and\ \citenamefont {Gisin}}]{Riedmatten2008}%
  \BibitemOpen
  \bibfield  {author} {\bibinfo {author} {\bibfnamefont {H.}~\bibnamefont
  {de~Riedmatten}}, \bibinfo {author} {\bibfnamefont {M.}~\bibnamefont
  {Afzelius}}, \bibinfo {author} {\bibfnamefont {M.~U.}\ \bibnamefont
  {Staudt}}, \bibinfo {author} {\bibfnamefont {C.}~\bibnamefont {Simon}},\ and\
  \bibinfo {author} {\bibfnamefont {N.}~\bibnamefont {Gisin}},\ }\href
  {https://www.nature.com/articles/nature07607} {\bibfield  {journal} {\bibinfo
   {journal} {Nature}\ }\textbf {\bibinfo {volume} {456}},\ \bibinfo {pages}
  {773} (\bibinfo {year} {2008})}\BibitemShut {NoStop}%
\bibitem [{\citenamefont {Ma}\ \emph {et~al.}(2021)\citenamefont {Ma},
  \citenamefont {Ma}, \citenamefont {Zhou}, \citenamefont {Li},\ and\
  \citenamefont {Guo}}]{Ma2021_rareearth}%
  \BibitemOpen
  \bibfield  {author} {\bibinfo {author} {\bibfnamefont {Y.}~\bibnamefont
  {Ma}}, \bibinfo {author} {\bibfnamefont {Y.-Z.}\ \bibnamefont {Ma}}, \bibinfo
  {author} {\bibfnamefont {Z.-Q.}\ \bibnamefont {Zhou}}, \bibinfo {author}
  {\bibfnamefont {C.-F.}\ \bibnamefont {Li}},\ and\ \bibinfo {author}
  {\bibfnamefont {G.-C.}\ \bibnamefont {Guo}},\ }\href
  {https://doi.org/10.1038/s41467-021-22706-y} {\bibfield  {journal} {\bibinfo
  {journal} {Nature Communications}\ }\textbf {\bibinfo {volume} {12}},\
  \bibinfo {pages} {2381} (\bibinfo {year} {2021})}\BibitemShut {NoStop}%
\bibitem [{\citenamefont {Ruf}\ \emph {et~al.}(2021)\citenamefont {Ruf},
  \citenamefont {Wan}, \citenamefont {Choi}, \citenamefont {Englund},\ and\
  \citenamefont {Hanson}}]{Maximilian2021}%
  \BibitemOpen
  \bibfield  {author} {\bibinfo {author} {\bibfnamefont {M.}~\bibnamefont
  {Ruf}}, \bibinfo {author} {\bibfnamefont {N.~H.}\ \bibnamefont {Wan}},
  \bibinfo {author} {\bibfnamefont {H.}~\bibnamefont {Choi}}, \bibinfo {author}
  {\bibfnamefont {D.}~\bibnamefont {Englund}},\ and\ \bibinfo {author}
  {\bibfnamefont {R.}~\bibnamefont {Hanson}},\ }\href
  {https://doi.org/10.1063/5.0056534} {\bibfield  {journal} {\bibinfo
  {journal} {Journal of Applied Physics}\ }\textbf {\bibinfo {volume} {130}},\
  \bibinfo {pages} {070901} (\bibinfo {year} {2021})}\BibitemShut {NoStop}%
\bibitem [{\citenamefont {Rozp\ifmmode~\mbox{\k{e}}\else \k{e}\fi{}dek}\ \emph
  {et~al.}(2019)\citenamefont {Rozp\ifmmode~\mbox{\k{e}}\else \k{e}\fi{}dek},
  \citenamefont {Yehia}, \citenamefont {Goodenough}, \citenamefont {Ruf},
  \citenamefont {Humphreys}, \citenamefont {Hanson}, \citenamefont {Wehner},\
  and\ \citenamefont {Elkouss}}]{Rozp2019}%
  \BibitemOpen
  \bibfield  {author} {\bibinfo {author} {\bibfnamefont {F.}~\bibnamefont
  {Rozp\ifmmode~\mbox{\k{e}}\else \k{e}\fi{}dek}}, \bibinfo {author}
  {\bibfnamefont {R.}~\bibnamefont {Yehia}}, \bibinfo {author} {\bibfnamefont
  {K.}~\bibnamefont {Goodenough}}, \bibinfo {author} {\bibfnamefont
  {M.}~\bibnamefont {Ruf}}, \bibinfo {author} {\bibfnamefont {P.~C.}\
  \bibnamefont {Humphreys}}, \bibinfo {author} {\bibfnamefont {R.}~\bibnamefont
  {Hanson}}, \bibinfo {author} {\bibfnamefont {S.}~\bibnamefont {Wehner}},\
  and\ \bibinfo {author} {\bibfnamefont {D.}~\bibnamefont {Elkouss}},\ }\href
  {https://doi.org/10.1103/PhysRevA.99.052330} {\bibfield  {journal} {\bibinfo
  {journal} {Phys. Rev. A}\ }\textbf {\bibinfo {volume} {99}},\ \bibinfo
  {pages} {052330} (\bibinfo {year} {2019})}\BibitemShut {NoStop}%
\bibitem [{\citenamefont {Bhaskar}\ \emph {et~al.}(2020)\citenamefont
  {Bhaskar}, \citenamefont {Riedinger}, \citenamefont {Machielse},
  \citenamefont {Levonian}, \citenamefont {Nguyen}, \citenamefont {Knall},
  \citenamefont {Park}, \citenamefont {Englund}, \citenamefont {Lon{\v c}ar},
  \citenamefont {Sukachev},\ and\ \citenamefont {Lukin}}]{Bhaskar2020}%
  \BibitemOpen
  \bibfield  {author} {\bibinfo {author} {\bibfnamefont {M.~K.}\ \bibnamefont
  {Bhaskar}}, \bibinfo {author} {\bibfnamefont {R.}~\bibnamefont {Riedinger}},
  \bibinfo {author} {\bibfnamefont {B.}~\bibnamefont {Machielse}}, \bibinfo
  {author} {\bibfnamefont {D.~S.}\ \bibnamefont {Levonian}}, \bibinfo {author}
  {\bibfnamefont {C.~T.}\ \bibnamefont {Nguyen}}, \bibinfo {author}
  {\bibfnamefont {E.~N.}\ \bibnamefont {Knall}}, \bibinfo {author}
  {\bibfnamefont {H.}~\bibnamefont {Park}}, \bibinfo {author} {\bibfnamefont
  {D.}~\bibnamefont {Englund}}, \bibinfo {author} {\bibfnamefont
  {M.}~\bibnamefont {Lon{\v c}ar}}, \bibinfo {author} {\bibfnamefont {D.~D.}\
  \bibnamefont {Sukachev}},\ and\ \bibinfo {author} {\bibfnamefont {M.~D.}\
  \bibnamefont {Lukin}},\ }\href {https://doi.org/10.1038/s41586-020-2103-5}
  {\bibfield  {journal} {\bibinfo  {journal} {Nature}\ }\textbf {\bibinfo
  {volume} {580}},\ \bibinfo {pages} {60} (\bibinfo {year} {2020})}\BibitemShut
  {NoStop}%
\bibitem [{\citenamefont {Hsiao}\ \emph {et~al.}(2018)\citenamefont {Hsiao},
  \citenamefont {Tsai}, \citenamefont {Chen}, \citenamefont {Lin},
  \citenamefont {Hung}, \citenamefont {Lee}, \citenamefont {Chen},
  \citenamefont {Chen}, \citenamefont {Yu},\ and\ \citenamefont
  {Chen}}]{Hsiao2018}%
  \BibitemOpen
  \bibfield  {author} {\bibinfo {author} {\bibfnamefont {Y.-F.}\ \bibnamefont
  {Hsiao}}, \bibinfo {author} {\bibfnamefont {P.-J.}\ \bibnamefont {Tsai}},
  \bibinfo {author} {\bibfnamefont {H.-S.}\ \bibnamefont {Chen}}, \bibinfo
  {author} {\bibfnamefont {S.-X.}\ \bibnamefont {Lin}}, \bibinfo {author}
  {\bibfnamefont {C.-C.}\ \bibnamefont {Hung}}, \bibinfo {author}
  {\bibfnamefont {C.-H.}\ \bibnamefont {Lee}}, \bibinfo {author} {\bibfnamefont
  {Y.-H.}\ \bibnamefont {Chen}}, \bibinfo {author} {\bibfnamefont {Y.-F.}\
  \bibnamefont {Chen}}, \bibinfo {author} {\bibfnamefont {I.~A.}\ \bibnamefont
  {Yu}},\ and\ \bibinfo {author} {\bibfnamefont {Y.-C.}\ \bibnamefont {Chen}},\
  }\href {https://doi.org/10.1103/PhysRevLett.120.183602} {\bibfield  {journal}
  {\bibinfo  {journal} {Phys. Rev. Lett.}\ }\textbf {\bibinfo {volume} {120}},\
  \bibinfo {pages} {183602} (\bibinfo {year} {2018})}\BibitemShut {NoStop}%
\bibitem [{\citenamefont {Wang}\ \emph {et~al.}(2019)\citenamefont {Wang},
  \citenamefont {Li}, \citenamefont {Zhang}, \citenamefont {Su}, \citenamefont
  {Zhou}, \citenamefont {Liao}, \citenamefont {Du}, \citenamefont {Yan},\ and\
  \citenamefont {Zhu}}]{Wang2019_RAMAN}%
  \BibitemOpen
  \bibfield  {author} {\bibinfo {author} {\bibfnamefont {Y.}~\bibnamefont
  {Wang}}, \bibinfo {author} {\bibfnamefont {J.}~\bibnamefont {Li}}, \bibinfo
  {author} {\bibfnamefont {S.}~\bibnamefont {Zhang}}, \bibinfo {author}
  {\bibfnamefont {K.}~\bibnamefont {Su}}, \bibinfo {author} {\bibfnamefont
  {Y.}~\bibnamefont {Zhou}}, \bibinfo {author} {\bibfnamefont {K.}~\bibnamefont
  {Liao}}, \bibinfo {author} {\bibfnamefont {S.}~\bibnamefont {Du}}, \bibinfo
  {author} {\bibfnamefont {H.}~\bibnamefont {Yan}},\ and\ \bibinfo {author}
  {\bibfnamefont {S.-L.}\ \bibnamefont {Zhu}},\ }\href
  {https://doi.org/10.1038/s41566-019-0368-8} {\bibfield  {journal} {\bibinfo
  {journal} {Nature Photonics}\ }\textbf {\bibinfo {volume} {13}},\ \bibinfo
  {pages} {346} (\bibinfo {year} {2019})}\BibitemShut {NoStop}%
\bibitem [{\citenamefont {Saglamyurek}\ \emph {et~al.}(2021)\citenamefont
  {Saglamyurek}, \citenamefont {Hrushevskyi}, \citenamefont {Rastogi},
  \citenamefont {Cooke}, \citenamefont {Smith},\ and\ \citenamefont
  {LeBlanc}}]{Saglamyurek2021}%
  \BibitemOpen
  \bibfield  {author} {\bibinfo {author} {\bibfnamefont {E.}~\bibnamefont
  {Saglamyurek}}, \bibinfo {author} {\bibfnamefont {T.}~\bibnamefont
  {Hrushevskyi}}, \bibinfo {author} {\bibfnamefont {A.}~\bibnamefont
  {Rastogi}}, \bibinfo {author} {\bibfnamefont {L.~W.}\ \bibnamefont {Cooke}},
  \bibinfo {author} {\bibfnamefont {B.~D.}\ \bibnamefont {Smith}},\ and\
  \bibinfo {author} {\bibfnamefont {L.~J.}\ \bibnamefont {LeBlanc}},\ }\href
  {https://doi.org/10.1088/1367-2630/abf1d9} {\bibfield  {journal} {\bibinfo
  {journal} {New Journal of Physics}\ }\textbf {\bibinfo {volume} {23}},\
  \bibinfo {pages} {043028} (\bibinfo {year} {2021})}\BibitemShut {NoStop}%
\bibitem [{\citenamefont {Knappe}\ \emph {et~al.}(2004)\citenamefont {Knappe},
  \citenamefont {Shah}, \citenamefont {Schwindt}, \citenamefont {Hollberg},
  \citenamefont {Kitching}, \citenamefont {Liew},\ and\ \citenamefont
  {Moreland}}]{Knappe2004}%
  \BibitemOpen
  \bibfield  {author} {\bibinfo {author} {\bibfnamefont {S.}~\bibnamefont
  {Knappe}}, \bibinfo {author} {\bibfnamefont {V.}~\bibnamefont {Shah}},
  \bibinfo {author} {\bibfnamefont {P.~D.~D.}\ \bibnamefont {Schwindt}},
  \bibinfo {author} {\bibfnamefont {L.}~\bibnamefont {Hollberg}}, \bibinfo
  {author} {\bibfnamefont {J.}~\bibnamefont {Kitching}}, \bibinfo {author}
  {\bibfnamefont {L.-A.}\ \bibnamefont {Liew}},\ and\ \bibinfo {author}
  {\bibfnamefont {J.}~\bibnamefont {Moreland}},\ }\href
  {https://doi.org/10.1063/1.1787942} {\bibfield  {journal} {\bibinfo
  {journal} {Applied Physics Letters}\ }\textbf {\bibinfo {volume} {85}},\
  \bibinfo {pages} {1460} (\bibinfo {year} {2004})}\BibitemShut {NoStop}%
\bibitem [{\citenamefont {Limes}\ \emph {et~al.}(2020)\citenamefont {Limes},
  \citenamefont {Foley}, \citenamefont {Kornack}, \citenamefont {Caliga},
  \citenamefont {McBride}, \citenamefont {Braun}, \citenamefont {Lee},
  \citenamefont {Lucivero},\ and\ \citenamefont {Romalis}}]{Limes2020}%
  \BibitemOpen
  \bibfield  {author} {\bibinfo {author} {\bibfnamefont {M.}~\bibnamefont
  {Limes}}, \bibinfo {author} {\bibfnamefont {E.}~\bibnamefont {Foley}},
  \bibinfo {author} {\bibfnamefont {T.}~\bibnamefont {Kornack}}, \bibinfo
  {author} {\bibfnamefont {S.}~\bibnamefont {Caliga}}, \bibinfo {author}
  {\bibfnamefont {S.}~\bibnamefont {McBride}}, \bibinfo {author} {\bibfnamefont
  {A.}~\bibnamefont {Braun}}, \bibinfo {author} {\bibfnamefont
  {W.}~\bibnamefont {Lee}}, \bibinfo {author} {\bibfnamefont {V.}~\bibnamefont
  {Lucivero}},\ and\ \bibinfo {author} {\bibfnamefont {M.}~\bibnamefont
  {Romalis}},\ }\href {https://doi.org/10.1103/PhysRevApplied.14.011002}
  {\bibfield  {journal} {\bibinfo  {journal} {Phys. Rev. Applied}\ }\textbf
  {\bibinfo {volume} {14}},\ \bibinfo {pages} {011002} (\bibinfo {year}
  {2020})}\BibitemShut {NoStop}%
\bibitem [{\citenamefont {Wu}\ \emph {et~al.}(2019)\citenamefont {Wu},
  \citenamefont {Pagel}, \citenamefont {Malek}, \citenamefont {Nguyen},
  \citenamefont {Zi}, \citenamefont {Scheirer},\ and\ \citenamefont
  {M{\"u}ller}}]{Wu2019}%
  \BibitemOpen
  \bibfield  {author} {\bibinfo {author} {\bibfnamefont {X.}~\bibnamefont
  {Wu}}, \bibinfo {author} {\bibfnamefont {Z.}~\bibnamefont {Pagel}}, \bibinfo
  {author} {\bibfnamefont {B.~S.}\ \bibnamefont {Malek}}, \bibinfo {author}
  {\bibfnamefont {T.~H.}\ \bibnamefont {Nguyen}}, \bibinfo {author}
  {\bibfnamefont {F.}~\bibnamefont {Zi}}, \bibinfo {author} {\bibfnamefont
  {D.~S.}\ \bibnamefont {Scheirer}},\ and\ \bibinfo {author} {\bibfnamefont
  {H.}~\bibnamefont {M{\"u}ller}},\ }\href
  {https://doi.org/10.1126/sciadv.aax0800} {\bibfield  {journal} {\bibinfo
  {journal} {Science Advances}\ }\textbf {\bibinfo {volume} {5}},\ \bibinfo
  {pages} {eaax0800} (\bibinfo {year} {2019})}\BibitemShut {NoStop}%
\bibitem [{\citenamefont {Hosseini}\ \emph
  {et~al.}(2011{\natexlab{a}})\citenamefont {Hosseini}, \citenamefont
  {Sparkes}, \citenamefont {Campbell}, \citenamefont {Lam},\ and\ \citenamefont
  {Buchler}}]{Hosseini2011_NatComm}%
  \BibitemOpen
  \bibfield  {author} {\bibinfo {author} {\bibfnamefont {M.}~\bibnamefont
  {Hosseini}}, \bibinfo {author} {\bibfnamefont {B.~M.}\ \bibnamefont
  {Sparkes}}, \bibinfo {author} {\bibfnamefont {G.}~\bibnamefont {Campbell}},
  \bibinfo {author} {\bibfnamefont {P.~K.}\ \bibnamefont {Lam}},\ and\ \bibinfo
  {author} {\bibfnamefont {B.~C.}\ \bibnamefont {Buchler}},\ }\href
  {https://doi.org/10.1038/ncomms1175} {\bibfield  {journal} {\bibinfo
  {journal} {Nature Communications}\ }\textbf {\bibinfo {volume} {2}},\
  \bibinfo {pages} {174} (\bibinfo {year} {2011}{\natexlab{a}})}\BibitemShut
  {NoStop}%
\bibitem [{\citenamefont {Katz}\ and\ \citenamefont
  {Firstenberg}(2018)}]{Katz2018}%
  \BibitemOpen
  \bibfield  {author} {\bibinfo {author} {\bibfnamefont {O.}~\bibnamefont
  {Katz}}\ and\ \bibinfo {author} {\bibfnamefont {O.}~\bibnamefont
  {Firstenberg}},\ }\href {https://doi.org/10.1038/s41467-018-04458-4}
  {\bibfield  {journal} {\bibinfo  {journal} {Nature Communications}\ }\textbf
  {\bibinfo {volume} {9}},\ \bibinfo {pages} {1} (\bibinfo {year}
  {2018})}\BibitemShut {NoStop}%
\bibitem [{\citenamefont {Kaczmarek}\ \emph {et~al.}(2018)\citenamefont
  {Kaczmarek}, \citenamefont {Ledingham}, \citenamefont {Brecht}, \citenamefont
  {Thomas}, \citenamefont {Thekkadath}, \citenamefont {Lazo-Arjona},
  \citenamefont {Munns}, \citenamefont {Poem}, \citenamefont {Feizpour},
  \citenamefont {Saunders}, \citenamefont {Nunn},\ and\ \citenamefont
  {Walmsley}}]{ORCA}%
  \BibitemOpen
  \bibfield  {author} {\bibinfo {author} {\bibfnamefont {K.~T.}\ \bibnamefont
  {Kaczmarek}}, \bibinfo {author} {\bibfnamefont {P.~M.}\ \bibnamefont
  {Ledingham}}, \bibinfo {author} {\bibfnamefont {B.}~\bibnamefont {Brecht}},
  \bibinfo {author} {\bibfnamefont {S.~E.}\ \bibnamefont {Thomas}}, \bibinfo
  {author} {\bibfnamefont {G.~S.}\ \bibnamefont {Thekkadath}}, \bibinfo
  {author} {\bibfnamefont {O.}~\bibnamefont {Lazo-Arjona}}, \bibinfo {author}
  {\bibfnamefont {J.~H.}\ \bibnamefont {Munns}}, \bibinfo {author}
  {\bibfnamefont {E.}~\bibnamefont {Poem}}, \bibinfo {author} {\bibfnamefont
  {A.}~\bibnamefont {Feizpour}}, \bibinfo {author} {\bibfnamefont {D.~J.}\
  \bibnamefont {Saunders}}, \bibinfo {author} {\bibfnamefont {J.}~\bibnamefont
  {Nunn}},\ and\ \bibinfo {author} {\bibfnamefont {I.~A.}\ \bibnamefont
  {Walmsley}},\ }\href {https://doi.org/10.1103/PhysRevA.97.042316} {\bibfield
  {journal} {\bibinfo  {journal} {Physical Review A}\ }\textbf {\bibinfo
  {volume} {97}},\ \bibinfo {pages} {1} (\bibinfo {year} {2018})}\BibitemShut
  {NoStop}%
\bibitem [{\citenamefont {Finkelstein}\ \emph {et~al.}(2018)\citenamefont
  {Finkelstein}, \citenamefont {Poem}, \citenamefont {Michel}, \citenamefont
  {Lahad},\ and\ \citenamefont {Firstenberg}}]{FLAME}%
  \BibitemOpen
  \bibfield  {author} {\bibinfo {author} {\bibfnamefont {R.}~\bibnamefont
  {Finkelstein}}, \bibinfo {author} {\bibfnamefont {E.}~\bibnamefont {Poem}},
  \bibinfo {author} {\bibfnamefont {O.}~\bibnamefont {Michel}}, \bibinfo
  {author} {\bibfnamefont {O.}~\bibnamefont {Lahad}},\ and\ \bibinfo {author}
  {\bibfnamefont {O.}~\bibnamefont {Firstenberg}},\ }\href
  {https://doi.org/10.1126/sciadv.aap8598} {\bibfield  {journal} {\bibinfo
  {journal} {Science Advances}\ }\textbf {\bibinfo {volume} {4}},\ \bibinfo
  {pages} {8598} (\bibinfo {year} {2018})}\BibitemShut {NoStop}%
\bibitem [{\citenamefont {Fleischhauer}\ and\ \citenamefont
  {Lukin}(2000)}]{Fleischhauer2000}%
  \BibitemOpen
  \bibfield  {author} {\bibinfo {author} {\bibfnamefont {M.}~\bibnamefont
  {Fleischhauer}}\ and\ \bibinfo {author} {\bibfnamefont {M.~D.}\ \bibnamefont
  {Lukin}},\ }\href {https://doi.org/10.1103/PhysRevLett.84.5094} {\bibfield
  {journal} {\bibinfo  {journal} {Phys. Rev. Lett.}\ }\textbf {\bibinfo
  {volume} {84}},\ \bibinfo {pages} {5094} (\bibinfo {year}
  {2000})}\BibitemShut {NoStop}%
\bibitem [{\citenamefont {Eisaman}\ \emph {et~al.}(2005)\citenamefont
  {Eisaman}, \citenamefont {Andr{\'e}}, \citenamefont {Massou}, \citenamefont
  {Fleischhauer}, \citenamefont {Zibrov},\ and\ \citenamefont
  {Lukin}}]{Eisaman2005}%
  \BibitemOpen
  \bibfield  {author} {\bibinfo {author} {\bibfnamefont {M.~D.}\ \bibnamefont
  {Eisaman}}, \bibinfo {author} {\bibfnamefont {A.}~\bibnamefont {Andr{\'e}}},
  \bibinfo {author} {\bibfnamefont {F.}~\bibnamefont {Massou}}, \bibinfo
  {author} {\bibfnamefont {M.}~\bibnamefont {Fleischhauer}}, \bibinfo {author}
  {\bibfnamefont {A.~S.}\ \bibnamefont {Zibrov}},\ and\ \bibinfo {author}
  {\bibfnamefont {M.~D.}\ \bibnamefont {Lukin}},\ }\href
  {https://doi.org/10.1038/nature04327} {\bibfield  {journal} {\bibinfo
  {journal} {Nature}\ }\textbf {\bibinfo {volume} {438}},\ \bibinfo {pages}
  {837} (\bibinfo {year} {2005})}\BibitemShut {NoStop}%
\bibitem [{\citenamefont {Kupchak}\ \emph {et~al.}(2015)\citenamefont
  {Kupchak}, \citenamefont {Mittiga}, \citenamefont {Jordaan}, \citenamefont
  {Namazi}, \citenamefont {N{\"o}lleke},\ and\ \citenamefont
  {Figueroa}}]{Connor2015}%
  \BibitemOpen
  \bibfield  {author} {\bibinfo {author} {\bibfnamefont {C.}~\bibnamefont
  {Kupchak}}, \bibinfo {author} {\bibfnamefont {T.}~\bibnamefont {Mittiga}},
  \bibinfo {author} {\bibfnamefont {B.}~\bibnamefont {Jordaan}}, \bibinfo
  {author} {\bibfnamefont {M.}~\bibnamefont {Namazi}}, \bibinfo {author}
  {\bibfnamefont {C.}~\bibnamefont {N{\"o}lleke}},\ and\ \bibinfo {author}
  {\bibfnamefont {E.}~\bibnamefont {Figueroa}},\ }\href
  {https://doi.org/10.1038/srep07658} {\bibfield  {journal} {\bibinfo
  {journal} {Scientific Reports}\ }\textbf {\bibinfo {volume} {5}},\ \bibinfo
  {pages} {7658} (\bibinfo {year} {2015})}\BibitemShut {NoStop}%
\bibitem [{\citenamefont {Ma}\ \emph {et~al.}(2017)\citenamefont {Ma},
  \citenamefont {Slattery},\ and\ \citenamefont {Tang}}]{Ma2017}%
  \BibitemOpen
  \bibfield  {author} {\bibinfo {author} {\bibfnamefont {L.}~\bibnamefont
  {Ma}}, \bibinfo {author} {\bibfnamefont {O.}~\bibnamefont {Slattery}},\ and\
  \bibinfo {author} {\bibfnamefont {X.}~\bibnamefont {Tang}},\ }\href
  {https://doi.org/10.1088/2040-8986/19/4/043001} {\bibfield  {journal}
  {\bibinfo  {journal} {Journal of Optics}\ }\textbf {\bibinfo {volume} {19}},\
  \bibinfo {pages} {043001} (\bibinfo {year} {2017})}\BibitemShut {NoStop}%
\bibitem [{\citenamefont {Reim}\ \emph {et~al.}(2011)\citenamefont {Reim},
  \citenamefont {Michelberger}, \citenamefont {Lee}, \citenamefont {Nunn},
  \citenamefont {Langford},\ and\ \citenamefont {Walmsley}}]{Reim2011}%
  \BibitemOpen
  \bibfield  {author} {\bibinfo {author} {\bibfnamefont {K.~F.}\ \bibnamefont
  {Reim}}, \bibinfo {author} {\bibfnamefont {P.}~\bibnamefont {Michelberger}},
  \bibinfo {author} {\bibfnamefont {K.~C.}\ \bibnamefont {Lee}}, \bibinfo
  {author} {\bibfnamefont {J.}~\bibnamefont {Nunn}}, \bibinfo {author}
  {\bibfnamefont {N.~K.}\ \bibnamefont {Langford}},\ and\ \bibinfo {author}
  {\bibfnamefont {I.~A.}\ \bibnamefont {Walmsley}},\ }\href
  {https://doi.org/10.1103/PhysRevLett.107.053603} {\bibfield  {journal}
  {\bibinfo  {journal} {Phys. Rev. Lett.}\ }\textbf {\bibinfo {volume} {107}},\
  \bibinfo {pages} {053603} (\bibinfo {year} {2011})}\BibitemShut {NoStop}%
\bibitem [{\citenamefont {Hosseini}\ \emph
  {et~al.}(2011{\natexlab{b}})\citenamefont {Hosseini}, \citenamefont
  {Campbell}, \citenamefont {Sparkes}, \citenamefont {Lam},\ and\ \citenamefont
  {Buchler}}]{Hosseini2011}%
  \BibitemOpen
  \bibfield  {author} {\bibinfo {author} {\bibfnamefont {M.}~\bibnamefont
  {Hosseini}}, \bibinfo {author} {\bibfnamefont {G.}~\bibnamefont {Campbell}},
  \bibinfo {author} {\bibfnamefont {B.~M.}\ \bibnamefont {Sparkes}}, \bibinfo
  {author} {\bibfnamefont {P.~K.}\ \bibnamefont {Lam}},\ and\ \bibinfo {author}
  {\bibfnamefont {B.~C.}\ \bibnamefont {Buchler}},\ }\href
  {https://doi.org/10.1038/nphys2021} {\bibfield  {journal} {\bibinfo
  {journal} {Nature Physics}\ }\textbf {\bibinfo {volume} {7}},\ \bibinfo
  {pages} {794} (\bibinfo {year} {2011}{\natexlab{b}})}\BibitemShut {NoStop}%
\bibitem [{\citenamefont {Gorshkov}\ \emph {et~al.}(2007)\citenamefont
  {Gorshkov}, \citenamefont {Andr\'e}, \citenamefont {Fleischhauer},
  \citenamefont {S\o{}rensen},\ and\ \citenamefont {Lukin}}]{Gorshkov2007}%
  \BibitemOpen
  \bibfield  {author} {\bibinfo {author} {\bibfnamefont {A.~V.}\ \bibnamefont
  {Gorshkov}}, \bibinfo {author} {\bibfnamefont {A.}~\bibnamefont {Andr\'e}},
  \bibinfo {author} {\bibfnamefont {M.}~\bibnamefont {Fleischhauer}}, \bibinfo
  {author} {\bibfnamefont {A.~S.}\ \bibnamefont {S\o{}rensen}},\ and\ \bibinfo
  {author} {\bibfnamefont {M.~D.}\ \bibnamefont {Lukin}},\ }\href
  {https://doi.org/10.1103/PhysRevLett.98.123601} {\bibfield  {journal}
  {\bibinfo  {journal} {Phys. Rev. Lett.}\ }\textbf {\bibinfo {volume} {98}},\
  \bibinfo {pages} {123601} (\bibinfo {year} {2007})}\BibitemShut {NoStop}%
\bibitem [{\citenamefont {Zhao}\ \emph {et~al.}(2009)\citenamefont {Zhao},
  \citenamefont {Chen}, \citenamefont {Bao}, \citenamefont {Strassel},
  \citenamefont {Chuu}, \citenamefont {Jin}, \citenamefont {Schmiedmayer},
  \citenamefont {Yuan}, \citenamefont {Chen},\ and\ \citenamefont
  {Pan}}]{Zhao2009}%
  \BibitemOpen
  \bibfield  {author} {\bibinfo {author} {\bibfnamefont {B.}~\bibnamefont
  {Zhao}}, \bibinfo {author} {\bibfnamefont {Y.-A.}\ \bibnamefont {Chen}},
  \bibinfo {author} {\bibfnamefont {X.-H.}\ \bibnamefont {Bao}}, \bibinfo
  {author} {\bibfnamefont {T.}~\bibnamefont {Strassel}}, \bibinfo {author}
  {\bibfnamefont {C.-S.}\ \bibnamefont {Chuu}}, \bibinfo {author}
  {\bibfnamefont {X.-M.}\ \bibnamefont {Jin}}, \bibinfo {author} {\bibfnamefont
  {J.}~\bibnamefont {Schmiedmayer}}, \bibinfo {author} {\bibfnamefont {Z.-S.}\
  \bibnamefont {Yuan}}, \bibinfo {author} {\bibfnamefont {S.}~\bibnamefont
  {Chen}},\ and\ \bibinfo {author} {\bibfnamefont {J.-W.}\ \bibnamefont
  {Pan}},\ }\href {https://doi.org/10.1038/nphys1153} {\bibfield  {journal}
  {\bibinfo  {journal} {Nature Physics}\ }\textbf {\bibinfo {volume} {5}},\
  \bibinfo {pages} {95} (\bibinfo {year} {2009})}\BibitemShut {NoStop}%
\bibitem [{\citenamefont {Namazi}\ \emph {et~al.}(2017)\citenamefont {Namazi},
  \citenamefont {Kupchak}, \citenamefont {Jordaan}, \citenamefont
  {Shahrokhshahi},\ and\ \citenamefont {Figueroa}}]{Namazi2017}%
  \BibitemOpen
  \bibfield  {author} {\bibinfo {author} {\bibfnamefont {M.}~\bibnamefont
  {Namazi}}, \bibinfo {author} {\bibfnamefont {C.}~\bibnamefont {Kupchak}},
  \bibinfo {author} {\bibfnamefont {B.}~\bibnamefont {Jordaan}}, \bibinfo
  {author} {\bibfnamefont {R.}~\bibnamefont {Shahrokhshahi}},\ and\ \bibinfo
  {author} {\bibfnamefont {E.}~\bibnamefont {Figueroa}},\ }\href
  {https://doi.org/10.1103/PhysRevApplied.8.034023} {\bibfield  {journal}
  {\bibinfo  {journal} {Phys. Rev. Applied}\ }\textbf {\bibinfo {volume} {8}},\
  \bibinfo {pages} {034023} (\bibinfo {year} {2017})}\BibitemShut {NoStop}%
\bibitem [{\citenamefont {Novikova}\ \emph {et~al.}(2012)\citenamefont
  {Novikova}, \citenamefont {Walsworth},\ and\ \citenamefont
  {Xiao}}]{Novikova2012}%
  \BibitemOpen
  \bibfield  {author} {\bibinfo {author} {\bibfnamefont {I.}~\bibnamefont
  {Novikova}}, \bibinfo {author} {\bibfnamefont {R.}~\bibnamefont
  {Walsworth}},\ and\ \bibinfo {author} {\bibfnamefont {Y.}~\bibnamefont
  {Xiao}},\ }\href {https://doi.org/https://doi.org/10.1002/lpor.201100021}
  {\bibfield  {journal} {\bibinfo  {journal} {Laser \& Photonics Reviews}\
  }\textbf {\bibinfo {volume} {6}},\ \bibinfo {pages} {333} (\bibinfo {year}
  {2012})}\BibitemShut {NoStop}%
\bibitem [{\citenamefont {Mager}(1970)}]{Mager1970}%
  \BibitemOpen
  \bibfield  {author} {\bibinfo {author} {\bibfnamefont {A.}~\bibnamefont
  {Mager}},\ }\href {https://doi.org/10.1109/TMAG.1970.1066714} {\bibfield
  {journal} {\bibinfo  {journal} {IEEE Transactions on Magnetics}\ }\textbf
  {\bibinfo {volume} {6}},\ \bibinfo {pages} {67} (\bibinfo {year}
  {1970})}\BibitemShut {NoStop}%
\bibitem [{\citenamefont {Altarev}\ \emph {et~al.}(2015)\citenamefont
  {Altarev}, \citenamefont {Bales}, \citenamefont {Beck}, \citenamefont
  {Chupp}, \citenamefont {Fierlinger}, \citenamefont {Fierlinger},
  \citenamefont {Kuchler}, \citenamefont {Lins}, \citenamefont {Marino},
  \citenamefont {Niessen}, \citenamefont {Petzoldt}, \citenamefont
  {Schl{\"a}pfer}, \citenamefont {Schnabel}, \citenamefont {Singh},
  \citenamefont {Stoepler}, \citenamefont {Stuiber}, \citenamefont {Sturm},
  \citenamefont {Taubenheim},\ and\ \citenamefont {Voigt}}]{Altarev2015}%
  \BibitemOpen
  \bibfield  {author} {\bibinfo {author} {\bibfnamefont {I.}~\bibnamefont
  {Altarev}}, \bibinfo {author} {\bibfnamefont {M.}~\bibnamefont {Bales}},
  \bibinfo {author} {\bibfnamefont {D.~H.}\ \bibnamefont {Beck}}, \bibinfo
  {author} {\bibfnamefont {T.}~\bibnamefont {Chupp}}, \bibinfo {author}
  {\bibfnamefont {K.}~\bibnamefont {Fierlinger}}, \bibinfo {author}
  {\bibfnamefont {P.}~\bibnamefont {Fierlinger}}, \bibinfo {author}
  {\bibfnamefont {F.}~\bibnamefont {Kuchler}}, \bibinfo {author} {\bibfnamefont
  {T.}~\bibnamefont {Lins}}, \bibinfo {author} {\bibfnamefont {M.~G.}\
  \bibnamefont {Marino}}, \bibinfo {author} {\bibfnamefont {B.}~\bibnamefont
  {Niessen}}, \bibinfo {author} {\bibfnamefont {G.}~\bibnamefont {Petzoldt}},
  \bibinfo {author} {\bibfnamefont {U.}~\bibnamefont {Schl{\"a}pfer}}, \bibinfo
  {author} {\bibfnamefont {A.}~\bibnamefont {Schnabel}}, \bibinfo {author}
  {\bibfnamefont {J.~T.}\ \bibnamefont {Singh}}, \bibinfo {author}
  {\bibfnamefont {R.}~\bibnamefont {Stoepler}}, \bibinfo {author}
  {\bibfnamefont {S.}~\bibnamefont {Stuiber}}, \bibinfo {author} {\bibfnamefont
  {M.}~\bibnamefont {Sturm}}, \bibinfo {author} {\bibfnamefont
  {B.}~\bibnamefont {Taubenheim}},\ and\ \bibinfo {author} {\bibfnamefont
  {J.}~\bibnamefont {Voigt}},\ }\href {https://doi.org/10.1063/1.4919366}
  {\bibfield  {journal} {\bibinfo  {journal} {Journal of Applied Physics}\
  }\textbf {\bibinfo {volume} {117}},\ \bibinfo {pages} {183903} (\bibinfo
  {year} {2015})}\BibitemShut {NoStop}%
\bibitem [{\citenamefont {Raymer}\ \emph {et~al.}(1985)\citenamefont {Raymer},
  \citenamefont {Walmsley}, \citenamefont {Mostowski},\ and\ \citenamefont
  {Sobolewska}}]{Raymer1985}%
  \BibitemOpen
  \bibfield  {author} {\bibinfo {author} {\bibfnamefont {M.~G.}\ \bibnamefont
  {Raymer}}, \bibinfo {author} {\bibfnamefont {I.~A.}\ \bibnamefont
  {Walmsley}}, \bibinfo {author} {\bibfnamefont {J.}~\bibnamefont
  {Mostowski}},\ and\ \bibinfo {author} {\bibfnamefont {B.}~\bibnamefont
  {Sobolewska}},\ }\href {https://doi.org/10.1103/PhysRevA.32.332} {\bibfield
  {journal} {\bibinfo  {journal} {Phys. Rev. A}\ }\textbf {\bibinfo {volume}
  {32}},\ \bibinfo {pages} {332} (\bibinfo {year} {1985})}\BibitemShut
  {NoStop}%
\bibitem [{\citenamefont {Lauk}\ \emph {et~al.}(2013)\citenamefont {Lauk},
  \citenamefont {O'Brien},\ and\ \citenamefont {Fleischhauer}}]{Lauk2013}%
  \BibitemOpen
  \bibfield  {author} {\bibinfo {author} {\bibfnamefont {N.}~\bibnamefont
  {Lauk}}, \bibinfo {author} {\bibfnamefont {C.}~\bibnamefont {O'Brien}},\ and\
  \bibinfo {author} {\bibfnamefont {M.}~\bibnamefont {Fleischhauer}},\ }\href
  {https://doi.org/10.1103/PhysRevA.88.013823} {\bibfield  {journal} {\bibinfo
  {journal} {Phys. Rev. A}\ }\textbf {\bibinfo {volume} {88}},\ \bibinfo
  {pages} {013823} (\bibinfo {year} {2013})}\BibitemShut {NoStop}%
\bibitem [{\citenamefont {WALTHER}\ \emph {et~al.}(2007)\citenamefont
  {WALTHER}, \citenamefont {EISAMAN}, \citenamefont {ANDR{\'E}}, \citenamefont
  {MASSOU}, \citenamefont {FLEISCHHAUER}, \citenamefont {ZIBROV},\ and\
  \citenamefont {LUKIN}}]{Walther2007}%
  \BibitemOpen
  \bibfield  {author} {\bibinfo {author} {\bibfnamefont {P.}~\bibnamefont
  {WALTHER}}, \bibinfo {author} {\bibfnamefont {M.~D.}\ \bibnamefont
  {EISAMAN}}, \bibinfo {author} {\bibfnamefont {A.}~\bibnamefont {ANDR{\'E}}},
  \bibinfo {author} {\bibfnamefont {F.}~\bibnamefont {MASSOU}}, \bibinfo
  {author} {\bibfnamefont {M.}~\bibnamefont {FLEISCHHAUER}}, \bibinfo {author}
  {\bibfnamefont {A.~S.}\ \bibnamefont {ZIBROV}},\ and\ \bibinfo {author}
  {\bibfnamefont {M.~D.}\ \bibnamefont {LUKIN}},\ }\href
  {https://doi.org/10.1142/S0219749907002773} {\bibfield  {journal} {\bibinfo
  {journal} {International Journal of Quantum Information}\ }\textbf {\bibinfo
  {volume} {05}},\ \bibinfo {pages} {51} (\bibinfo {year} {2007})}\BibitemShut
  {NoStop}%
\bibitem [{\citenamefont {Zhang}\ \emph {et~al.}(2014)\citenamefont {Zhang},
  \citenamefont {Guo}, \citenamefont {Chen}, \citenamefont {Yuan},
  \citenamefont {Ou},\ and\ \citenamefont {Zhang}}]{Zhang2014}%
  \BibitemOpen
  \bibfield  {author} {\bibinfo {author} {\bibfnamefont {K.}~\bibnamefont
  {Zhang}}, \bibinfo {author} {\bibfnamefont {J.}~\bibnamefont {Guo}}, \bibinfo
  {author} {\bibfnamefont {L.~Q.}\ \bibnamefont {Chen}}, \bibinfo {author}
  {\bibfnamefont {C.}~\bibnamefont {Yuan}}, \bibinfo {author} {\bibfnamefont
  {Z.~Y.}\ \bibnamefont {Ou}},\ and\ \bibinfo {author} {\bibfnamefont
  {W.}~\bibnamefont {Zhang}},\ }\href
  {https://doi.org/10.1103/PhysRevA.90.033823} {\bibfield  {journal} {\bibinfo
  {journal} {Phys. Rev. A}\ }\textbf {\bibinfo {volume} {90}},\ \bibinfo
  {pages} {033823} (\bibinfo {year} {2014})}\BibitemShut {NoStop}%
\bibitem [{\citenamefont {Wolters}\ \emph {et~al.}(2017)\citenamefont
  {Wolters}, \citenamefont {Buser}, \citenamefont {Horsley}, \citenamefont
  {B\'eguin}, \citenamefont {J\"ockel}, \citenamefont {Jahn}, \citenamefont
  {Warburton},\ and\ \citenamefont {Treutlein}}]{Wolters2017}%
  \BibitemOpen
  \bibfield  {author} {\bibinfo {author} {\bibfnamefont {J.}~\bibnamefont
  {Wolters}}, \bibinfo {author} {\bibfnamefont {G.}~\bibnamefont {Buser}},
  \bibinfo {author} {\bibfnamefont {A.}~\bibnamefont {Horsley}}, \bibinfo
  {author} {\bibfnamefont {L.}~\bibnamefont {B\'eguin}}, \bibinfo {author}
  {\bibfnamefont {A.}~\bibnamefont {J\"ockel}}, \bibinfo {author}
  {\bibfnamefont {J.-P.}\ \bibnamefont {Jahn}}, \bibinfo {author}
  {\bibfnamefont {R.~J.}\ \bibnamefont {Warburton}},\ and\ \bibinfo {author}
  {\bibfnamefont {P.}~\bibnamefont {Treutlein}},\ }\href
  {https://doi.org/10.1103/PhysRevLett.119.060502} {\bibfield  {journal}
  {\bibinfo  {journal} {Phys. Rev. Lett.}\ }\textbf {\bibinfo {volume} {119}},\
  \bibinfo {pages} {060502} (\bibinfo {year} {2017})}\BibitemShut {NoStop}%
\bibitem [{\citenamefont {Blow}\ and\ \citenamefont {Wood}(1989)}]{Blow1989}%
  \BibitemOpen
  \bibfield  {author} {\bibinfo {author} {\bibfnamefont {K.}~\bibnamefont
  {Blow}}\ and\ \bibinfo {author} {\bibfnamefont {D.}~\bibnamefont {Wood}},\
  }\href {https://doi.org/10.1109/3.40655} {\bibfield  {journal} {\bibinfo
  {journal} {IEEE Journal of Quantum Electronics}\ }\textbf {\bibinfo {volume}
  {25}},\ \bibinfo {pages} {2665} (\bibinfo {year} {1989})}\BibitemShut
  {NoStop}%
\bibitem [{\citenamefont {Novikova}\ \emph {et~al.}(2007)\citenamefont
  {Novikova}, \citenamefont {Gorshkov}, \citenamefont {Phillips}, \citenamefont
  {S\o{}rensen}, \citenamefont {Lukin},\ and\ \citenamefont
  {Walsworth}}]{Novikova2007}%
  \BibitemOpen
  \bibfield  {author} {\bibinfo {author} {\bibfnamefont {I.}~\bibnamefont
  {Novikova}}, \bibinfo {author} {\bibfnamefont {A.~V.}\ \bibnamefont
  {Gorshkov}}, \bibinfo {author} {\bibfnamefont {D.~F.}\ \bibnamefont
  {Phillips}}, \bibinfo {author} {\bibfnamefont {A.~S.}\ \bibnamefont
  {S\o{}rensen}}, \bibinfo {author} {\bibfnamefont {M.~D.}\ \bibnamefont
  {Lukin}},\ and\ \bibinfo {author} {\bibfnamefont {R.~L.}\ \bibnamefont
  {Walsworth}},\ }\href {https://doi.org/10.1103/PhysRevLett.98.243602}
  {\bibfield  {journal} {\bibinfo  {journal} {Phys. Rev. Lett.}\ }\textbf
  {\bibinfo {volume} {98}},\ \bibinfo {pages} {243602} (\bibinfo {year}
  {2007})}\BibitemShut {NoStop}%
\bibitem [{\citenamefont {Phillips}\ \emph {et~al.}(2008)\citenamefont
  {Phillips}, \citenamefont {Gorshkov},\ and\ \citenamefont
  {Novikova}}]{Phillips2008}%
  \BibitemOpen
  \bibfield  {author} {\bibinfo {author} {\bibfnamefont {N.~B.}\ \bibnamefont
  {Phillips}}, \bibinfo {author} {\bibfnamefont {A.~V.}\ \bibnamefont
  {Gorshkov}},\ and\ \bibinfo {author} {\bibfnamefont {I.}~\bibnamefont
  {Novikova}},\ }\href {https://doi.org/10.1103/PhysRevA.78.023801} {\bibfield
  {journal} {\bibinfo  {journal} {Phys. Rev. A}\ }\textbf {\bibinfo {volume}
  {78}},\ \bibinfo {pages} {023801} (\bibinfo {year} {2008})}\BibitemShut
  {NoStop}%
\bibitem [{\citenamefont {Klein}(2009)}]{KleinThesis}%
  \BibitemOpen
  \bibfield  {author} {\bibinfo {author} {\bibfnamefont {M.}~\bibnamefont
  {Klein}},\ }\emph {\bibinfo {title} {Slow and Stored Light in Atomic Vapor
  Cells}},\ \href@noop {} {Ph.D. thesis},\ \bibinfo  {school} {Harvard
  University} (\bibinfo {year} {2009})\BibitemShut {NoStop}%
\bibitem [{\citenamefont {Peyronel}(2013)}]{peyronel2013quantum}%
  \BibitemOpen
  \bibfield  {author} {\bibinfo {author} {\bibfnamefont {T.}~\bibnamefont
  {Peyronel}},\ }\emph {\bibinfo {title} {Quantum nonlinear optics using cold
  atomic ensembles}},\ \href@noop {} {Ph.D. thesis},\ \bibinfo  {school}
  {Massachusetts Institute of Technology} (\bibinfo {year} {2013})\BibitemShut
  {NoStop}%
\bibitem [{\citenamefont {Vurgaftman}\ and\ \citenamefont
  {Bashkansky}(2013)}]{Vurgaftman2013}%
  \BibitemOpen
  \bibfield  {author} {\bibinfo {author} {\bibfnamefont {I.}~\bibnamefont
  {Vurgaftman}}\ and\ \bibinfo {author} {\bibfnamefont {M.}~\bibnamefont
  {Bashkansky}},\ }\href {https://doi.org/10.1103/PhysRevA.87.063836}
  {\bibfield  {journal} {\bibinfo  {journal} {Phys. Rev. A}\ }\textbf {\bibinfo
  {volume} {87}},\ \bibinfo {pages} {063836} (\bibinfo {year}
  {2013})}\BibitemShut {NoStop}%
\bibitem [{\citenamefont {Rotondaro}\ and\ \citenamefont
  {Perram}(1997)}]{Matthew1997}%
  \BibitemOpen
  \bibfield  {author} {\bibinfo {author} {\bibfnamefont {M.~D.}\ \bibnamefont
  {Rotondaro}}\ and\ \bibinfo {author} {\bibfnamefont {G.~P.}\ \bibnamefont
  {Perram}},\ }\href
  {https://doi.org/https://doi.org/10.1016/S0022-4073(96)00147-1} {\bibfield
  {journal} {\bibinfo  {journal} {Journal of Quantitative Spectroscopy and
  Radiative Transfer}\ }\textbf {\bibinfo {volume} {57}},\ \bibinfo {pages}
  {497} (\bibinfo {year} {1997})}\BibitemShut {NoStop}%
\bibitem [{\citenamefont {Pogorelov}\ \emph {et~al.}(2021)\citenamefont
  {Pogorelov}, \citenamefont {Feldker}, \citenamefont {Marciniak},
  \citenamefont {Postler}, \citenamefont {Jacob}, \citenamefont
  {Krieglsteiner}, \citenamefont {Podlesnic}, \citenamefont {Meth},
  \citenamefont {Negnevitsky}, \citenamefont {Stadler}, \citenamefont
  {H\"ofer}, \citenamefont {W\"achter}, \citenamefont {Lakhmanskiy},
  \citenamefont {Blatt}, \citenamefont {Schindler},\ and\ \citenamefont
  {Monz}}]{Pogorelov2021}%
  \BibitemOpen
  \bibfield  {author} {\bibinfo {author} {\bibfnamefont {I.}~\bibnamefont
  {Pogorelov}}, \bibinfo {author} {\bibfnamefont {T.}~\bibnamefont {Feldker}},
  \bibinfo {author} {\bibfnamefont {C.~D.}\ \bibnamefont {Marciniak}}, \bibinfo
  {author} {\bibfnamefont {L.}~\bibnamefont {Postler}}, \bibinfo {author}
  {\bibfnamefont {G.}~\bibnamefont {Jacob}}, \bibinfo {author} {\bibfnamefont
  {O.}~\bibnamefont {Krieglsteiner}}, \bibinfo {author} {\bibfnamefont
  {V.}~\bibnamefont {Podlesnic}}, \bibinfo {author} {\bibfnamefont
  {M.}~\bibnamefont {Meth}}, \bibinfo {author} {\bibfnamefont {V.}~\bibnamefont
  {Negnevitsky}}, \bibinfo {author} {\bibfnamefont {M.}~\bibnamefont
  {Stadler}}, \bibinfo {author} {\bibfnamefont {B.}~\bibnamefont {H\"ofer}},
  \bibinfo {author} {\bibfnamefont {C.}~\bibnamefont {W\"achter}}, \bibinfo
  {author} {\bibfnamefont {K.}~\bibnamefont {Lakhmanskiy}}, \bibinfo {author}
  {\bibfnamefont {R.}~\bibnamefont {Blatt}}, \bibinfo {author} {\bibfnamefont
  {P.}~\bibnamefont {Schindler}},\ and\ \bibinfo {author} {\bibfnamefont
  {T.}~\bibnamefont {Monz}},\ }\href
  {https://doi.org/10.1103/PRXQuantum.2.020343} {\bibfield  {journal} {\bibinfo
   {journal} {PRX Quantum}\ }\textbf {\bibinfo {volume} {2}},\ \bibinfo {pages}
  {020343} (\bibinfo {year} {2021})}\BibitemShut {NoStop}%
\bibitem [{\citenamefont {Salazar-Serrano}\ \emph {et~al.}(2015)\citenamefont
  {Salazar-Serrano}, \citenamefont {Valencia},\ and\ \citenamefont
  {Torres}}]{Salazar2015}%
  \BibitemOpen
  \bibfield  {author} {\bibinfo {author} {\bibfnamefont {L.~J.}\ \bibnamefont
  {Salazar-Serrano}}, \bibinfo {author} {\bibfnamefont {A.}~\bibnamefont
  {Valencia}},\ and\ \bibinfo {author} {\bibfnamefont {J.~P.}\ \bibnamefont
  {Torres}},\ }\href {https://doi.org/10.1063/1.4914834} {\bibfield  {journal}
  {\bibinfo  {journal} {Review of Scientific Instruments}\ }\textbf {\bibinfo
  {volume} {86}},\ \bibinfo {pages} {033109} (\bibinfo {year}
  {2015})}\BibitemShut {NoStop}%
\bibitem [{\citenamefont {Dai}\ \emph {et~al.}(2015)\citenamefont {Dai},
  \citenamefont {Jiang}, \citenamefont {Hang}, \citenamefont {Bi},\ and\
  \citenamefont {Ma}}]{Dai2015}%
  \BibitemOpen
  \bibfield  {author} {\bibinfo {author} {\bibfnamefont {X.}~\bibnamefont
  {Dai}}, \bibinfo {author} {\bibfnamefont {Y.}~\bibnamefont {Jiang}}, \bibinfo
  {author} {\bibfnamefont {C.}~\bibnamefont {Hang}}, \bibinfo {author}
  {\bibfnamefont {Z.}~\bibnamefont {Bi}},\ and\ \bibinfo {author}
  {\bibfnamefont {L.}~\bibnamefont {Ma}},\ }\href
  {https://doi.org/10.1364/OE.23.005134} {\bibfield  {journal} {\bibinfo
  {journal} {Opt. Express}\ }\textbf {\bibinfo {volume} {23}},\ \bibinfo
  {pages} {5134} (\bibinfo {year} {2015})}\BibitemShut {NoStop}%
\bibitem [{\citenamefont {Wei}\ \emph {et~al.}(2020)\citenamefont {Wei},
  \citenamefont {Wu}, \citenamefont {Hsiao}, \citenamefont {Tsai},\ and\
  \citenamefont {Chen}}]{Wei2020}%
  \BibitemOpen
  \bibfield  {author} {\bibinfo {author} {\bibfnamefont {Y.-C.}\ \bibnamefont
  {Wei}}, \bibinfo {author} {\bibfnamefont {B.-H.}\ \bibnamefont {Wu}},
  \bibinfo {author} {\bibfnamefont {Y.-F.}\ \bibnamefont {Hsiao}}, \bibinfo
  {author} {\bibfnamefont {P.-J.}\ \bibnamefont {Tsai}},\ and\ \bibinfo
  {author} {\bibfnamefont {Y.-C.}\ \bibnamefont {Chen}},\ }\href
  {https://doi.org/10.1103/PhysRevA.102.063720} {\bibfield  {journal} {\bibinfo
   {journal} {Phys. Rev. A}\ }\textbf {\bibinfo {volume} {102}},\ \bibinfo
  {pages} {063720} (\bibinfo {year} {2020})}\BibitemShut {NoStop}%
\bibitem [{\citenamefont {Ahlrichs}\ \emph {et~al.}(2013)\citenamefont
  {Ahlrichs}, \citenamefont {Berkemeier}, \citenamefont {Sprenger},\ and\
  \citenamefont {Benson}}]{Ahlrichs2013}%
  \BibitemOpen
  \bibfield  {author} {\bibinfo {author} {\bibfnamefont {A.}~\bibnamefont
  {Ahlrichs}}, \bibinfo {author} {\bibfnamefont {C.}~\bibnamefont
  {Berkemeier}}, \bibinfo {author} {\bibfnamefont {B.}~\bibnamefont
  {Sprenger}},\ and\ \bibinfo {author} {\bibfnamefont {O.}~\bibnamefont
  {Benson}},\ }\href {https://doi.org/10.1063/1.4846316} {\bibfield  {journal}
  {\bibinfo  {journal} {Applied Physics Letters}\ }\textbf {\bibinfo {volume}
  {103}},\ \bibinfo {pages} {241110} (\bibinfo {year} {2013})}\BibitemShut
  {NoStop}%
\bibitem [{\citenamefont {G\"undo\ifmmode~\breve{g}\else \u{g}\fi{}an}\ \emph
  {et~al.}(2012)\citenamefont {G\"undo\ifmmode~\breve{g}\else \u{g}\fi{}an},
  \citenamefont {Ledingham}, \citenamefont {Almasi}, \citenamefont
  {Cristiani},\ and\ \citenamefont {de~Riedmatten}}]{Mustafa2012}%
  \BibitemOpen
  \bibfield  {author} {\bibinfo {author} {\bibfnamefont {M.}~\bibnamefont
  {G\"undo\ifmmode~\breve{g}\else \u{g}\fi{}an}}, \bibinfo {author}
  {\bibfnamefont {P.~M.}\ \bibnamefont {Ledingham}}, \bibinfo {author}
  {\bibfnamefont {A.}~\bibnamefont {Almasi}}, \bibinfo {author} {\bibfnamefont
  {M.}~\bibnamefont {Cristiani}},\ and\ \bibinfo {author} {\bibfnamefont
  {H.}~\bibnamefont {de~Riedmatten}},\ }\href
  {https://doi.org/10.1103/PhysRevLett.108.190504} {\bibfield  {journal}
  {\bibinfo  {journal} {Phys. Rev. Lett.}\ }\textbf {\bibinfo {volume} {108}},\
  \bibinfo {pages} {190504} (\bibinfo {year} {2012})}\BibitemShut {NoStop}%
\bibitem [{\citenamefont {Specht}\ \emph {et~al.}(2011)\citenamefont {Specht},
  \citenamefont {N{\"o}lleke}, \citenamefont {Reiserer}, \citenamefont
  {Uphoff}, \citenamefont {Figueroa}, \citenamefont {Ritter},\ and\
  \citenamefont {Rempe}}]{Specht2011}%
  \BibitemOpen
  \bibfield  {author} {\bibinfo {author} {\bibfnamefont {H.~P.}\ \bibnamefont
  {Specht}}, \bibinfo {author} {\bibfnamefont {C.}~\bibnamefont {N{\"o}lleke}},
  \bibinfo {author} {\bibfnamefont {A.}~\bibnamefont {Reiserer}}, \bibinfo
  {author} {\bibfnamefont {M.}~\bibnamefont {Uphoff}}, \bibinfo {author}
  {\bibfnamefont {E.}~\bibnamefont {Figueroa}}, \bibinfo {author}
  {\bibfnamefont {S.}~\bibnamefont {Ritter}},\ and\ \bibinfo {author}
  {\bibfnamefont {G.}~\bibnamefont {Rempe}},\ }\href
  {https://doi.org/10.1038/nature09997} {\bibfield  {journal} {\bibinfo
  {journal} {Nature}\ }\textbf {\bibinfo {volume} {473}},\ \bibinfo {pages}
  {190} (\bibinfo {year} {2011})}\BibitemShut {NoStop}%
\bibitem [{\citenamefont {Yan}\ \emph {et~al.}(2018)\citenamefont {Yan},
  \citenamefont {Liu}, \citenamefont {Yan},\ and\ \citenamefont
  {Jia}}]{Yan2018}%
  \BibitemOpen
  \bibfield  {author} {\bibinfo {author} {\bibfnamefont {Z.}~\bibnamefont
  {Yan}}, \bibinfo {author} {\bibfnamefont {Y.}~\bibnamefont {Liu}}, \bibinfo
  {author} {\bibfnamefont {J.}~\bibnamefont {Yan}},\ and\ \bibinfo {author}
  {\bibfnamefont {X.}~\bibnamefont {Jia}},\ }\href
  {https://doi.org/10.1103/PhysRevA.97.013856} {\bibfield  {journal} {\bibinfo
  {journal} {Physical Review A}\ }\textbf {\bibinfo {volume} {97}},\ \bibinfo
  {pages} {1} (\bibinfo {year} {2018})}\BibitemShut {NoStop}%
\bibitem [{\citenamefont {Saunders}\ \emph {et~al.}(2016)\citenamefont
  {Saunders}, \citenamefont {Munns}, \citenamefont {Champion}, \citenamefont
  {Qiu}, \citenamefont {Kaczmarek}, \citenamefont {Poem}, \citenamefont
  {Ledingham}, \citenamefont {Walmsley},\ and\ \citenamefont
  {Nunn}}]{Saunders2016}%
  \BibitemOpen
  \bibfield  {author} {\bibinfo {author} {\bibfnamefont {D.~J.}\ \bibnamefont
  {Saunders}}, \bibinfo {author} {\bibfnamefont {J.~H.}\ \bibnamefont {Munns}},
  \bibinfo {author} {\bibfnamefont {T.~F.}\ \bibnamefont {Champion}}, \bibinfo
  {author} {\bibfnamefont {C.}~\bibnamefont {Qiu}}, \bibinfo {author}
  {\bibfnamefont {K.~T.}\ \bibnamefont {Kaczmarek}}, \bibinfo {author}
  {\bibfnamefont {E.}~\bibnamefont {Poem}}, \bibinfo {author} {\bibfnamefont
  {P.~M.}\ \bibnamefont {Ledingham}}, \bibinfo {author} {\bibfnamefont {I.~A.}\
  \bibnamefont {Walmsley}},\ and\ \bibinfo {author} {\bibfnamefont
  {J.}~\bibnamefont {Nunn}},\ }\href
  {https://doi.org/10.1103/PhysRevLett.116.090501} {\bibfield  {journal}
  {\bibinfo  {journal} {Physical Review Letters}\ }\textbf {\bibinfo {volume}
  {116}},\ \bibinfo {pages} {24} (\bibinfo {year} {2016})},\ \Eprint
  {https://arxiv.org/abs/1510.04625} {1510.04625} \BibitemShut {NoStop}%
\bibitem [{\citenamefont {Ma}\ \emph {et~al.}(2022)\citenamefont {Ma},
  \citenamefont {Lei}, \citenamefont {Yan}, \citenamefont {Li}, \citenamefont
  {Chai}, \citenamefont {Yan}, \citenamefont {Jia}, \citenamefont {Xie},\ and\
  \citenamefont {Peng}}]{Ma2021}%
  \BibitemOpen
  \bibfield  {author} {\bibinfo {author} {\bibfnamefont {L.}~\bibnamefont
  {Ma}}, \bibinfo {author} {\bibfnamefont {X.}~\bibnamefont {Lei}}, \bibinfo
  {author} {\bibfnamefont {J.}~\bibnamefont {Yan}}, \bibinfo {author}
  {\bibfnamefont {R.}~\bibnamefont {Li}}, \bibinfo {author} {\bibfnamefont
  {T.}~\bibnamefont {Chai}}, \bibinfo {author} {\bibfnamefont {Z.}~\bibnamefont
  {Yan}}, \bibinfo {author} {\bibfnamefont {X.}~\bibnamefont {Jia}}, \bibinfo
  {author} {\bibfnamefont {C.}~\bibnamefont {Xie}},\ and\ \bibinfo {author}
  {\bibfnamefont {K.}~\bibnamefont {Peng}},\ }\href
  {https://doi.org/10.1038/s41467-022-30077-1} {\bibfield  {journal} {\bibinfo
  {journal} {Nature Communications}\ }\textbf {\bibinfo {volume} {13}},\
  \bibinfo {pages} {2368} (\bibinfo {year} {2022})}\BibitemShut {NoStop}%
\bibitem [{\citenamefont {Dideriksen}\ \emph {et~al.}(2021)\citenamefont
  {Dideriksen}, \citenamefont {Schmieg}, \citenamefont {Zugenmaier},\ and\
  \citenamefont {Polzik}}]{Dideriksen2021}%
  \BibitemOpen
  \bibfield  {author} {\bibinfo {author} {\bibfnamefont {K.~B.}\ \bibnamefont
  {Dideriksen}}, \bibinfo {author} {\bibfnamefont {R.}~\bibnamefont {Schmieg}},
  \bibinfo {author} {\bibfnamefont {M.}~\bibnamefont {Zugenmaier}},\ and\
  \bibinfo {author} {\bibfnamefont {E.~S.}\ \bibnamefont {Polzik}},\ }\href
  {https://doi.org/10.1038/s41467-021-24033-8} {\bibfield  {journal} {\bibinfo
  {journal} {Nature Communications}\ }\textbf {\bibinfo {volume} {12}},\
  \bibinfo {pages} {3699} (\bibinfo {year} {2021})}\BibitemShut {NoStop}%
\bibitem [{\citenamefont {Bao}\ \emph {et~al.}(2012)\citenamefont {Bao},
  \citenamefont {Reingruber}, \citenamefont {Dietrich}, \citenamefont {Rui},
  \citenamefont {D{\"u}ck}, \citenamefont {Strassel}, \citenamefont {Li},
  \citenamefont {Liu}, \citenamefont {Zhao},\ and\ \citenamefont
  {Pan}}]{Bao2012}%
  \BibitemOpen
  \bibfield  {author} {\bibinfo {author} {\bibfnamefont {X.-H.}\ \bibnamefont
  {Bao}}, \bibinfo {author} {\bibfnamefont {A.}~\bibnamefont {Reingruber}},
  \bibinfo {author} {\bibfnamefont {P.}~\bibnamefont {Dietrich}}, \bibinfo
  {author} {\bibfnamefont {J.}~\bibnamefont {Rui}}, \bibinfo {author}
  {\bibfnamefont {A.}~\bibnamefont {D{\"u}ck}}, \bibinfo {author}
  {\bibfnamefont {T.}~\bibnamefont {Strassel}}, \bibinfo {author}
  {\bibfnamefont {L.}~\bibnamefont {Li}}, \bibinfo {author} {\bibfnamefont
  {N.-L.}\ \bibnamefont {Liu}}, \bibinfo {author} {\bibfnamefont
  {B.}~\bibnamefont {Zhao}},\ and\ \bibinfo {author} {\bibfnamefont {J.-W.}\
  \bibnamefont {Pan}},\ }\href {https://doi.org/10.1038/nphys2324} {\bibfield
  {journal} {\bibinfo  {journal} {Nature Physics}\ }\textbf {\bibinfo {volume}
  {8}},\ \bibinfo {pages} {517} (\bibinfo {year} {2012})}\BibitemShut {NoStop}%
\bibitem [{\citenamefont {Willis}\ \emph {et~al.}(2010)\citenamefont {Willis},
  \citenamefont {Becerra}, \citenamefont {Orozco},\ and\ \citenamefont
  {Rolston}}]{Willis2010}%
  \BibitemOpen
  \bibfield  {author} {\bibinfo {author} {\bibfnamefont {R.~T.}\ \bibnamefont
  {Willis}}, \bibinfo {author} {\bibfnamefont {F.~E.}\ \bibnamefont {Becerra}},
  \bibinfo {author} {\bibfnamefont {L.~A.}\ \bibnamefont {Orozco}},\ and\
  \bibinfo {author} {\bibfnamefont {S.~L.}\ \bibnamefont {Rolston}},\ }\href
  {https://doi.org/10.1103/PhysRevA.82.053842} {\bibfield  {journal} {\bibinfo
  {journal} {Phys. Rev. A}\ }\textbf {\bibinfo {volume} {82}},\ \bibinfo
  {pages} {053842} (\bibinfo {year} {2010})}\BibitemShut {NoStop}%
\bibitem [{\citenamefont {Heller}\ \emph {et~al.}(2021)\citenamefont {Heller},
  \citenamefont {Lowinski}, \citenamefont {Theophilo}, \citenamefont
  {Padrón-Brito},\ and\ \citenamefont {de~Riedmatten}}]{Lukas2021}%
  \BibitemOpen
  \bibfield  {author} {\bibinfo {author} {\bibfnamefont {L.}~\bibnamefont
  {Heller}}, \bibinfo {author} {\bibfnamefont {J.}~\bibnamefont {Lowinski}},
  \bibinfo {author} {\bibfnamefont {K.}~\bibnamefont {Theophilo}}, \bibinfo
  {author} {\bibfnamefont {A.}~\bibnamefont {Padrón-Brito}},\ and\ \bibinfo
  {author} {\bibfnamefont {H.}~\bibnamefont {de~Riedmatten}},\ }\href
  {https://doi.org/https://doi.org/10.48550/arXiv.2111.08598} {\bibfield
  {journal} {\bibinfo  {journal} {arXiv:2111.08598}\ } (\bibinfo {year}
  {2021})}\BibitemShut {NoStop}%
\bibitem [{\citenamefont {Buser}\ \emph {et~al.}(2022)\citenamefont {Buser},
  \citenamefont {Mottola}, \citenamefont {Cotting}, \citenamefont {Wolters},\
  and\ \citenamefont {Treutlein}}]{Buser2022}%
  \BibitemOpen
  \bibfield  {author} {\bibinfo {author} {\bibfnamefont {G.}~\bibnamefont
  {Buser}}, \bibinfo {author} {\bibfnamefont {R.}~\bibnamefont {Mottola}},
  \bibinfo {author} {\bibfnamefont {B.}~\bibnamefont {Cotting}}, \bibinfo
  {author} {\bibfnamefont {J.}~\bibnamefont {Wolters}},\ and\ \bibinfo {author}
  {\bibfnamefont {P.}~\bibnamefont {Treutlein}},\ }\href
  {https://doi.org/10.1103/PRXQuantum.3.020349} {\bibfield  {journal} {\bibinfo
   {journal} {PRX Quantum}\ }\textbf {\bibinfo {volume} {3}},\ \bibinfo {pages}
  {020349} (\bibinfo {year} {2022})}\BibitemShut {NoStop}%
\bibitem [{\citenamefont {Seri}\ \emph {et~al.}(2017)\citenamefont {Seri},
  \citenamefont {Lenhard}, \citenamefont {Riel\"ander}, \citenamefont
  {G\"undo\ifmmode~\breve{g}\else \u{g}\fi{}an}, \citenamefont {Ledingham},
  \citenamefont {Mazzera},\ and\ \citenamefont {de~Riedmatten}}]{Seri2017}%
  \BibitemOpen
  \bibfield  {author} {\bibinfo {author} {\bibfnamefont {A.}~\bibnamefont
  {Seri}}, \bibinfo {author} {\bibfnamefont {A.}~\bibnamefont {Lenhard}},
  \bibinfo {author} {\bibfnamefont {D.}~\bibnamefont {Riel\"ander}}, \bibinfo
  {author} {\bibfnamefont {M.}~\bibnamefont {G\"undo\ifmmode~\breve{g}\else
  \u{g}\fi{}an}}, \bibinfo {author} {\bibfnamefont {P.~M.}\ \bibnamefont
  {Ledingham}}, \bibinfo {author} {\bibfnamefont {M.}~\bibnamefont {Mazzera}},\
  and\ \bibinfo {author} {\bibfnamefont {H.}~\bibnamefont {de~Riedmatten}},\
  }\href {https://doi.org/10.1103/PhysRevX.7.021028} {\bibfield  {journal}
  {\bibinfo  {journal} {Phys. Rev. X}\ }\textbf {\bibinfo {volume} {7}},\
  \bibinfo {pages} {021028} (\bibinfo {year} {2017})}\BibitemShut {NoStop}%
\end{thebibliography}%

\end{document}